\documentclass[useAMS,usenatbib]{mn2e}
\usepackage{epsfig}
\usepackage{graphicx}
\usepackage{times}
\usepackage{color}
\usepackage{amssymb}
\usepackage{mathabx}
\def\cm3{cm$^{-3}$}
\def\kms{km~s$^{-1}$}

\def\rsun{R$_{\odot}$}

\def\msun{M$_{\odot}$}

\def\one{{\,\sc i}}

\def\beq{\begin{equation}}
\def\eeq{\end{equation}}

\def\ergs{\,erg\,s$^{-1}$}
\def\foe{10$^{51}$\,erg}

\def\aj{AJ}

\def\apj{ApJ}

\def\apjs{ApJS}
\def\apjl{ApJL}
\def\aap{A\&A}

\def\aaps{A\&AS}
\def\mnras{MNRAS}

\def\solphys{Solar Physics}

\def\cmfflux{{\sc cmf\_flux}}
\def\cmfgen{{\sc cmfgen}}

\def\v1d{{\sc v1d}}

\newcommand{\iso}[2]{\ensuremath{^{#1}\rm{#2}}}

\title[Radiative transfer of SNe IIb/Ib/Ic]{Inferring supernova IIb/Ib/Ic ejecta properties
from light curves and spectra:  Correlations from radiative-transfer models}
\author
[Luc Dessart et al.]
{\vspace{0.3cm} Luc Dessart,$^1$\thanks{email: Luc.Dessart@oca.eu}
D. John Hillier,$^{2}$
Stan Woosley,$^3$
Eli Livne,$^4$
Roni Waldman,$^4$
\newauthor
Sung-Chul Yoon,$^5$
and Norbert Langer.$^6$
\newauthor
\\
$^{1}$:
Laboratoire Lagrange, Universit\'e C\^{o}te d'Azur, Observatoire de la C\^{o}te
d'Azur, CNRS, Boulevard de l'Observatoire, CS 34229, 06304 Nice cedex 4, France. \\
$^2$: Department of Physics and Astronomy \& Pittsburgh Particle Physics,
Astrophysics, and Cosmology Center (PITT PACC),  University of Pittsburgh, \\
3941 O'Hara Street, Pittsburgh, PA 15260, USA. \\
$^3$: Department of Astronomy and Astrophysics, University of California,
Santa Cruz, CA 95064, USA. \\
$^4$: Racah Institute of Physics, The Hebrew University, Jerusalem 91904, Israel. \\
$^5$: Department of Physics and Astronomy, Seoul National University, Gwanak-ro 1, Gwanak-gu, Seoul, 151-742, Republic of Korea \\
$^6$: Argelander-Institut f\"{u}r Astronomie, Universit{\"a}t Bonn, Auf dem H\"{u}gel 71, 53121, Bonn, Germany \\
}

\date{Accepted . Received }

\pagerange{\pageref{firstpage}--\pageref{lastpage}} \pubyear{2014}

\begin{document}

\maketitle

\label{firstpage}

\begin{abstract}
We present 1-D non-Local-Thermodynamic-Equilibrium time-dependent
radiative-transfer simulations for a large grid of supernovae (SNe) IIb/Ib/Ic
that result from the terminal explosion of the mass donor in a close-binary system.
Our sample covers ejecta masses $M_{\rm e}$ of 1.7--5.2\,\msun, kinetic energies $E_{\rm kin}$
of 0.6--5.0$\times$10$^{51}$\,erg, and \iso{56}Ni masses of 0.05--0.30\,\msun.
We find a strong correlation between the \iso{56}Ni mass and the photometric properties at maximum,
and between the rise time to bolometric maximum and the post-maximum decline rate.
We confirm the small scatter in ($V-R$) at 10\,d past $R$-band maximum.
The quantity $V_{\rm m} \equiv \sqrt{2E_{\rm kin}/M_{\rm e}}$ is comparable to the Doppler velocity
measured from He\one\,5875\,\AA\ at maximum in SNe IIb/Ib, although some scatter arises from the uncertain level
of chemical mixing. The O\one\,7772\,\AA\ line may be used for SNe Ic, but the correspondence deteriorates
with higher ejecta mass/energy.
We identify a temporal reversal of the Doppler velocity at maximum absorption in the $\sim$\,1.05\,$\mu$m feature
in all models.
The reversal is due to He\one\ alone and could serve as a test for the presence of helium in SNe Ic.
Because of variations in composition and ionisation, the ejecta opacity shows substantial variations
with both velocity and time. This is in part the origin of the offset between our model light curves
and the predictions from the Arnett model.
\end{abstract}


\section{Introduction}

The origin of supernovae (SNe) IIb/Ib/Ic remains somewhat elusive. While the close-binary evolution
scenario offers an attractive solution to both core-collapse SN statistics
(see, e.g., \citealt{podsiadlowski_92,eldridge_etal_08,smith_etal_11})
and inferred ejecta properties \citep{ensman_woosley_88,woosley_etal_95}, it is not clear today what fraction
arises from the explosion of stars that evolve in isolation.
The diversity of massive close binaries can qualitatively explain the observed diversity of SNe IIb
\citep{claeys+13}, but for moderate main-sequence masses, the binary channel seems to favour
the production of SNe Ib \citep{yoon_etal_10}.
The distinction between SNe Ib and Ic, which is observational
\citep{wheeler_levreault_85,harkness_ib_87,filippenko_87M_90,wheeler_ibc_87}, is challenged by the presence
of broad lines, causing blending/overlap, and the difficulty of exciting He\,\one\ \citep{lucy_91}.
Non-thermal processes and mixing complicates the interpretation of observations \citep{d12_snibc}.
More work is needed to understand these events adequately.
In our approach, we try to address these issues by modelling the SN radiation. Our goal is to complement,
and also to confront to, the independent inferences based on SN-subtype distribution,
host properties (see, e.g., \citealt{anderson_james_08,anderson_james_09,anderson_etal_10,anderson_etal_12,
arcavi+10,modjaz_etal_11,sanders_etal_12,kelly_kirshner_12,crowther_13}),
or pre-SN star properties \citep{yoon_presn_12,eldridge_presn_13,groh_13_presn,kim_snibc_prog}.

Most simulations of SN IIb/Ib/Ic radiation to date have been limited
to grey/multi-frequency radiation hydrodynamics, which delivers bolometric and/or multi-band light curves
(see, e.g., \citealt{blinnikov_94_93j,bersten_etal_12_11dh}),
and to steady-state radiative transfer (see, e.g., \citealt{sauer_etal_06}).
In contrast, our method provides the
emergent flux as a function of wavelength and time by means of a solution to the time-dependent
non-local thermodynamic equilibrium (non-LTE) radiative transfer problem that takes as initial conditions
a physical model of the star and its explosion \citep{dh10_87a,HD12} --- we give a brief summary of our numerical
approach with \cmfgen\ and on the atomic data used in the calculations in Appendix~\ref{sect_cmfgen}.
We can thus attempt to directly link SN signatures to the progenitor structure and
the explosion properties. By treating the problem in non-LTE, we can include the time-dependent
and non-thermal terms that appear in the statistical-equilibrium equations. This is a preriquisite for
the description of H and He, and therefore for the understanding of SNe IIb, Ib, and Ic \citep{dessart_11_wr,d12_snibc}.

\citet{D15_SNIbc_I} presented results for three SN ejecta models stemming from the explosion of the mass donor
in moderate-mass massive stars evolved in a close-binary system with an initial orbital period
of $\approx$\,4\,d \citep{yoon_etal_10}.  These selected SN models were of type IIb, Ib, and Ic and served
to investigate the properties of the radiative transfer in these ejecta. Here, we broaden the scope and consider
the entire grid of 27 models. The numerical approach is described in
\citet{D15_SNIbc_I}. Numerous properties of  these pre-SN models and the corresponding ejecta are provided in
tabulated form in the appendix. Our grid comprises models 3p0, 3p65, 4p64, 5p11, and 6p5 (where the name
refers to the pre-SN mass of the primary star),  which were evolved at a metallicity of 0.02 
(models 3p0, 3p65, 5p11, and 6p5) or 0.004 (model 4p64). The main-sequence masses for these systems were
18\,\msun\ $\oplus$ 17\,\msun\ (model 3p0;  $P_{\rm orb,init}=$\,3\,d),
16\,\msun\ $\oplus$ 14\,\msun\ (model 3p65; $P_{\rm orb,init}=$\,5\,d),
18\,\msun\ $\oplus$ 12\,\msun\ (model 4p64; $P_{\rm orb,init}=$\,5\,d),
60\,\msun\ $\oplus$ 40\,\msun\ (model 5p11; $P_{\rm orb,init}=$\,7\,d),
and 25\,\msun\ $\oplus$ 24\,\msun\ (model 6p5; $P_{\rm orb,init}=$\,6\,d).
Upon reaching iron core collapse, the models were
exploded by means of a piston to produce four different asymptotic ejecta kinetic energies
$E_{\rm kin}$.
We adopt the following nomenclature:
\begin{itemize}
\item Suffix C:  $E_{\rm kin} = 0.6\times\,10^{51}$\,erg. Other models are scaled
in energy by a factor of about 2, 4, and 8.
\item Suffix A:  $E_{\rm kin} = 1.2\times\,10^{51}$\,erg. We take this as the standard
core-collapse SN ejecta kinetic energy at infinity.
\item Suffix B:  $E_{\rm kin} = 2.4\times\,10^{51}$\,erg.
\item Suffix D or G:  $E_{\rm kin} = 5.0\times\,10^{51}$\,erg. The difference between the two
is whether the piston that injects the energy is placed at the edge of the iron core or
where the entropy rises to 4\,k$_{\rm B}$ per baryon (which is further out).
\end{itemize}

The ejecta mass $M_{\rm e}$ for models 3p0, 3p65, 4p64, 5p11, and 6p5
depends on explosion energy (as well as on mixing, but only very slightly) and is 1.71--1.73, 2.18--2.23,
3.11--3.21, 3.54--3.63, and 4.95--5.18\,\msun, respectively.
All these simulations leave behind a neutron star, with a mass in the range 1.27--1.57\,\msun\
(see Table~\ref{tab_ejecta_glob_appendix}).

\begin{figure}
\epsfig{file=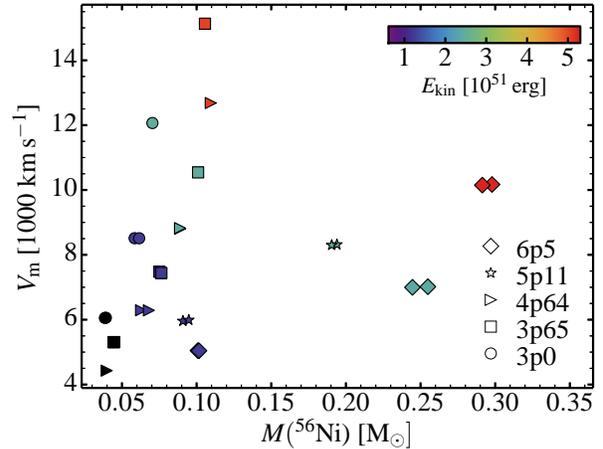,width=8.5cm}
\caption{Distribution of $M$(\iso{56}Ni), ejecta kinetic energy $E_{\rm kin}$, and expansion rate
$V_{\rm m}\equiv \sqrt{2E_{\rm kin}/M_{\rm e}}$ for our grid of  models (pre-SN masses
are shown with a different symbol).
For a given progenitor, both the expansion rate $V_{\rm m}$ and the \iso{56}Ni mass increase
with explosion energy.
\label{fig_presn}}
\end{figure}

Figure~\ref{fig_presn} shows the distribution of  $M$(\iso{56}Ni),
``representative" expansion rate
$V_{\rm m}\equiv \sqrt{2E_{\rm kin}/M_{\rm e}}$,  $E_{\rm kin}$, and pre-SN mass for the full grid of models.
The \iso{56}Ni mass lies between 0.05 and 0.3\,\msun, although most of our simulations produce $\lesssim$\,0.1\,\msun.
While the kinetic energy is specified by the user as a parameter of the explosion simulation, the \iso{56}Ni
mass is a byproduct of the explosion, controlled physically by the explosion power and energy as well
as the progenitor core structure and composition.
For a given progenitor model, the larger the explosion energy, the larger the \iso{56}Ni mass.
If we consider the full model set, the maximum \iso{56}Ni yield tends to increase for larger mass progenitors because
they have a larger density above the iron core.
Finally, to account for multi-dimensional effects associated with the explosion mechanism
\citep{fryxell_etal_91,wongwathanarat_15_3d},
we enforce two levels of mixing in these models, one moderate (suffix x1) and one strong (suffix x2) --- see
\citet{D15_SNIbc_I} for details.

The SN type associated with each model was discussed in \citet{D15_SNIbc_I}.
Ejecta models 3p0, 3p65, 4p64 contain some residual hydrogen in the outermost parts
and helium represents at least 50\% of their composition. These models make a type IIb
for all explosion energies and mixing values used here.
Ejecta model 6p5 is hydrogen deficient and  helium represents $\approx$\,35\% of its composition.
This model makes a type Ib for all explosion energies and mixing values used here.
Because it is hydrogen deficient and poor in helium, ejecta model 5p11 makes a type Ic
for all explosion energies and mixing values used here.
Here, we do not yet discuss the suitability of these models to match SNe IIb/Ib/Ic observations.

We select this grid of models so that we encompass a range of mass and composition.
We adopt four different explosion energies to cover a range around the
representative core-collapse SN value of 10$^{51}$\,erg.
This ignores the probable correlation between explosion energy and progenitor mass/structure ---
modelling of the neutrino-driven explosion is necessary to produce a more physically consistent ejecta model
(see, e.g., \citealt{tuguldur_ccsn_15}).
In our model set, lower explosion energies are probably more suitable for the lower mass progenitors;
higher explosion energies may not explode by neutrino power alone but may require some other mechanism,
perhaps related to core rotation \citep{ugliano_ccsn_12}.
The correlations we extract from our
results obviously reflect the properties for our set of progenitor/explosion models and should therefore be
considered as such. In Nature, SNe Ibc may be associated with lower/higher ejecta masses and energies,
different levels of mixing, or may stem from binary massive stars that evolved differently from the main
sequence (through variations in mass loss rates or initial rotation, angular momentum transport etc.). The hope
with this model sample and extracted trends is to provide a framework to interpret observations.

To complement the previous study of \citet{D15_SNIbc_I},
we investigate the trends  that emerge from our entire grid of models, in particular the correlations
arising from variations in ejecta kinetic energy $E_{\rm kin}$, ejecta mass $M_{\rm e}$, \iso{56}Ni mass,
and progenitor composition.
In turn, we discuss our results for the bolometric luminosity light curves (Section~\ref{sect_lbol}),
the multi-band light curves (Section~\ref{sect_phot}),
the colour evolution  (Section~\ref{sect_colour}),
and some spectral properties (Section~\ref{sect_spec}).
We then confront our results to other works, and in particular discuss the shortcomings of the Arnett model
for SNe Ib/c (Section~\ref{sect_disc}).
Finally, we present our conclusions (Section~\ref{sect_conc}).

\begin{figure}
\epsfig{file=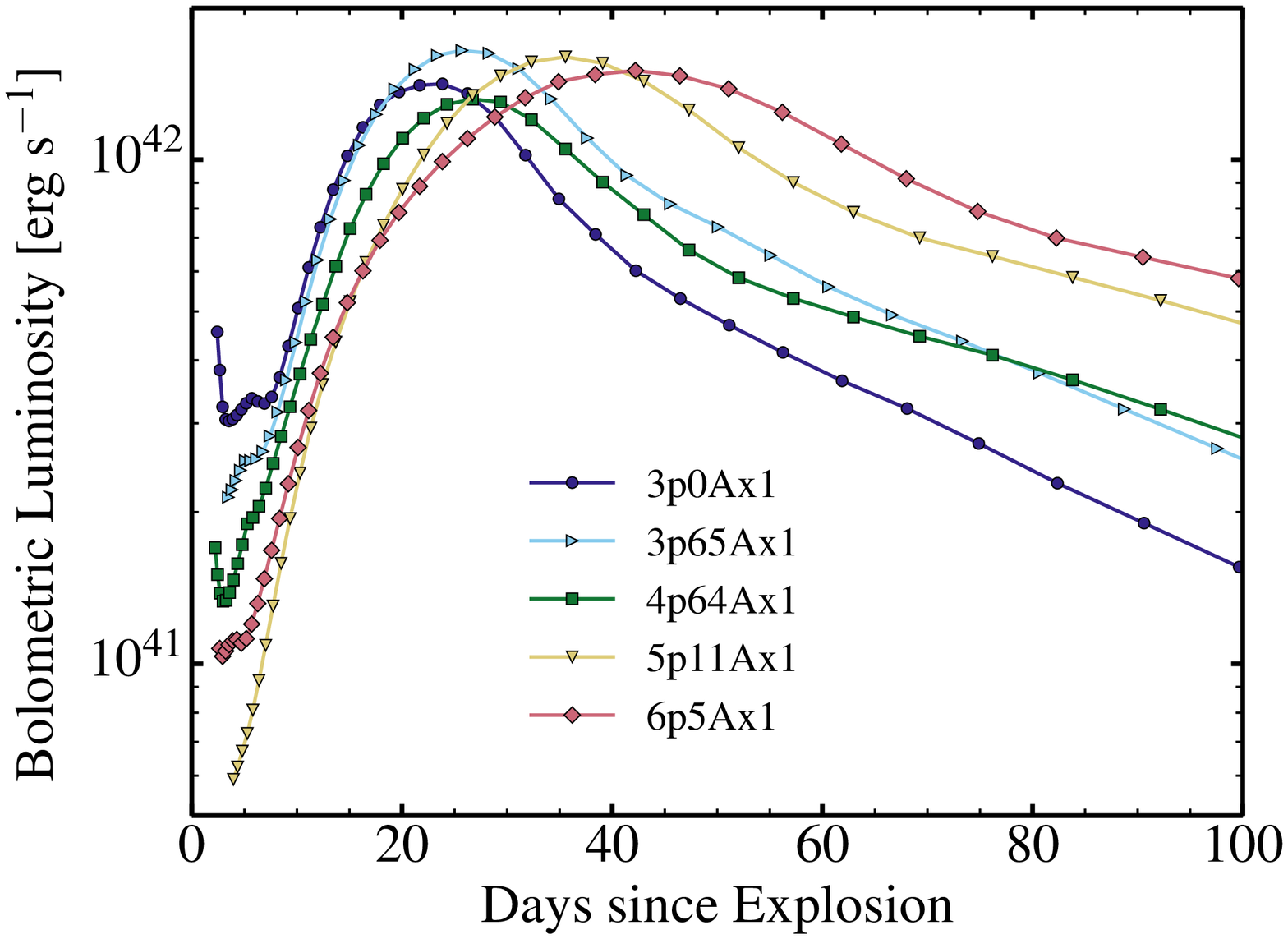,width=8.5cm}
\epsfig{file=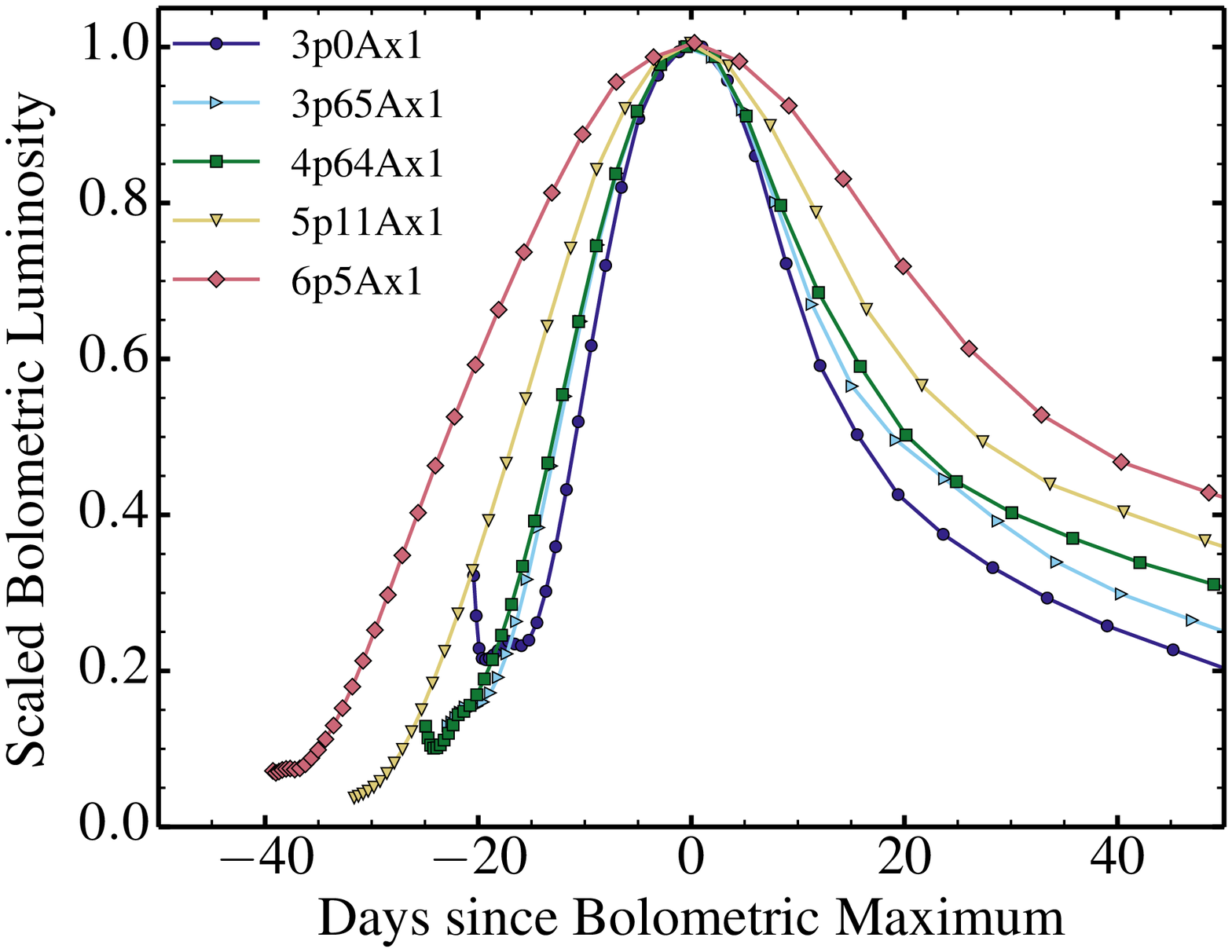,width=8.5cm}
\epsfig{file=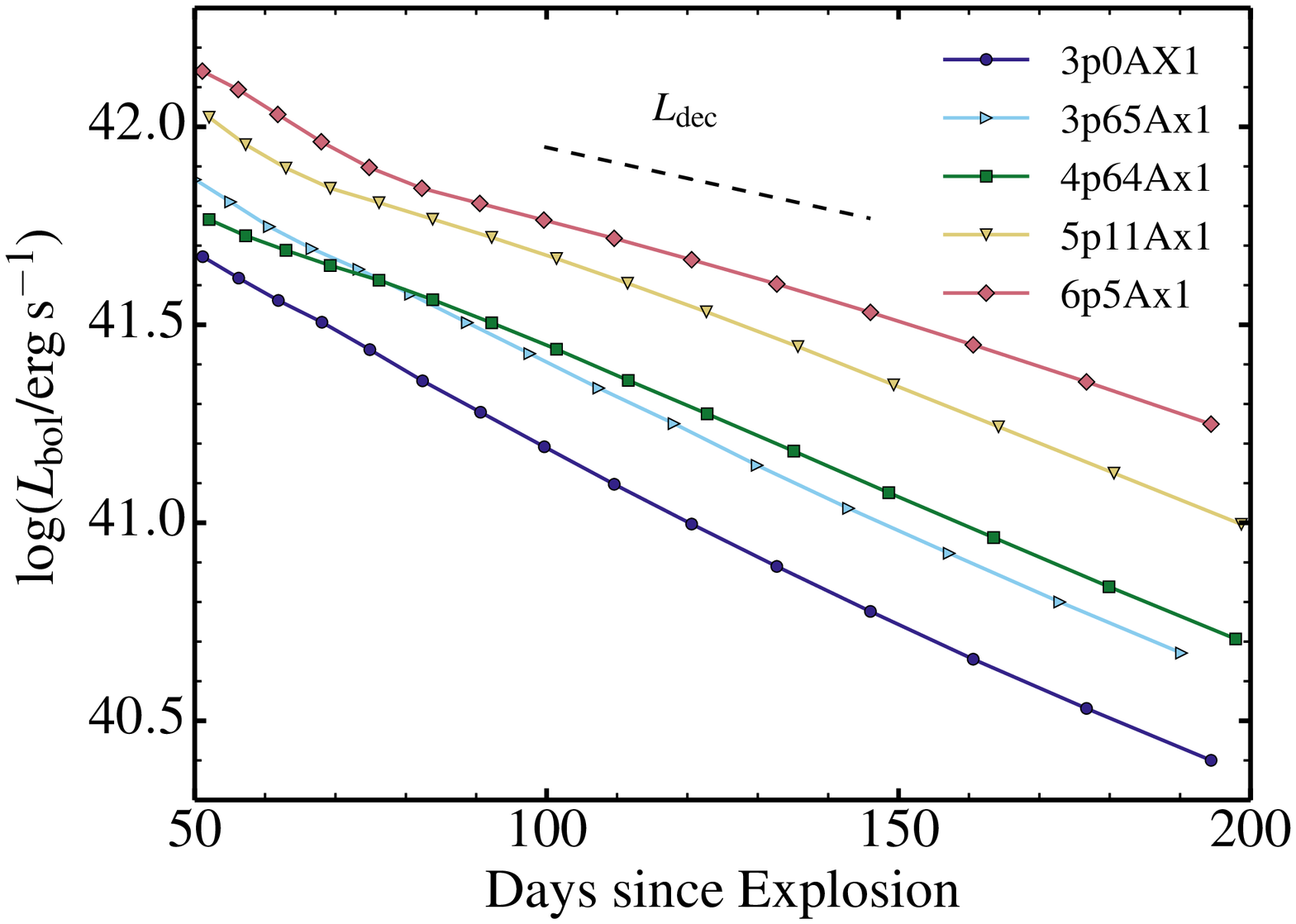,width=8.5cm}
\caption{
{\it Top:} Illustration of the bolometric light curves for the models characterised by
a 1.2$\times$\foe\ explosion energy but covering a range of ejecta masses, from 1.73 (model 3p0Ax1)
to 2.23 (model 3p65Ax1), 3.11 (model 4p64Ax1), 3.54 (model 5p11Ax1),
and 4.97\,\msun\ (model 6p5Ax1).
{\it Middle:} Same as top, but with respect to the time of bolometric maximum and
with all light curves normalised to the value at maximum.
{\it Bottom:} Same as top, but with respect to the time of bolometric maximum and
limited to the late time evolution. We overlay the power $L_{\rm dec}$ corresponding to
the decay of 0.15\,\msun\ of \iso{56}Ni (dashed line).
The larger the mass, the longer the rise to maximum, the
smaller the post-maximum brightness decline rate, and the broader the light curve around bolometric maximum.
\label{fig_lbol_A}}
\end{figure}

\begin{figure}
\epsfig{file=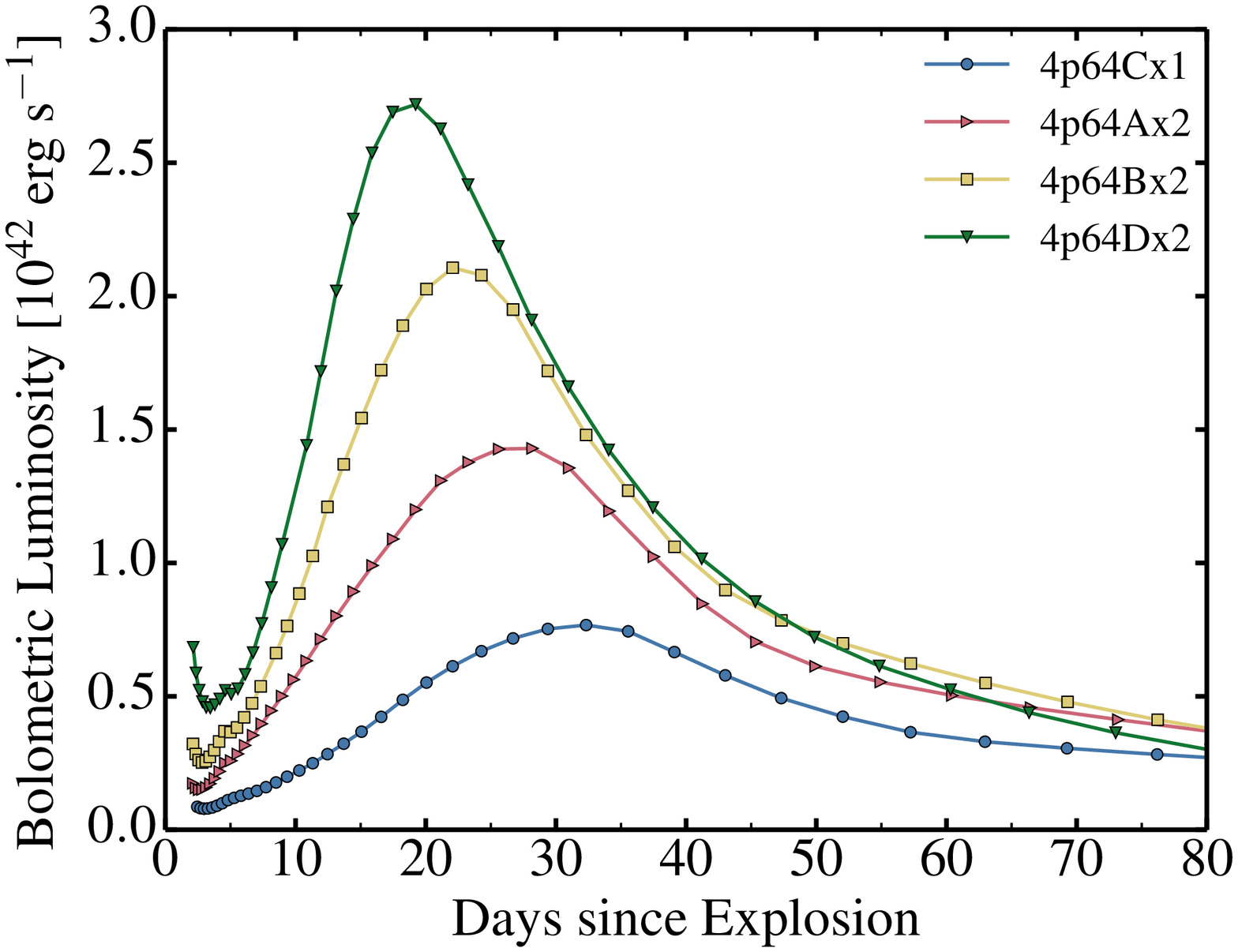,width=8.5cm}
\epsfig{file=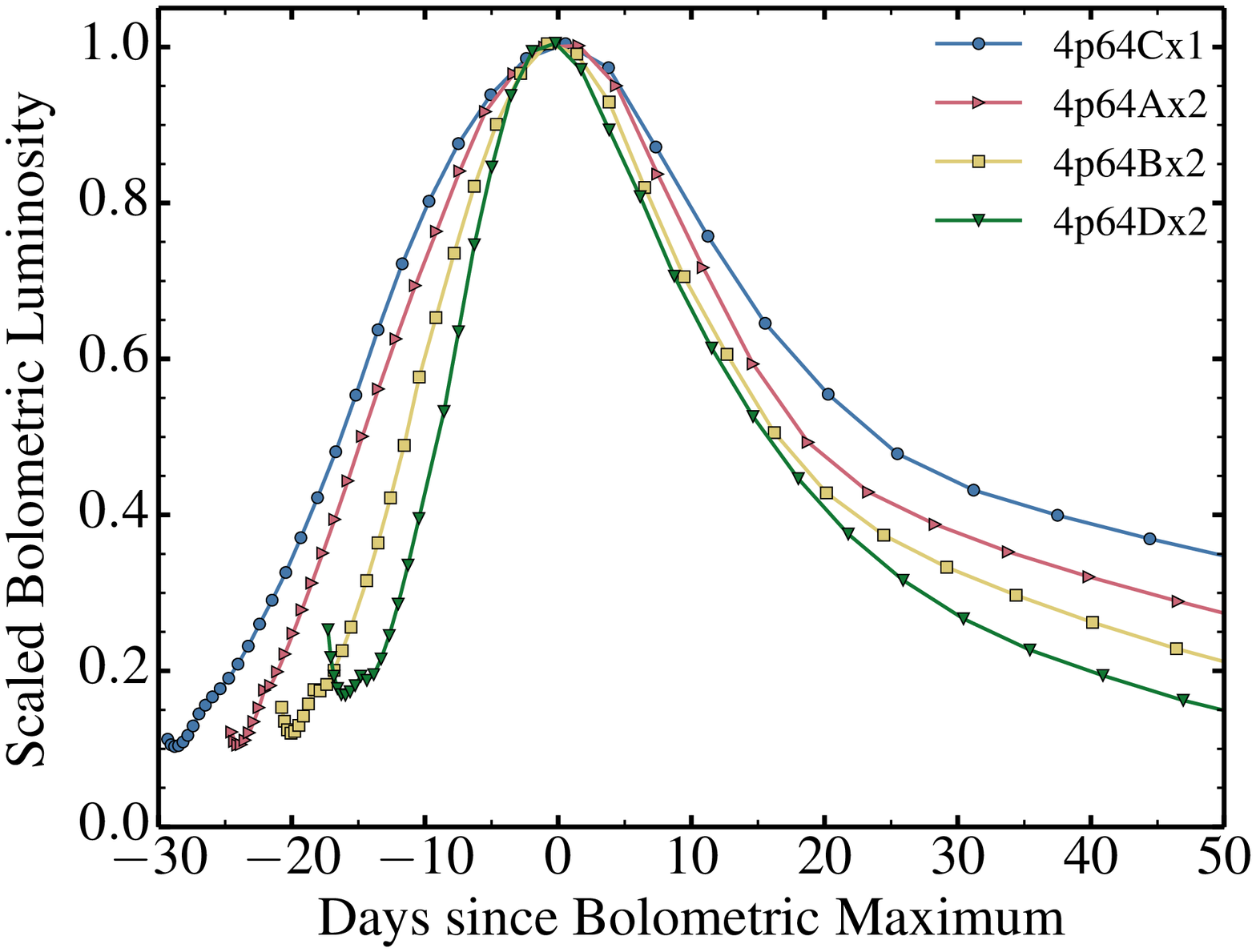,width=8.5cm}
\epsfig{file=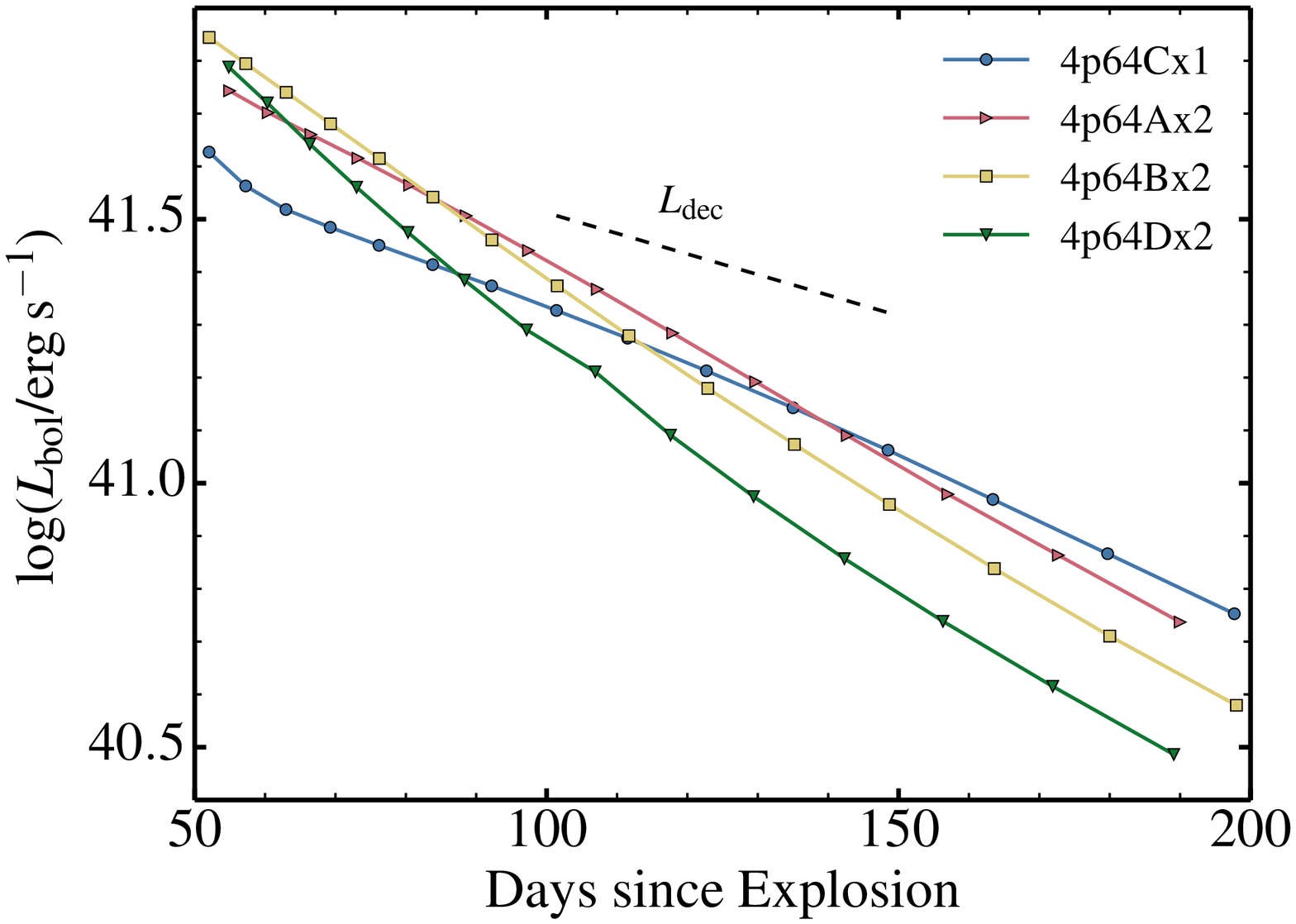,width=8.5cm}
\caption{
Same as Figure~\ref{fig_lbol_A}, but now for the 4p64 models characterised by
different explosion energies (in our nomenclature, $E_{\rm kin}$ grows from
0.6, to 1.2, 2.4, and 5.0$\times$\foe\ as we step through models C, A, B, and D).
The dashed line (bottom panel) corresponds to the decay power of 0.055\,\msun\ of \iso{56}Ni.
Higher explosion energies shorten the rise time to bolometric maximum, increase the post-maximum luminosity decline rate,
and produce a narrower light curve around bolometric maximum.
In these models, higher explosion energies tend to correlate with the \iso{56}Ni mass (Figure~\ref{fig_presn}),
which exacerbates the trend seen in the top panel (i.e. higher peak luminosity for higher ejecta kinetic energy).
\label{fig_lbol_4p64}}
\end{figure}

\section{Bolometric properties}
\label{sect_lbol}

  Figure~\ref{fig_lbol_A} shows the bolometric light curves for models that share a common
  ejecta kinetic energy of $\approx$\,1.2$\times$\foe, but cover a range of ejecta masses from 1.73 to 4.97\,\msun.
  Over that range, the rise time to maximum increases monotonically from $\approx$\,23 to $\approx$\,42\,d.
  The bolometric luminosity at maximum is comparable between models, within the range 1.32--1.65$\times$\,10$^{42}$\ergs,
  and  does not vary monotonically with pre-SN mass.
  The non-monotonic behaviour arises because the different models have different \iso{56}Ni masses
  (the range is from 0.058 to 0.099\,\msun),
  and because the lower the ejecta mass, the greater the bolometric maximum (all else being the same).
  Finally, the post-maximum decline decreases steadily as the ejecta mass increases.
  The bolometric magnitude drop between maximum and 15\,d later decreases from 0.72 to 0.22\,mag
  from model 3p0 to 6p5. This brightness decline correlates with the width of the light curve, which appears
  very symmetric when plotted with respect to the time of maximum (middle panel of  Figure~\ref{fig_lbol_A}).

  This trend persists to later times, but it then has a different origin. During the photospheric phase, the width
  of the light curve is controlled by the trapping of radiation energy, which is stored in optical/UV photons.
  At nebular times, the decline rate is controlled by the trapping of $\gamma$-rays from radioactive decay
  (bottom panel of  Figure~\ref{fig_lbol_A}). The opacity affecting low and high energy photons is
  fundamentally distinct. In particular, the total opacity to low-energy photons is strongly dependent on
  ionisation, while the $\gamma$-ray opacity is primarily sensitive to the total number of electrons.

  Figure~\ref{fig_lbol_4p64} shows the
  diversity of light curves for a given pre-SN model exploded with different energies.
  As the ejecta kinetic energy increases from 0.62 to 1.22, 2.46, and 5.13$\times$\foe, the rise
  time to bolometric maximum decreases from 31.8 to 19.4\,d, and the post-maximum decline
  increases from 0.45 to 0.72\,mag. The luminosity at bolometric maximum is larger for
  models with a higher ejecta kinetic energy, a feature exacerbated by the shorter rise time and
  the larger \iso{56}Ni mass (a factor of about two between models 4p64Cx1 and 4p64Dx2).
  An increase in ejecta mass or a decrease in ejecta kinetic energy has a comparable impact
  on the light curve width (Figures~\ref{fig_lbol_A} and \ref{fig_lbol_4p64}).

   In most of our models, the early light curve comprises a short post-breakout ``plateau"
prior to the rise to maximum (see also \citealt{dessart_11_wr}). This post-breakout
plateau is brighter for larger progenitor radii and explosion energy (see also \citealt{bersten_etal_12_11dh}).
It is shorter for lower mass ejecta (and enhanced \iso{56}Ni mixing; \citealt{d12_snibc}).
These various properties are function of the relative contributions of the shock-deposited and decay energies,
and how these energy sources are distributed within the ejecta.

\begin{figure}
\epsfig{file=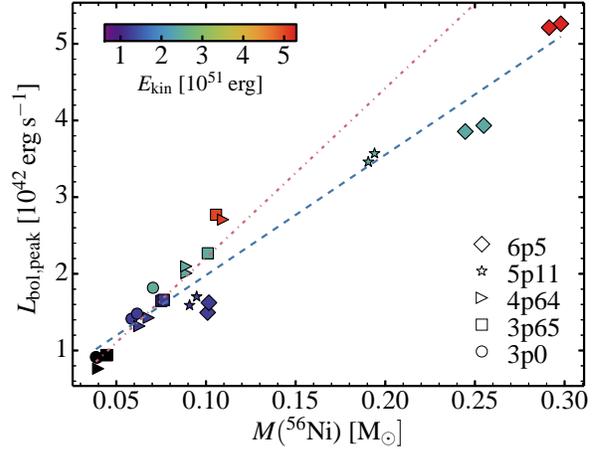,width=8.5cm}
\caption{Maximum bolometric luminosity versus \iso{56}Ni mass for our grid of models.
Symbols correspond to different pre-SN progenitor mass. The colour coding refers to
the model kinetic energy.
The dashed line corresponds to $L_{\rm bol, peak}/10^{42}\,{\rm erg}\,{\rm s}^{-1} = 15.718 (M(^{56}{\rm Ni})/{\rm M}_\odot) +0.41$.
The dash-dotted curve is the $L_{\rm bol, peak}$ versus $M(^{56}{\rm Ni})/{\rm M}_\odot$ relation
for the SN Ia models of \citet{blondin_etal_13} --- the minimum \iso{56}Ni mass for that SN Ia model set
is actually 0.18\,\msun.
\label{fig_lbol_ni56}}
\end{figure}

Figure~\ref{fig_lbol_ni56} shows the distribution of the luminosity at bolometric maximum versus
\iso{56}Ni for the entire grid of models. This figure shows that \iso{56}Ni is, as expected, the key power source
behind these SN models. For example, the doubling of the \iso{56}Ni mass leads to roughly the doubling
of the peak luminosity. However, there is some scatter, which arises from the relatively large range of ejecta masses.
For a given \iso{56}Ni mass and ejecta kinetic energy, the larger mass ejecta appear under-luminous at maximum
because they radiate essentially the same total decay energy but over a longer time.
A least-square polynomial fit to the distribution of $N$ model points gives
 $$L_{\rm bol, peak}/10^{42}\,{\rm erg}\,{\rm s}^{-1} = 15.718 (M(^{56}{\rm Ni})/{\rm M}_\odot) +0.41 \, .$$
We approximate the dispersion by the quantity
 $ \sqrt{\left[   \sum_i \left(  Y_{i, \rm model} - Y_{i, \rm fit}     \right)^2 \right]/ (N-2) } $,
 where $i$ runs from 1 to $N$ and $Y_i$ is $L_{\rm bol, peak}$ for model $i$.
 Here, this dispersion is 0.30$\times$\,10$^{42}$\,\ergs.
 This relation is significantly flatter than for SNe Ia (see, e.g., \citealt{blondin_etal_13}), probably because
 of the larger range of ejecta masses for our SNe IIb/Ib/Ic models.

\section{Photometric properties}
\label{sect_phot}

The photometric properties discussed in \citet{D15_SNIbc_I} are supported by the larger grid of models
and so not all properties will be discussed again. We focus on the properties of the $R$ band,
which we find to be analogous to that of the bolometric luminosity
(Tables~\ref{tab_lc_mod1_appendix}-\ref{tab_lc_mod3_appendix}).

\begin{figure}
\epsfig{file=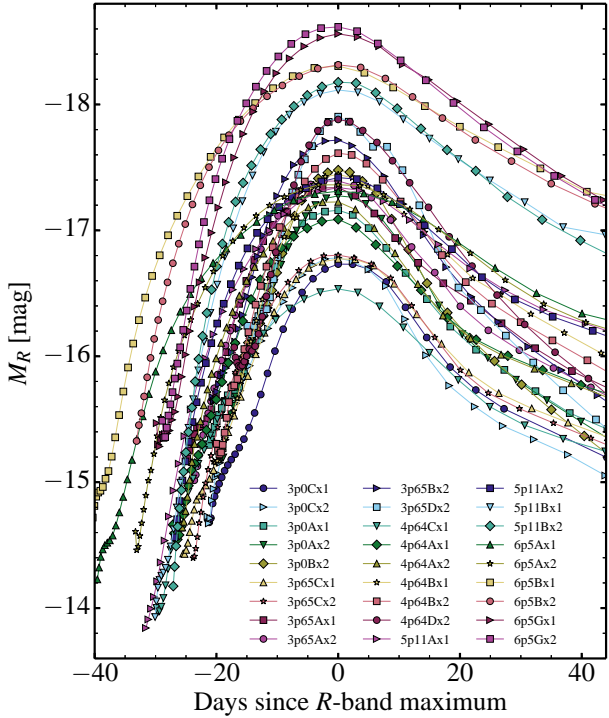,width=8.5cm}
\vspace{-0.8cm}
\caption{$R$-band light curves with respect to the time of $R$-band maximum
for the full set of models.
\label{fig_lc_r}
}
\end{figure}

Figure~\ref{fig_lc_r} shows the whole set of $R$-band light curves around maximum.
The origin of the $x$ axis is the time of $R$-band maximum. This reduces the strong overlap
caused by the wide range in rise time. It also better reveals the scatter in
light curve peaks and widths.
One notable difference with the bolometric light curves is the lack of an obvious
post-breakout plateau in the $R$ band.

\begin{figure}
\epsfig{file=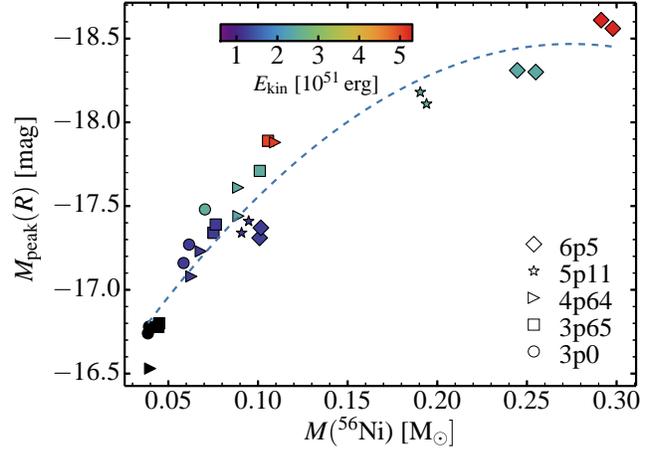,width=8.5cm}
\caption{Peak $R$-band magnitude versus \iso{56}Ni mass for our grid of models.
The dashed line corresponds to $(M_{\rm peak}(R)/{\rm mag}) = -16.21-16.44\,M(^{56}{\rm Ni})/{\rm M}_\odot) + 29.93 (M(^{56}{\rm Ni})/{\rm M}_\odot)^2$
\label{fig_maxr_ni56}}
\end{figure}

\begin{figure}
\epsfig{file=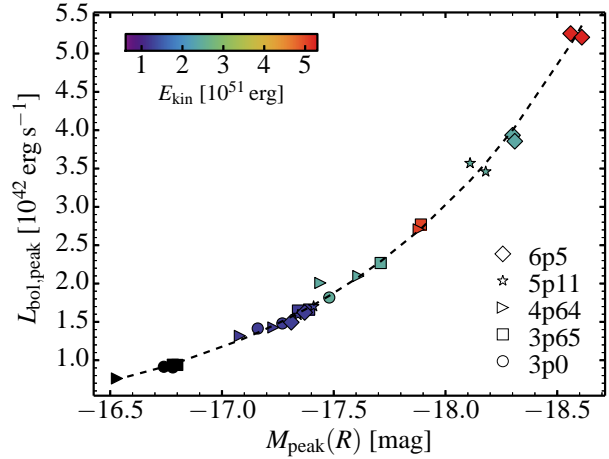,width=8.5cm}
\caption{Maximum bolometric luminosity versus maximum $R$-band magnitude for our grid of models.
The dashed curve corresponds to $\log (L_{\rm bol, peak}/10^{42}\,{\rm erg}\,{\rm s}^{-1}) = -0.41 (M_{\rm peak}(R)/{\rm mag}) -6.92$.
\label{fig_maxr_lbol}}
\end{figure}

As for the bolometric luminosity, the maximum $R$-band magnitude correlates
with the \iso{56}Ni mass.
The correlation is strong at low \iso{56}Ni mass
and flattens out at large \iso{56}Ni mass (Figure~\ref{fig_maxr_ni56}).
A least-square polynomial fit that gives a rough match to the distribution
of model results is
$$\frac{M_{\rm peak}(R)}{{\rm mag}} = -16.21-16.44\,\frac{M(^{56}{\rm Ni})}{{\rm M}_\odot}
+ 29.93 \big(\frac{M(^{56}{\rm Ni})}{{\rm M}_\odot}\big)^2 \, ,$$
with a dispersion of 0.16\,mag.
At large \iso{56}Ni mass, the fitted curve starts declining, which is unphysical, so this and other fitted formula ought
to be used with circumspection.
Since the $R$-band magnitude can be more easily inferred from observations than the bolometric luminosity,
we also show how the two compare  in Figure~\ref{fig_maxr_lbol}.
The relation that closely holds between the two is
$$\log (L_{\rm bol, peak}/10^{42}\,{\rm erg}\,{\rm s}^{-1}) = -0.41 (M_{\rm peak}(R)/{\rm mag}) -6.92 \, ,$$
with a dispersion of 0.017 (in the log) --- corresponding to a dispersion of 3-4\% in  $L_{\rm bol, peak}$.

\begin{figure}
\epsfig{file=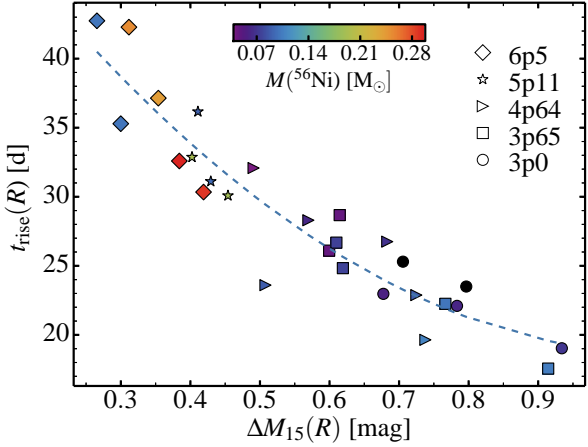,width=8.5cm}
\caption{Rise time to bolometric maximum versus post-maximum decline rate $\Delta M_{15}(R)$.
for our grid of models.
The dashed curve corresponds to $t_{\rm rise}/{\rm d} = 57.08 -71.17 \Delta M_{15}(R) + 32.98\Delta M_{15}^2(R) $
\label{fig_dm15_trise}}
\end{figure}

Figure~\ref{fig_dm15_trise} shows that there is a very strong correlation between the
post-maximum decline rate $\Delta M_{15}(R)$ and the rise time $t_{\rm rise}(R)$ to $R$-band maximum.
Using a least-square polynomial fit, our results follow closely the relation
$$t_{\rm rise}/{\rm d} = 57.08 -71.17 \Delta M_{15}(R) + 32.98\Delta M_{15}^2(R) \, ,$$
with a dispersion of 2.36\,d.
This relation holds across a wide range of \iso{56}Ni mass, ejecta masses, and ejecta kinetic energies.
It simply reflects the fact that the brightening rate to maximum is comparable to the early decline rate after maximum,
a property that is expected in the diffusion regime controlling the light curve.
This relation may help constrain the explosion time of observed SNe IIb/Ib/Ic, especially those lacking early time
observations.

\begin{figure}
\epsfig{file=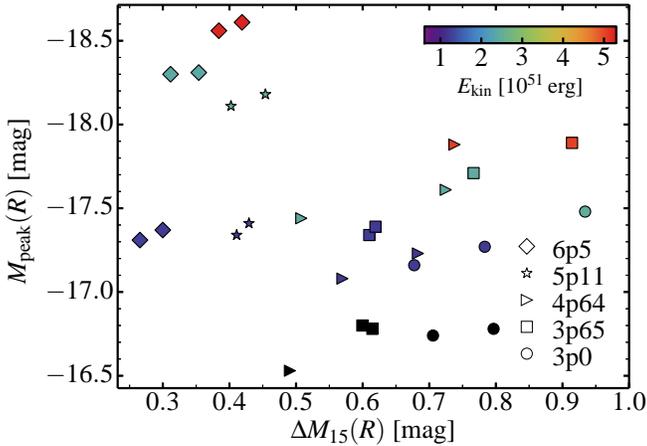,width=8.5cm}
\caption{Distribution of $R$-band peak magnitude versus post-maximum decline $\Delta M_{15}(R)$.
The large scatter stems from the complicated dependence of $\Delta M_{15}(R)$ on
$E_{\rm kin}$, $M_{\rm e}$ and \iso{56}Ni mass.
\label{fig_dm15_maxr}}
\end{figure}

The distribution of $\Delta M_{15}(R)$ and
$M_{\rm peak}(R)$ values shows a large scatter (Figure~\ref{fig_dm15_maxr}) --- the same holds
if we compare to $M_{\rm peak}(V)$ or $M_{\rm peak}({\rm bol})$.
The peak magnitude is primarily controlled by the \iso{56}Ni mass, while the post-maximum decline
is sensitive to $E_{\rm kin}$, $M_{\rm e}$, as well as \iso{56}Ni mass.
An example of this complicated sensitivity is that the decay energy tends to raise, or at least maintain,
the ionisation and therefore influences the ejecta optical depth. This matters throughout the photospheric
phase, which lasts for up to about 2--3 weeks after maximum.

\section{Colour properties}
\label{sect_colour}

Photometric variations can arise from colour changes.
These changes are, however, quite modest in our simulations because the emergent radiation
falls primarily within the optical range \citep{D15_SNIbc_I}.
The pre-SN radius of our models is below $\sim$\,10\,\rsun, which causes a significant cooling
when the ejecta expands to a SN-like radius in the first week after explosion. Consequently, none
of our models appear blue early on. Instead, they show relatively red colours throughout their evolution,
with only modest variations through the early post-breakout plateau, the rise to maximum, and the
post-maximum phase.

\begin{figure}
\epsfig{file=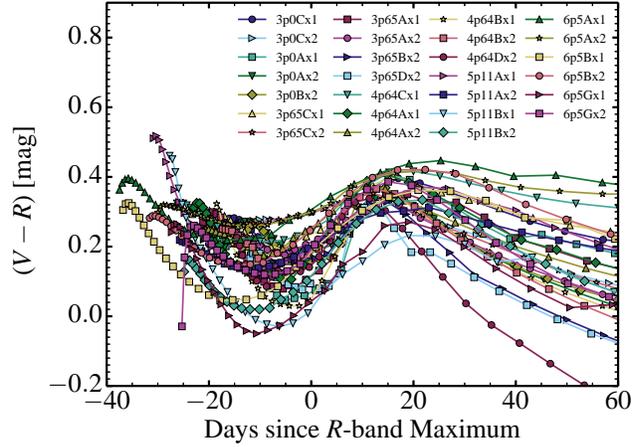,width=8.5cm}
\caption{$V-R$ colour light curve with respect to $R$-band maximum for the full set of models.
\label{fig_vmr}}
\end{figure}

Figure~\ref{fig_vmr} shows the evolution of the ($V-R$) light curve for the whole set of models.
There is a first phase of reddening prior to \iso{56}Ni decay heating in the spectrum formation
region, followed by a hardening on the rise to maximum when \iso{56}Ni decay heating is
strong, followed by a reddening phase as heating ebbs and line blanketing from metals strengthens.

\begin{figure}
\epsfig{file=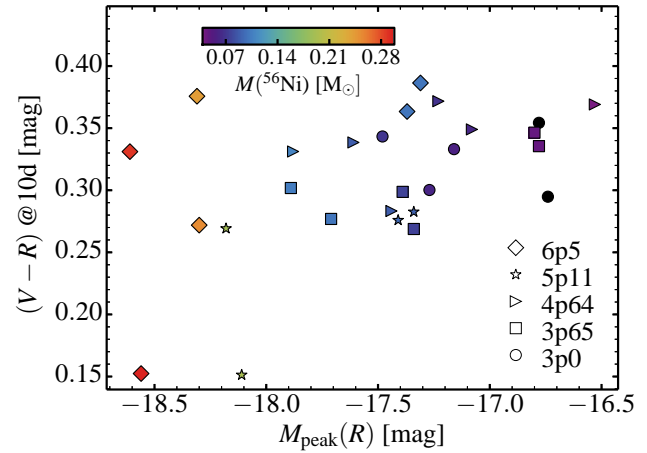,width=8.5cm}
\caption{Plot emphasises the degeneracy of ($V-R$) colour at 10\,d after $R$-band maximum --- exceptions
to the trend are the high mass high explosion energy moderately-mixed models 6p5Gx1 and 5p11Bx1.
\label{fig_maxr_vmr}}
\end{figure}

The ($V-R$) light curve shows a narrow range of values at 10\,d after $R$-band maximum (Figure~\ref{fig_maxr_vmr}).
This property was identified in observations by \citet{drout_etal_11} (see also \citealt{bianco_etal_14}).
In our models, the spectrum forms in the inner ejecta at this epoch.
It is influenced by a comparable decay-heating rate, and the composition is similar in all models, with a dominance of C and O
at the electron-scattering photosphere at that time.
This uniformity of photospheric properties between models is likely responsible for the degeneracy
in the ($V-R$) colour early after peak.

There are two outliers with a bluer colour than the rest of the sample.
These models correspond to higher-mass higher-energy ejecta with a high \iso{56}Ni mass
and characterised by a weaker mixing.
Weaker mixing favours higher temperatures in the
inner ejecta, causing redder colours early on, but bluer colours around maximum
(as discussed in \citet{d12_snibc}, weaker mixing has numerous other implications,
visible in the bolometric light curve, in the multi-band light curves, and in the spectra).
For the full grid of models, the ($V-R$) colour at 10\,d after $R$-band maximum has a mean value of
0.319\,mag and a standard deviation of 0.053\,mag. Taking out the two outliers from the set, the mean
($V-R$) colour value is 0.33\,mag and the standard deviation is 0.035\,mag.

\section{Spectral properties}
\label{sect_spec}

With about 30 model sequences, each containing about 50 time steps, we have a total of 1500 individual
spectra. The amount of information contained in such a set of spectra is very large.
The salient signatures for models of type IIb, Ib, and Ic were discussed in \citet{D15_SNIbc_I}.
In this section, we discuss a few important results that emerge from the whole set.

\subsection{H$\alpha$}

\begin{figure}
\epsfig{file=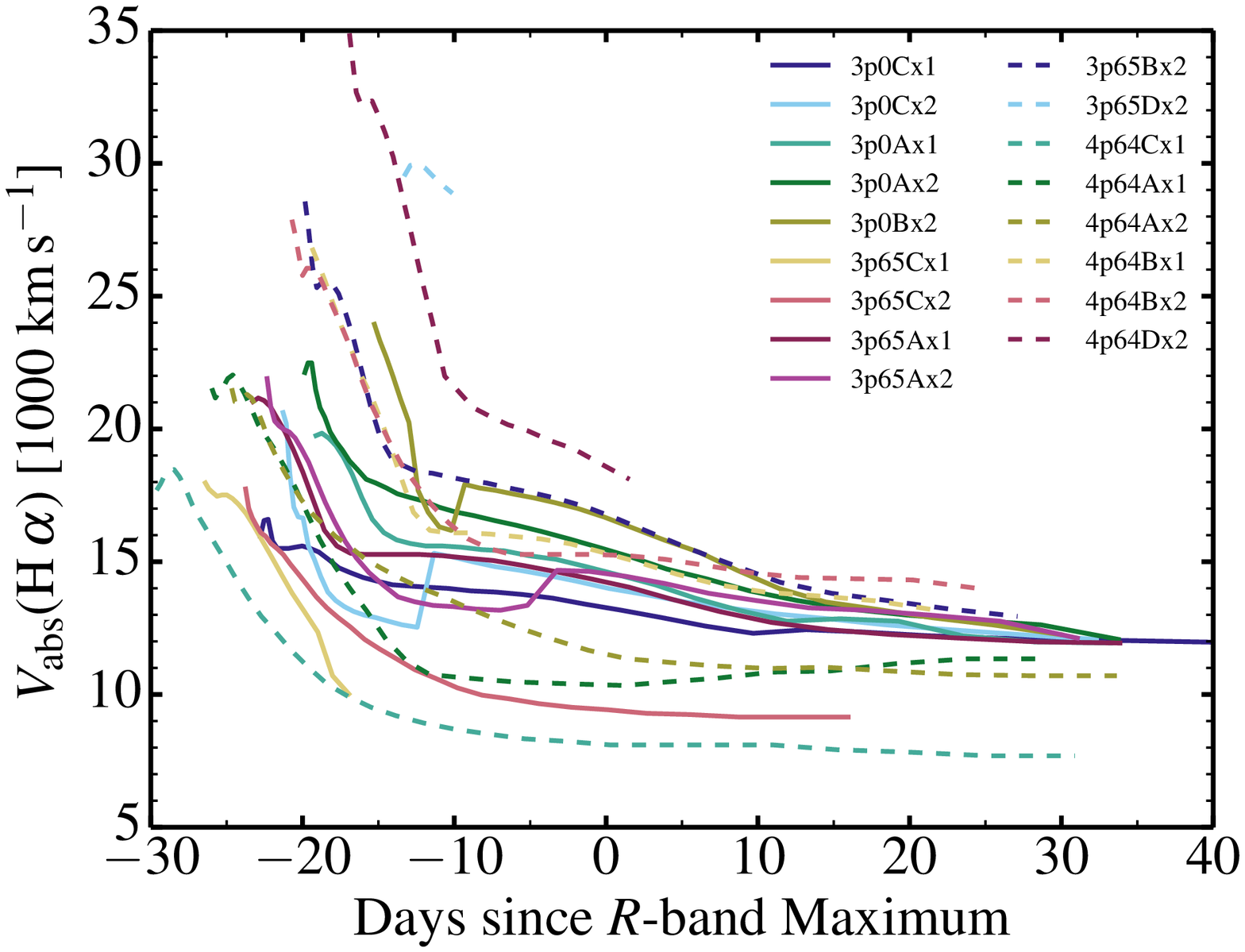,width=8.5cm}
\epsfig{file=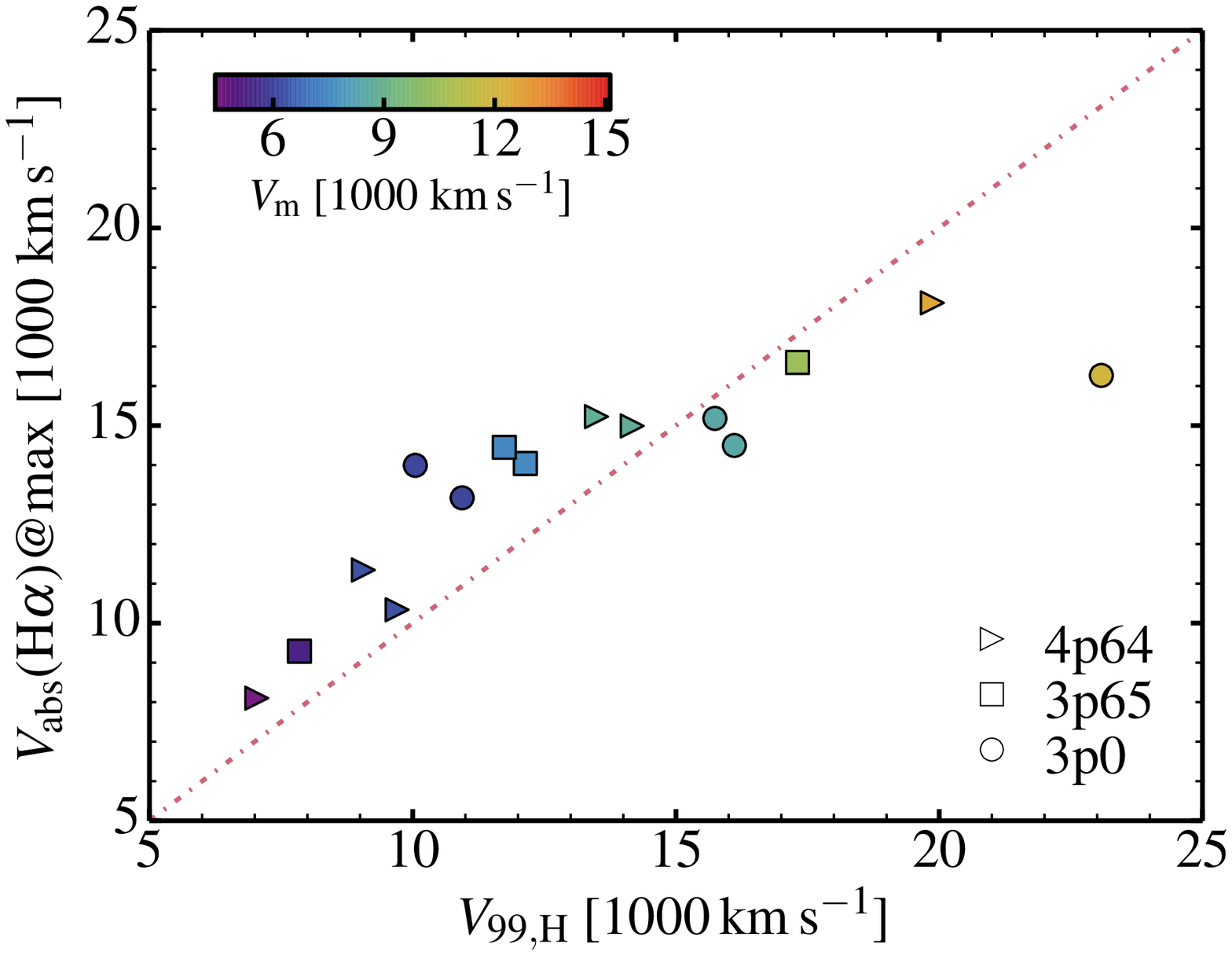,width=8.5cm}
\caption{{\it Top:} Evolution of the Doppler velocity associated with the maximum absorption in
H$\alpha$, noted $V_{\rm abs}$(H$\alpha$), for the SNe IIb models in our sample.
{\it Bottom:} Comparison of $V_{\rm abs}$(H$\alpha$) at $R$-band maximum versus
the inner velocity of the shell that contains 99\% of the total hydrogen mass, noted
 $V_{\rm 99}$(H) (integration done inwards from the outermost ejecta location).
 If both quantities were equal, they would lie along the dash-dotted line.
\label{fig_vha}
}
\end{figure}

    In our SN IIb models (3p0, 3p65, and 4p64), H$\alpha$ is present as a strong line at
 early times. Its strength decreases with time, such that
 by the time of maximum, the line may only be visible as a weak absorption.
 The top panel of Figure~\ref{fig_vha} shows the evolution of the Doppler velocity at maximum absorption
 in H$\alpha$ for our SNe IIb models. In some high-energy explosion models, we lose
 track of the feature after 1-2 weeks, but in most cases, we can follow $V_{\rm abs}$(H$\alpha$)
 from early times when it is very large, until light curve maximum when its evolution tends to flatten.
 This asymptotic velocity, which is a factor of $\approx$\,2 smaller than the initial values, falls in the range
 8000-15000\,\kms, and corresponds to the velocity at the base of the shell that
 contains hydrogen --- a smaller asymptotic velocity suggests a larger mass for the H-rich shell
 (bottom panel of Figure~\ref{fig_vha} and Tables~\ref{tab_ejecta_glob_appendix}--~\ref{tab_ejecta_mass_appendix}).
 Lower values are not possible since the deeper layers are hydrogen deficient.
 In SNe IIb, one can therefore search for that stationary notch to locate the
 minimum velocity of the hydrogen-rich shell in the ejecta.
 Interestingly, this property of SN IIb spectra is the only unambiguous, and therefore robust,
 signature of chemical stratification in SNe IIb/Ib/Ic. It is also observed \citep{liu_snibc_15}.

\subsection{He\one\ lines}

\begin{figure}
\epsfig{file=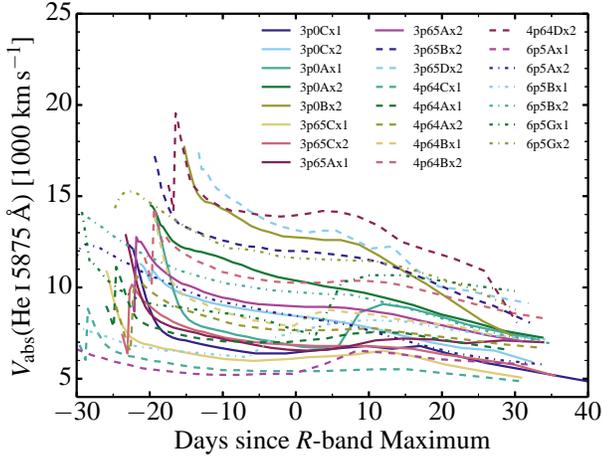,width=8.5cm}
\caption{Evolution of the Doppler velocity associated with the maximum absorption in He\one\,5875\,\AA\
for the SNe IIb and Ib models in our sample.
\label{fig_vhei_5875}
}
\end{figure}

Figure~\ref{fig_vhei_5875} shows the evolution of  the Doppler velocity
at maximum absorption in He\one\,5875\,\AA, noted $V_{\rm abs}$(He\one\,5875\,\AA),
for the SNe IIb/Ib in our model set. Qualitatively, the evolution is similar to that of
H$\alpha$, but shifted to lower values. This arises from the fact that in the corresponding
ejecta models, helium is more abundant at depth, where the density is larger, so the
maximum line optical depth is reached at lower velocity.

\begin{figure}
\epsfig{file=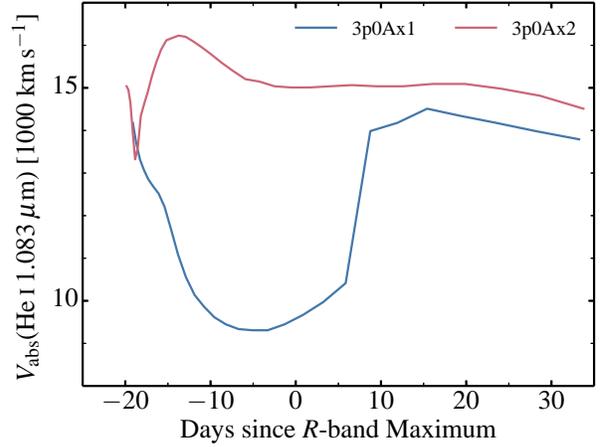,width=8.5cm}
\epsfig{file=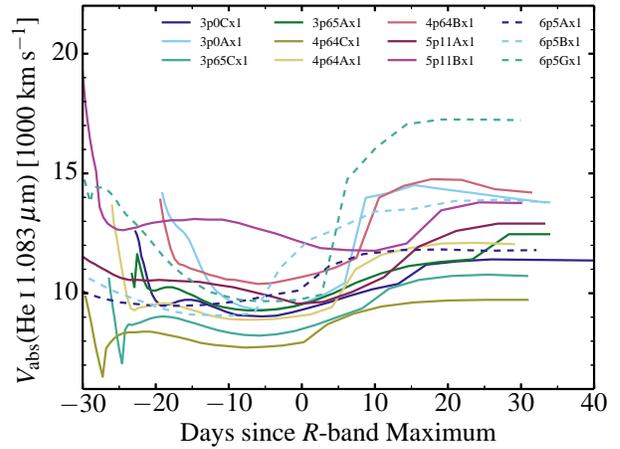,width=8.5cm}
\epsfig{file=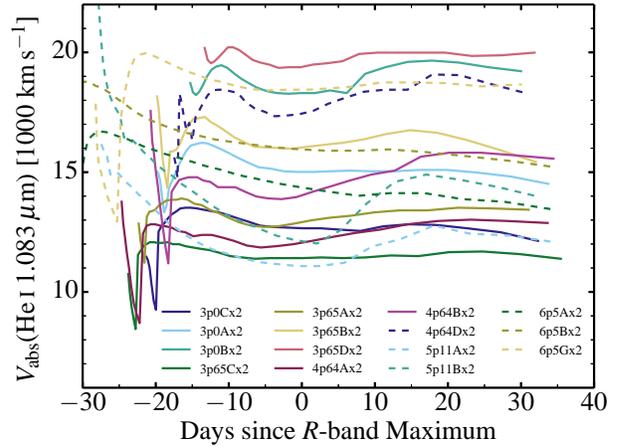,width=8.5cm}
\caption{
{\it Top:} Evolution of the Doppler velocity associated with the maximum absorption in He\one\,1.083\,$\mu$m
for models 3p0Ax1 and 3p0Ax2.
{\it Middle:} Same as top, but now for all models with moderate mixing (x1).
{\it Bottom:} Same as top, but now for all models with strong mixing (x2).
In models with moderate mixing, the velocity at maximum absorption decreases near the time
of bolometric maximum and increases after maximum due to non-local energy deposition and non-thermal effects.
In models with strong mixing, this reversal is weaker, or even absent (e.g., in a low-mass ejecta model like 3p0Ax2).
The hook at early times corresponds to the epoch when we switch on non-thermal processes in the \cmfgen\ calculation.
\label{fig_vhei_10830}
}
\end{figure}

Just like for H$\alpha$, the values for $V_{\rm abs}$(He\one\,5875\,\AA)
tend to level out near maximum. In lower mass ejecta with weaker mixing,
there is even a reversal and  $V_{\rm abs}$(He\one\,5875\,\AA) starts increasing
again (the line broadens) after $R$-band maximum.
This effect is more strongly present in the He\one\,1.083\,$\mu$m line,
although for strong mixing, the reversal is reduced and may even vanish (Figure~\ref{fig_vhei_10830}).
Observationally, the reversal in $V_{\rm abs}$(He\one\,5875\,\AA) is generally not seen in SNe IIb/Ib
\citet{liu_snibc_15} --- an exception is SN\,2011dh \citep{ergon_14_11dh}.

As discussed in \citet{D15_SNIbc_I}, this flattening and subsequent reversal is a natural consequence of the sensitivity
of He\one\ lines to non-thermal processes. Around maximum, as the $\gamma$-ray mean free path
becomes comparable to the SN radius (see, e.g., figures~1--2 of \citealt{d12_snibc}),
the absorption of $\gamma$-rays in the outer ejecta is enhanced. The outer ejecta layers thus become
increasingly influenced by non-thermal effects. In general, the outer  layers will tend to be more helium
rich in SNe IIb/Ib/Ic ejecta, because these correspond to the outer edge of the helium core.
Consequently, around $R$-band maximum, while the bulk of the spectrum forms in the CO-rich regions at the base
of the ejecta, the He\one\,1.083\,$\mu$m line may start to strengthen and broaden again.

This feature is potentially important. Other species like C, O, Ca are less dependent on
non-thermal processes to be excited and to produce lines. They are also more abundant
at depth. In contrast, helium is both very sensitive to non-thermal processes and more
abundant in the outer ejecta. So, only helium can cause a broadening of the absorption feature
seen at $\sim$\,1.05\,$\mu$m in SNe IIb/Ib/Ic spectra. In our models, this broadening is seen
even in SN Ic models (i.e., models 5p11Ax1, 5p11Ax2, 5p11Bx1, and 5p11Bx2)
and it therefore provides unambiguous evidence for the presence
of helium in the model. The behaviour of $V_{\rm abs}$(He\one\,1.083\,$\mu$m)
could therefore be used to assess the presence of helium in SN Ic ejecta.

High quality near-IR spectra of SNe IIb/Ib/Ic are lacking so a dedicated
program to seek such observations is necessary.

\subsection{Estimate of the ejecta expansion rate}

In \citet{D15_SNIbc_I}, we discussed the ambiguity that surrounds the notion of
a photosphere in SNe IIb/Ib/Ic. A more meaningful quantity to constrain is the expansion
rate $V_{\rm m}\equiv \sqrt{2E_{\rm kin}/M_{\rm e}}$ since it relates
to fundamental quantities characterising a SN ejecta, and it is also used
in simplified light-curve modelling. The question is then what line measurement
constrains $V_{\rm m}$ with reasonable accuracy?

\begin{figure}
\epsfig{file=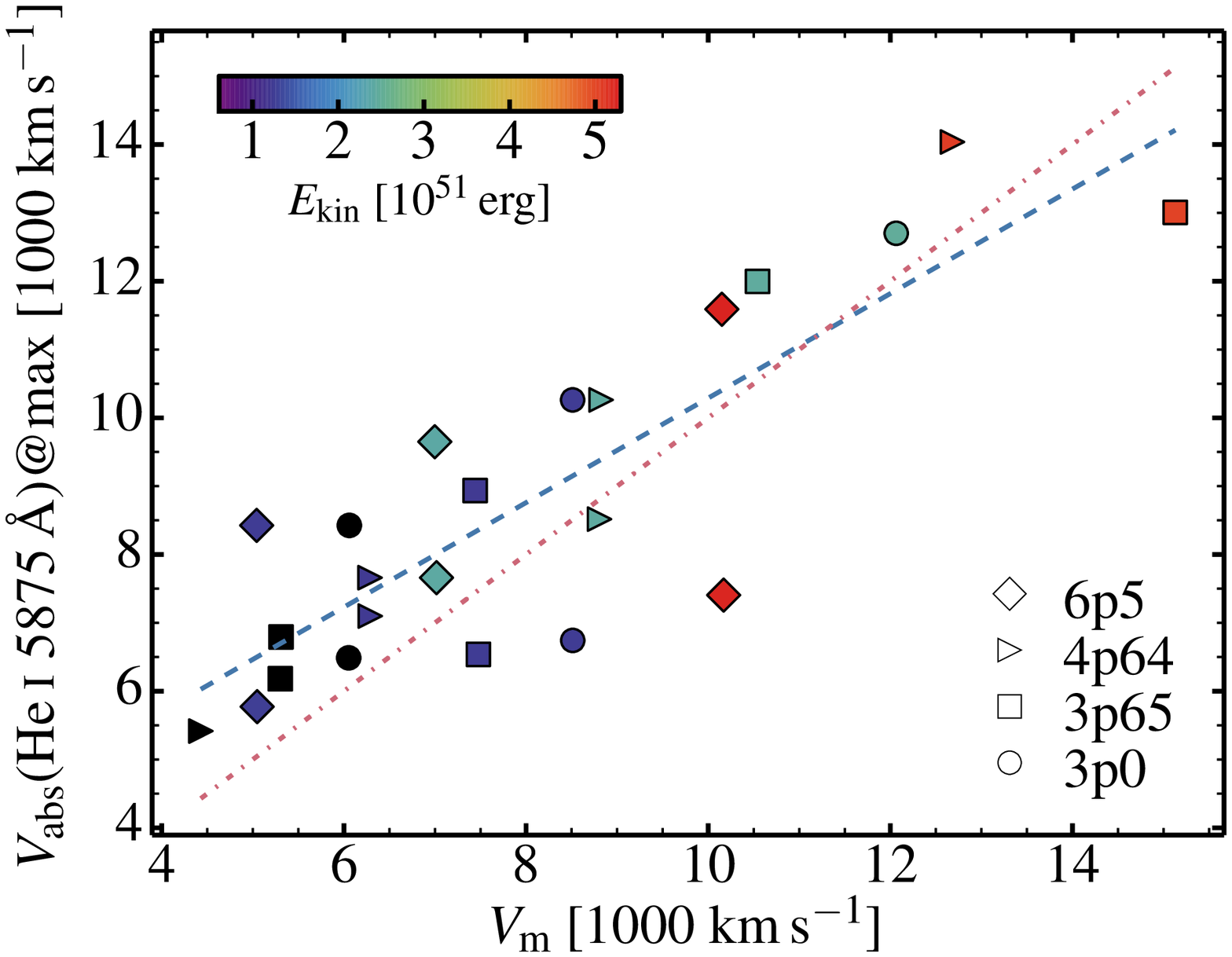,width=8.5cm}
\caption{Doppler velocity at maximum absorption in the line He\one\,5875\,\AA\ at $R$-band maximum
versus the quantity $V_{\rm m}\equiv \sqrt{2E_{\rm kin}/M_{\rm e}}$ for our SNe IIb/Ib models
(when two identical symbols lie at the same $V_{\rm m}$, the upper symbol corresponds to the model with stronger mixing).
The dashed curve is a fit to the distribution of values and has the form
$V_{\rm abs}$(He\one\,5875\,\AA)/1000\,km\,s$^{-1}= 2.64 + 0.765 V_{\rm m}$/1000\,km\,s$^{-1}$.
All values would lie along the dash-dotted curve if the two quantities plotted were equal (i.e., $V_{\rm abs} = V_{\rm m}$).
\label{fig_vm_hei}}
\end{figure}

Figure~\ref{fig_vm_hei} compares the Doppler velocity at maximum absorption
in the line He\one\,5875\,\AA\ at $R$-band maximum to the expansion rate $V_{\rm m}$
for our set of SNe IIb/Ib models. A least-square polynomial fit to the distribution of
points yields
$$\frac{V_{\rm abs}({\rm He\,{\textsc i}\,5875\,\AA)}}{{\rm 1000\,km\,s}^{-1}}= 2.64 + 0.765 \frac{V_{\rm m}}{{\rm 1000\,km\,s}^{-1}}\, , $$
with a dispersion of 1370\,\kms.
The two quantities are not equal (if so, they would lie on the dash-dotted curve), but
they are close. Some of the scatter stems from the different magnitude of mixing between
models with suffix x1 and x2. Stronger mixing systematically produces broader He\one\ lines.
Unfortunately, we do not know the mixing process with  much certainty so reducing the scatter is non trivial.
This sensitivity introduces an uncertainty for the estimate of the expansion rate in SNe IIb/Ib,
and therefore for the determination of $E_{\rm kin}$ and $M_{\rm e}$.

\begin{figure}
\epsfig{file=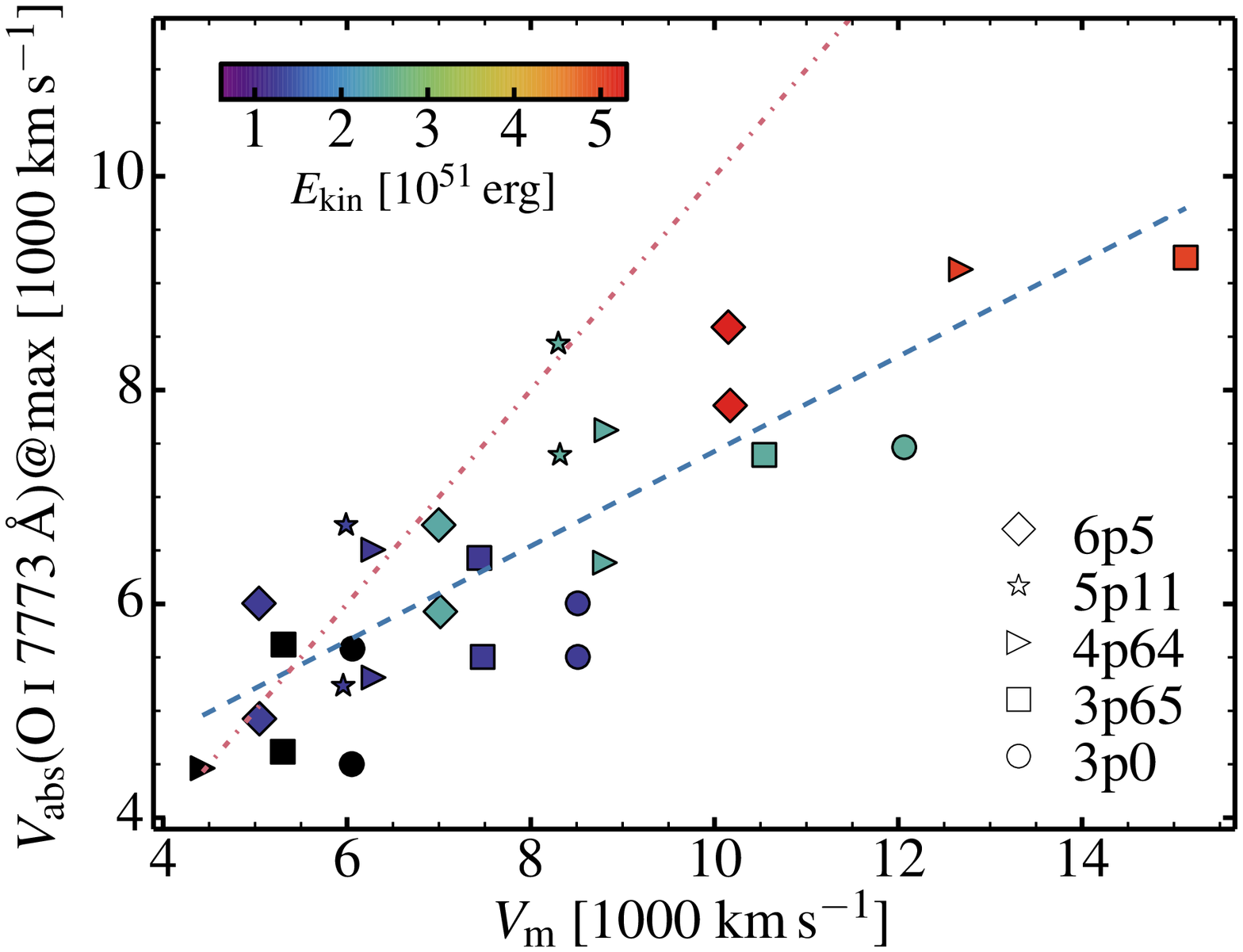,width=8.5cm}
\caption{Doppler velocity at maximum absorption in the line O\one\,7772\,\AA\ at $R$-band maximum
versus the quantity $V_{\rm m}\equiv \sqrt{2E_{\rm kin}/M_{\rm e}}$ for our grid of models.
(when two identical symbols lie at the same $V_{\rm m}$, the upper symbol corresponds to the model with stronger mixing).
The dashed curve is a fit to the distribution of values and has the form
$V_{\rm abs}$(O\one\,7772\,\AA)/1000\,km\,s$^{-1}= 2.99 + 0.443 V_{\rm m}$/1000\,km\,s$^{-1}$.
All values would lie along the dash-dotted curve if
the two quantities plotted were equal (i.e., $V_{\rm abs} = V_{\rm m}$).
\label{fig_vm_oi}}
\end{figure}

Figure~\ref{fig_vm_oi} compares the Doppler velocity at maximum absorption in the line
O\one\,7772\,\AA\ at $R$-band maximum to $V_{\rm m}$ for our entire set of models.
A least-square polynomial fit to the distribution of points yields

$$\frac{V_{\rm abs}({\rm O\,{\textsc i}\,7772\,\AA})}{{\rm 1000\,km\,s}^{-1}}= 2.99 + 0.443 \frac{V_{\rm m}}{{\rm 1000\,km\,s}^{-1}} \, ,$$
with a dispersion of 780\,\kms.
The slope is much flatter, i.e., $V_{\rm abs}$(O\one\,7772\,\AA) tends to be lower than $V_{\rm m}$,
and the offset is greater for increasing ejecta mass and kinetic energy. This probably arises because
the oxygen abundance increases inwards in the ejecta, biasing the line optical depth to lower velocities.
Also, for higher mass, the peak is delayed so at maximum, the outer ejecta is cold, optically thin,
and contributes little to the emergent spectrum. At higher explosion energy, the same result obtains
but because the outer ejecta expands faster.
Still, for model 5p11, $V_{\rm abs}$(O\one\,7772\,\AA) offers a satisfactory means to constrain
$V_{\rm m}$ to within 10-20\%.

So, we suggest to constrain the expansion rate $V_{\rm m}$ with He\one\,5875\,\AA\ in SNe IIb/Ib and
with O\one\,7772\,\AA\ in SNe Ic.

\section{Comparison to other work}
\label{sect_disc}

In this section, we compare some results with those estimated using different approaches.
In particular, we discuss how our grid of models compares with the energy constraint set by the
time-integrated form of the first law of thermodynamics (Section~\ref{sect_katz}). We then
compare our results to the Arnett model \citep{arnett_82},
including the so-called Arnett rule (Section~\ref{sect_arnett}).

\subsection{Constraints from energy conservation}
\label{sect_katz}

  \citet{katz_13_56ni} propose to use the energy equation to constrain the \iso{56}Ni mass.
But we can also use it to gain insights into the physics of SN IIb/Ib/Ic radiation, for example,
to check the long-term energy conservation of the \cmfgen\ sequence.

\begin{figure}
\epsfig{file=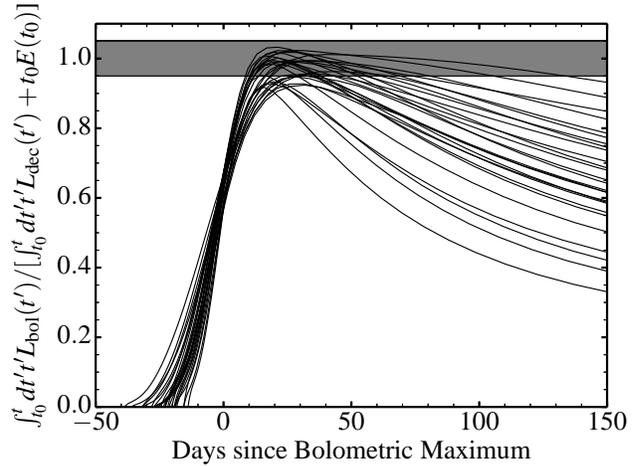,width=8.5cm}
\caption{Illustration of the variation of the ratio
$\int_{t_0}^t dt' t' L_{\rm bol}(t')$ /[$\int_{t_0}^t dt' t' L_{\rm dec}(t') + t_0E(t_0)]$
versus time since bolometric maximum (we use $t_0 \gtrsim 3$\,d)  for our grid of SNe IIb/Ib/Ic models.
The shaded area corresponds to offsets of $\pm$5\% from unity.
For some models, the ratio remains below unity because of the early escape of $\gamma$-rays.
\label{fig_katz}
}
\end{figure}

\begin{figure}
\epsfig{file=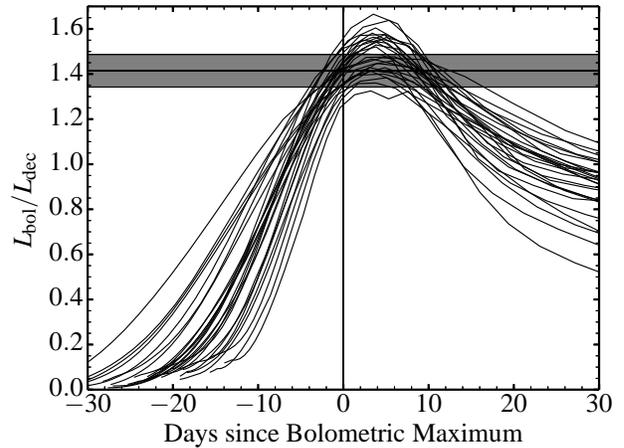,width=8.5cm}
\caption{Variation of the ratio $L_{\rm bol}/L_{\rm dec}$
versus time since bolometric maximum for our grid of SNe IIb/Ib/Ic models.
The ratio at bolometric maximum has a mean of  1.41 and a standard deviation of  $\sigma=$0.072.
The shaded area corresponds to the mean $\pm\sigma$.
\label{fig_arnett}
}
\end{figure}

Considering heating from radioactive decay
and cooling through expansion and radiation, one can solve the internal energy equation out to late
times to find
\begin{equation}
\int_{t_0}^t dt' t' E(t')  =  \int_{t_0}^t dt' t' L_{\rm dec}(t') - \int_{t_0}^t dt' t' L_{\rm bol}(t') \, ,
\end{equation}
where $E(t')$ is the total radiation energy trapped within the ejecta at $t'$ and $L_{\rm dec}$ is
the total decay power (Katz et al. use instead the total decay energy deposited within the ejecta, but
this quantity is not known directly).
At late times, as the ejecta becomes optically thin, there is essentially no stored radiation energy
so we neglect the associated term $t E(t)$.
When evaluating this ratio for the \cmfgen\ simulation, we use $t_0=$\,3\,d to skip
the first few steps of the time sequences. In that case, we include the term
$E(t_0)$, which is of the order of $\lesssim$\,10$^{48}$\,erg at that time in our models.

\begin{figure*}
\epsfig{file=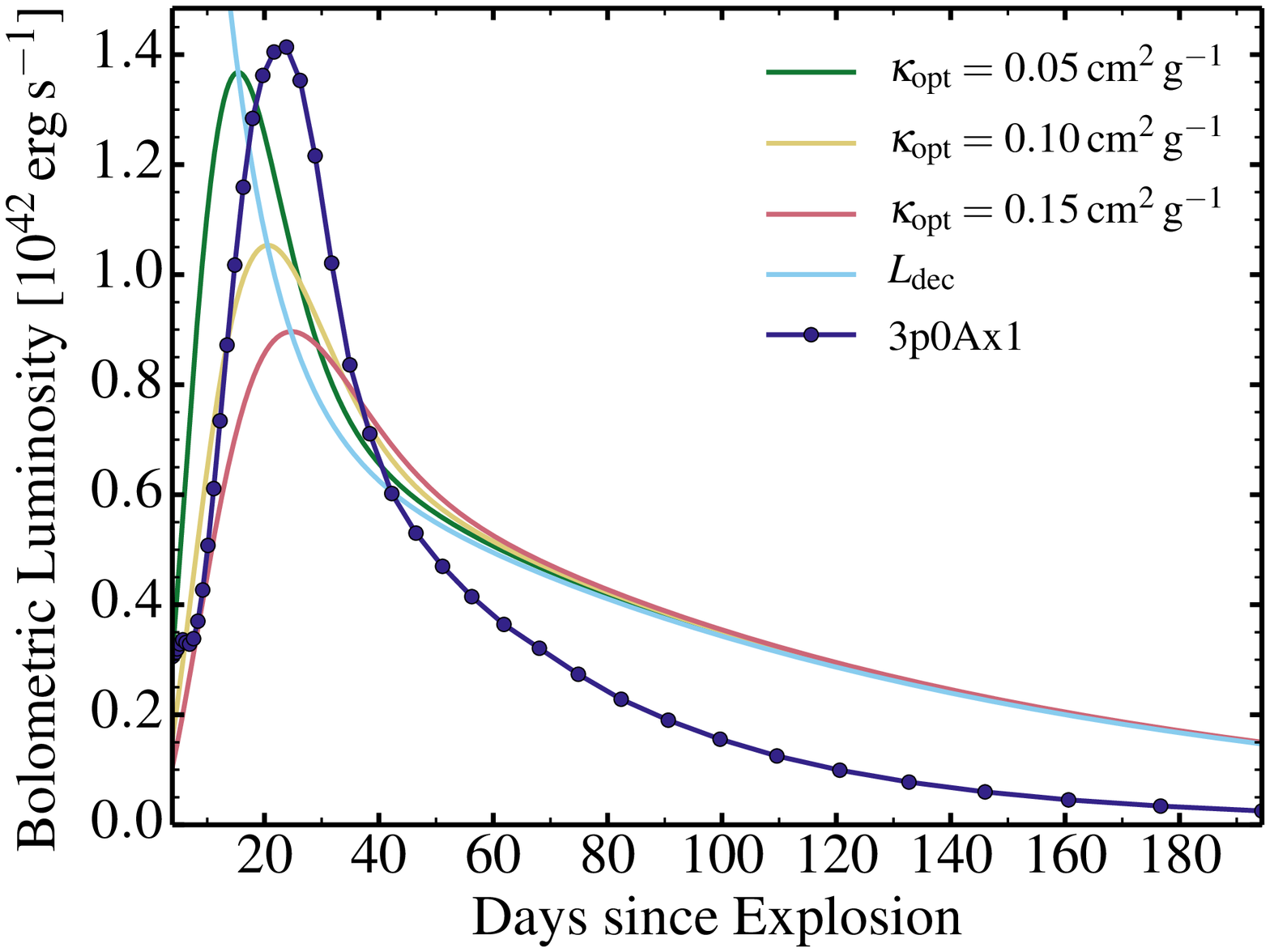,width=8.5cm}
\epsfig{file=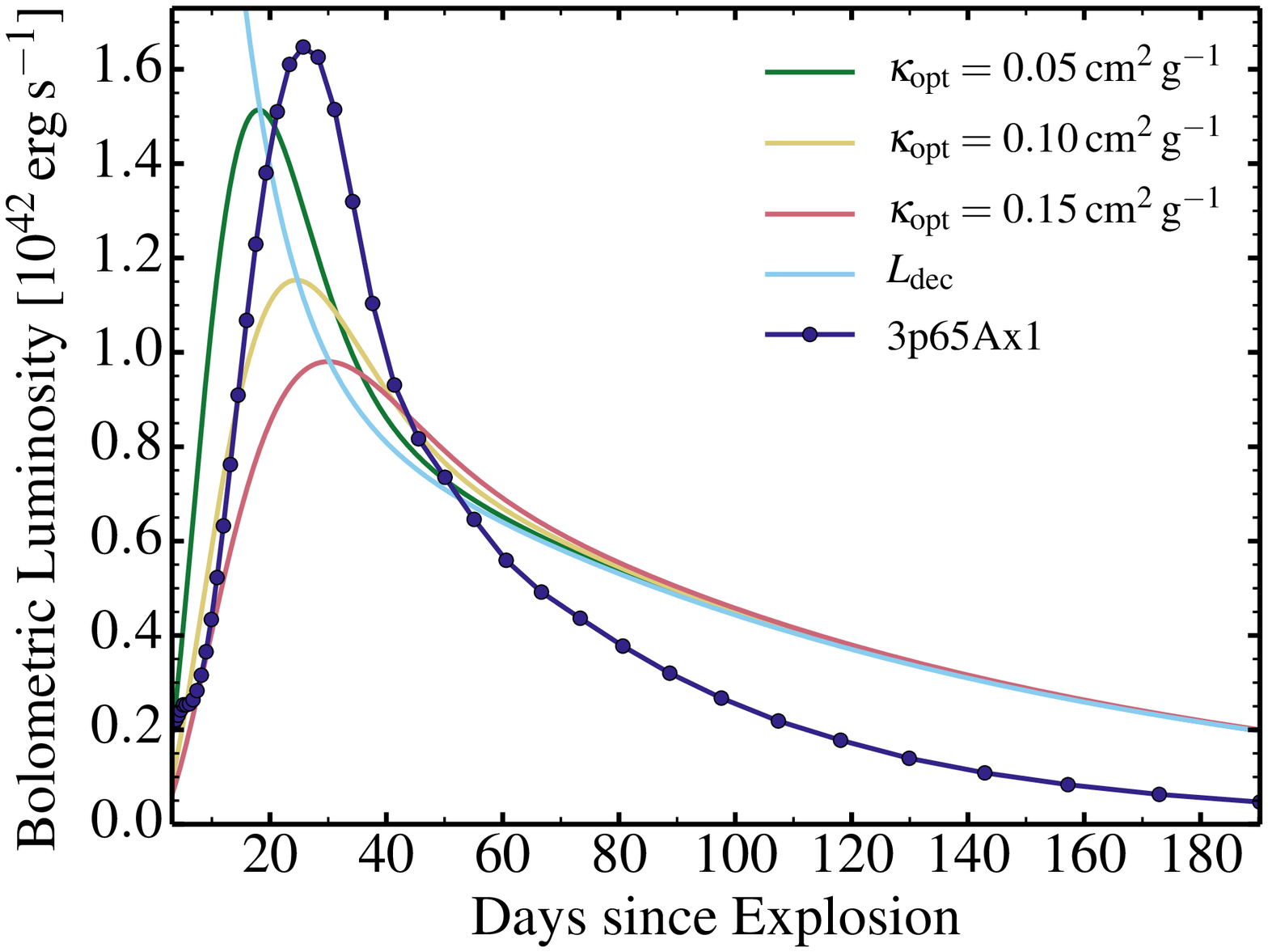,width=8.5cm}
\epsfig{file=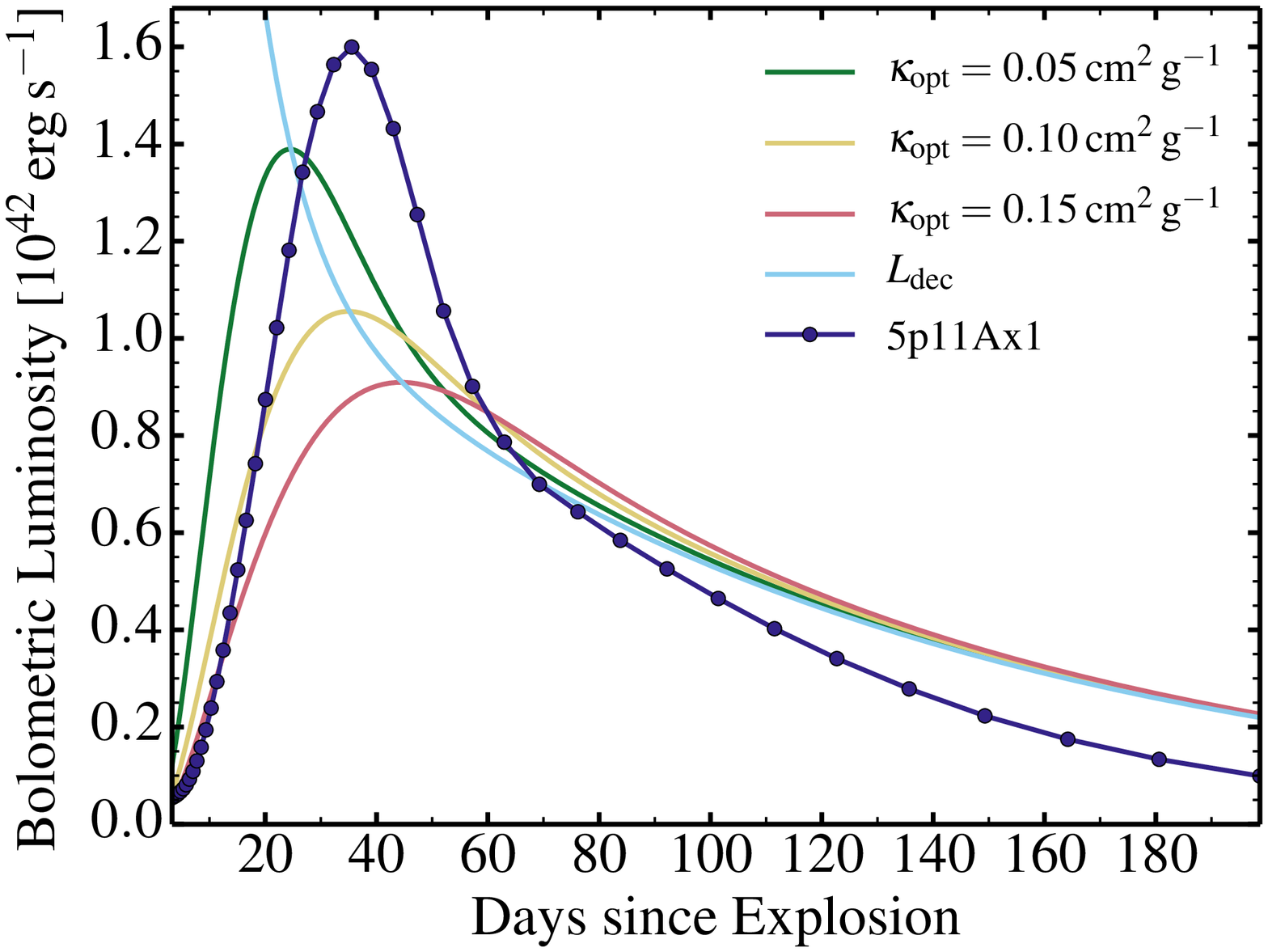,width=8.5cm}
\epsfig{file=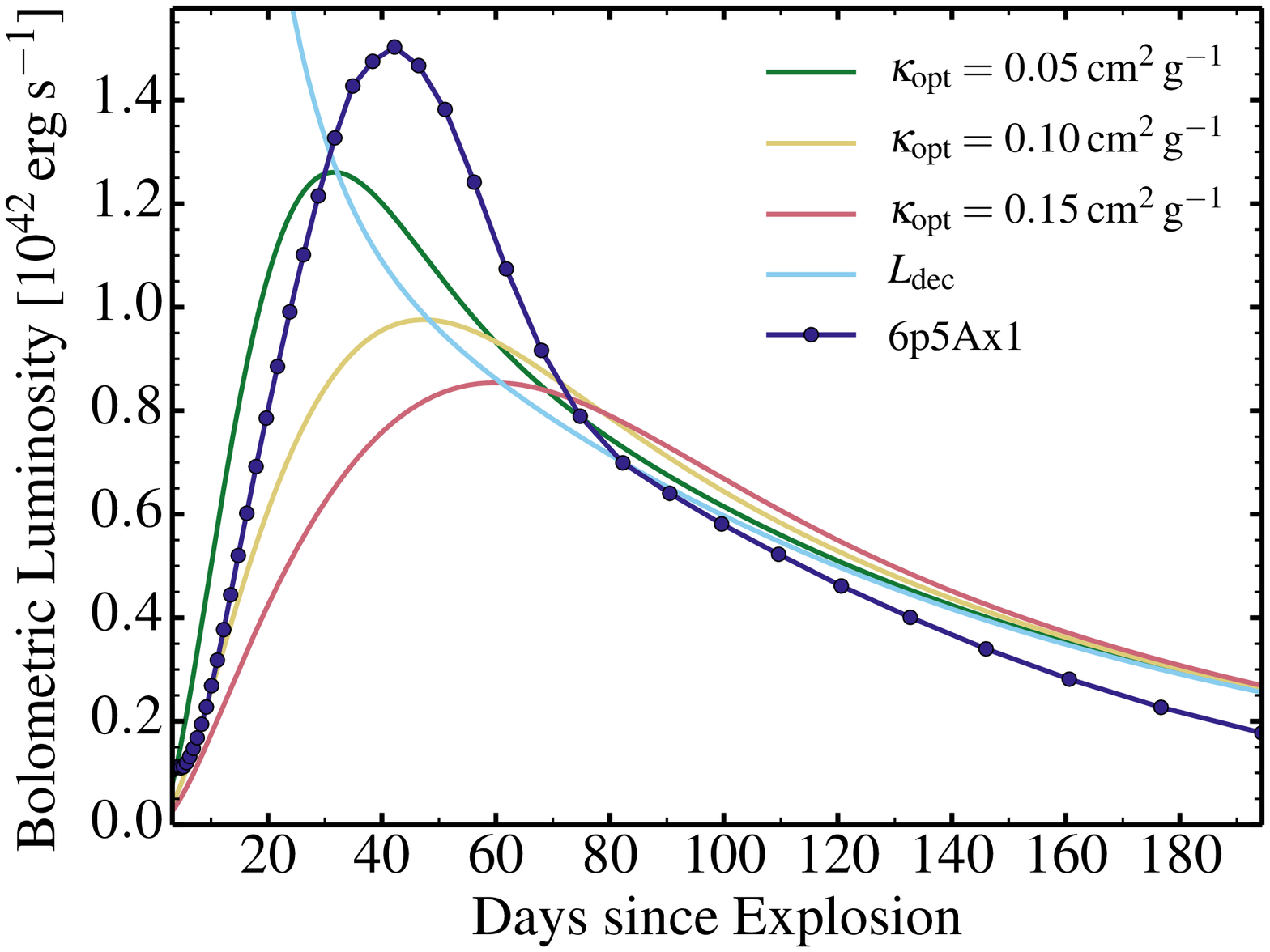,width=8.5cm}
\caption{
Comparison between the bolometric luminosity light curve for models
3p0Ax1, 3p65Ax1, 5p11Ax1, and 6p5Ax1, the total \iso{56}Ni decay power,
and the predictions of  the Arnett model \citep{arnett_82, valenti_08_03jd}
(we use the formulation of \citealt{valenti_08_03jd} but corrected for
an error in the expression of $B(t)$ and $s$; see our Appendix~\ref{appendix_valenti}.)
using three different values
of $\kappa_{\rm opt}$ (to save space, we omit model 4p64Ax1).
In each case, we use the ejecta kinetic energy, mass and \iso{56}Ni mass
of the \cmfgen\ model.
The value $\kappa_{\rm opt}=$\,0.05\,cm$^2$\,g$^{-1}$ used by \citet{drout_etal_11}
for their SNe Ibc study underestimates the rise time in all 5 cases, while the Arnett model overestimates
the \iso{56}Ni mass by up to a factor of two (Figure~\ref{fig_arnett}).
\label{fig_lc_valenti}
}
\end{figure*}

In \citet{D15_SNIbc_I}, we found that the ratio $\int_{t_0}^t dt' t' L_{\rm bol}(t') [\int_{t_0}^t dt' t' L_{\rm dec}(t')+t_0E(t_0)]$
was within $\sim$\,1\% of unity at 10-20\,d after maximum for our three selected models. In these cases, $\gamma$-ray escape
was negligible up to 10-20\,d after maximum, and the convergence of the ratio to unity demonstrated instead the
good energy conservation of the \cmfgen\ model.

Figure~\ref{fig_katz} shows this ratio for the full grid of models. As before, many models reach unity at 10-20\,d after
bolometric maximum. However, some do not, because
$\gamma$-rays from radioactive decay start escaping the ejecta before it is optically-thin to optical photons.
This is the case for lower mass ejecta and/or higher energy explosions. Stronger mixing also exacerbates
the effect since it biases the \iso{56}Ni distribution toward higher-velocity lower-density regions.
The impact of $\gamma$-ray escape can be seen directly from the post-maximum decline rate
(Figs.~\ref{fig_lbol_A}--\ref{fig_lbol_4p64}).

\subsection{The Arnett model}
\label{sect_arnett}

A common expedient in the SN community is to apply the Arnett model \citep{arnett_82}, originally
developed for SNe Ia, to all Type I SN light curves.

One feature of this model is the prediction that the bolometric luminosity
at maximum is close to the instantaneous decay rate at that time.
Hence, knowledge of the distance and reddening to a given SN Ia gives the \iso{56}Ni mass
needed to match the peak luminosity.
For SNe Ia, detailed radiative-transfer calculations suggest that this relation is fairly
accurate (see, e.g., \citealt{blondin_etal_13}), despite the various simplifications.

In \citet{D15_SNIbc_I}, we found that the agreement is poorer when applied to SNe IIb/Ib/Ic.
When the comparison is extended to our full grid of models, the disagreement is not reduced
(Figure~\ref{fig_arnett}). Applying Arnett rule to our SN IIb/Ib/Ic models would lead to
an overestimate of the \iso{56}Ni mass by as much as 50\%.
In our simulations, $L_{\rm bol}/L_{\rm dec}$ has a mean of  1.41 and a standard deviation of  0.072.
The offset found in \citet{D15_SNIbc_I} for a few models is therefore present in the larger set of models.

The adopted mixing plays a minor role (see Paper\one, Section~6, where the influence of mixing
is discussed for model 3p65A, 5p11A, and 6p5A, as well as \citet{d12_snibc} for an extended discussion of
the impact of mixing in SNe Ibc light curves and spectra).
In our models, enhanced mixing shortens the rise to bolometric maximum
and enhances weakly the peak luminosity (all else being the same).
For example, for model 3p65Ax1 (moderate mixing), $L_{\rm bol}/L_{\rm dec}$ is 1.5. In model 3p65Ax2 (strong mixing),
the bolometric
maximum is a fraction of a percent higher but $L_{\rm bol}/L_{\rm dec}$ is only 1.42. The reduction in the offset between
$L_{\rm bol}$ and $L_{\rm dec}$ is related to the shorter rise time of 1.5\,d ---
the instantaneous decay power for the \iso{56}Ni chain drops by 5\% in 1.5\,d at 20-25\,d after explosion.
As a reminder, in the absence of mixing, models also show a 40-50\% offset between $L_{\rm bol}$ and $L_{\rm dec}$
at bolometric maximum (see Figure~13 of \citealt{dessart_11_wr}).

Let us now compare the \cmfgen\ bolometric light curve to the Arnett light-curve model
(we use the formulation of \citealt{valenti_08_03jd}, but we corrected
the expressions of $B(t)$ and $s$; see our Appendix~\ref{appendix_valenti}.)
This model has four unknowns: $M_{\rm e}$, $E_{\rm kin}$, M(\iso{56}Ni), and the mass-absorption
coefficient for low-energy photons $\kappa_{\rm opt}$.
We focus on the first part of the light curve, around maximum, and assume full $\gamma$-ray trapping.
Using the values of $M_{\rm e}$, $E_{\rm kin}$, and $M$(\iso{56}Ni) in a \cmfgen\ model,
we can compare the Arnett model light curve to the \cmfgen\ result.
The standard implementation of the Arnett model assumes
a fixed opacity, while \cmfgen\ works from physical ejecta models and computes the opacity
frequency-by-frequency and at all ejecta locations. This microphysics is treated entirely differently and there is
no simple way to import the \cmfgen\ opacity into the Arnett model. So, we show the predictions
of the Arnett model for three different values of $\kappa_{\rm opt}$.

Figure~\ref{fig_lc_valenti} shows that the central peak of the light curve from the Arnett model has the same qualitative shape
but with an offset in all cases.\footnote{There is an obvious offset at late times because we assume full $\gamma$-ray
trapping in the Arnett model but allow for $\gamma$-ray escape in the \cmfgen\ calculations}
 The maximum value is systematically under-estimated (see above). The post-breakout plateau is absent.
The width of the light curve is better matched for models with a smaller $\kappa_{\rm opt}$,
but the rise time to bolometric maximum is then underestimated.
Larger values of $\kappa_{\rm opt}$ help resolving these deficiencies but the offset at bolometric
maximum is increased.
We can improve the match of the Arnett model to the bolometric light curve peak of model 3p65Ax1
computed by \cmfgen\ if we shift the Arnett light curve by 8\,d and increase the \iso{56}Ni mass by 10\%.
The time shift is needed to account for the post-breakout plateau predicted in our \cmfgen\ simulation and
absent in the Arnett model. This manipulation is very artificial and this exercise merely illustrative.

\begin{figure}
\epsfig{file=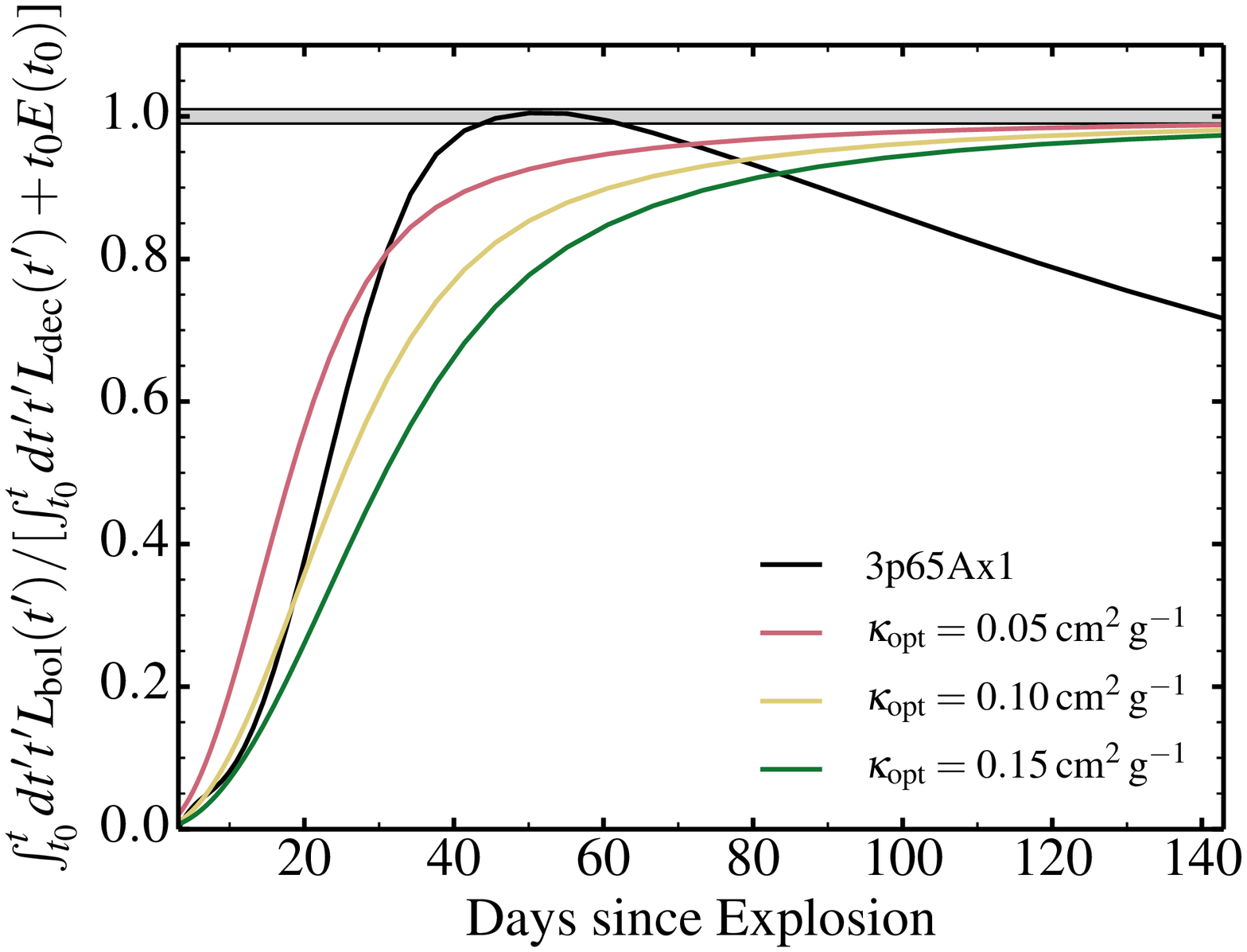,width=8.5cm}
\epsfig{file=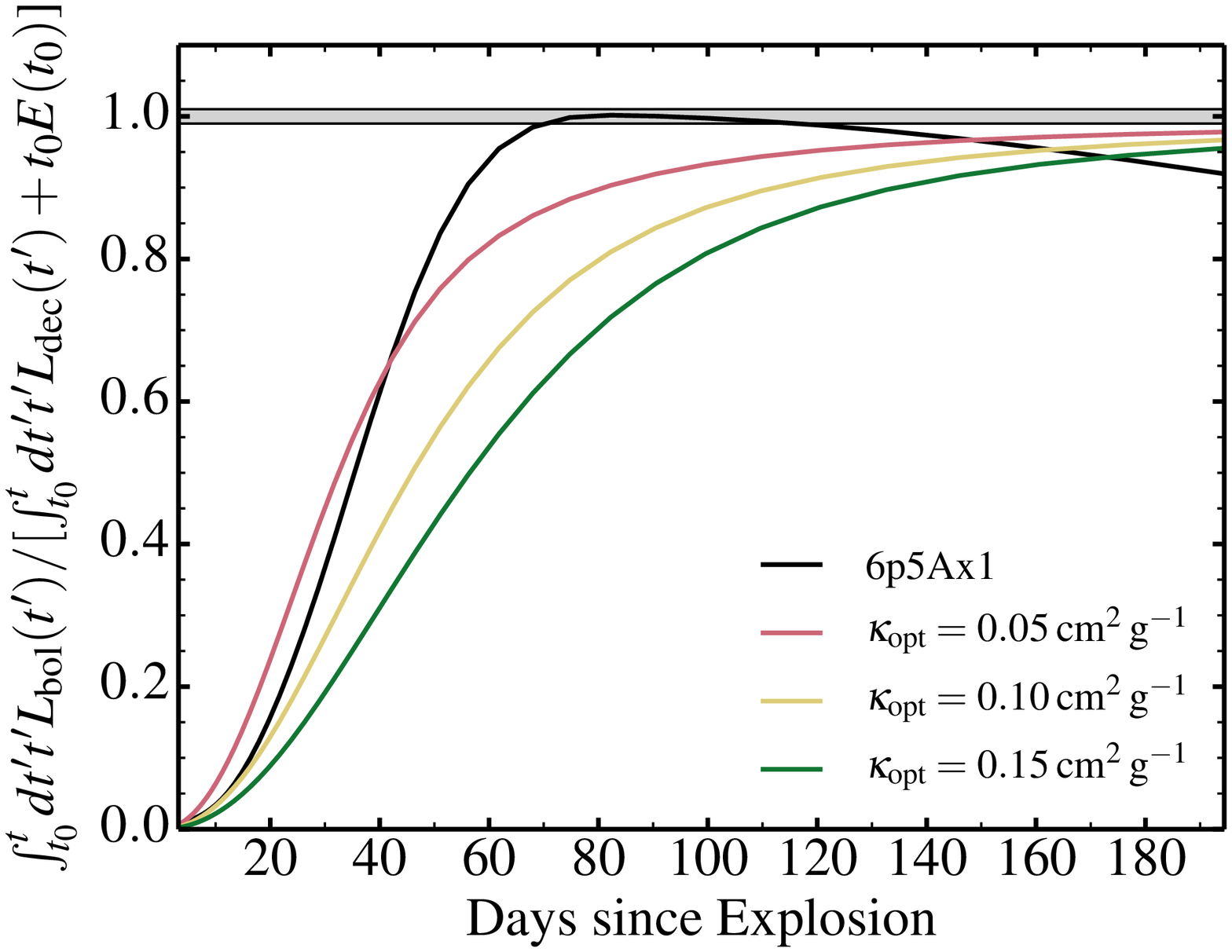,width=8.5cm}
\caption{Illustration of the variation of the ratio
$\int_{t_0}^t dt' t' L_{\rm bol}(t')$ /[$\int_{t_0}^t dt' t' L_{\rm dec}(t') + t_0E(t_0)]$
versus time since bolometric maximum (we use $t_0 \gtrsim 3$\,d)  for our models 3p65Ax1
and 6p5Ax1 together with the predictions from Arnett's model using three different values
of $\kappa_{\rm opt}$ (for these curves, we  use $t_0=0$\,d).
The ejecta kinetic energy, mass and \iso{56}Ni mass are taken from the corresponding \cmfgen\ model.
In the present implementation of the Arnett model we assume full $\gamma$-ray trapping
so the curves rise to unity at late times.
When unity is reached in the \cmfgen\ model, the Arnett model is at around 0.9 and the offset
is larger for larger values of $\kappa_{\rm opt}$.
One reason for the offset between the standard Arnett model predictions and the more detailed calculations
with \cmfgen\ is the simplified treatment of the opacity.
\label{fig_lc_valenti_katz}
}
\end{figure}

If the SN light curve is entirely powered by \iso{56}Ni power, then the Arnett model should give a good
match to the energy criterion expressed in \citet{katz_13_56ni}.
Figure~\ref{fig_lc_valenti_katz} reproduces Figure~\ref{fig_katz} but for models 3p65Ax1 and 6p5Ax1
together with their Arnett model counterpart --- we use the same $M_{\rm e}$, $E_{\rm kin}$, and M(\iso{56}Ni)
as the \cmfgen\ model and cover three different values of $\kappa_{\rm opt}$.
When the \cmfgen\ model reaches unity, the Arnett model is at $\sim$0.9, and even less for higher values
of $\kappa_{\rm opt}$.

\begin{figure}
\epsfig{file=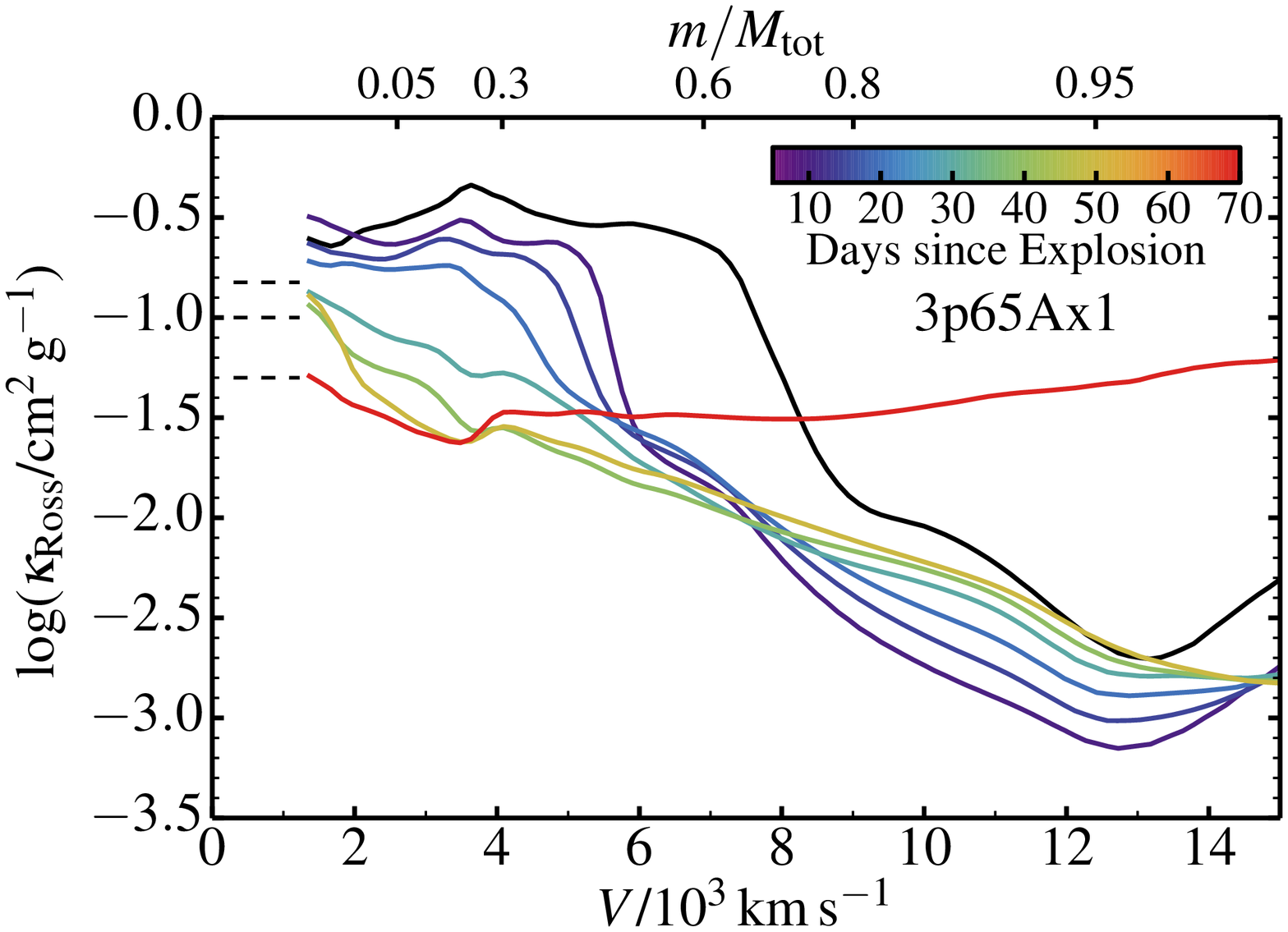,width=8.5cm}
\epsfig{file=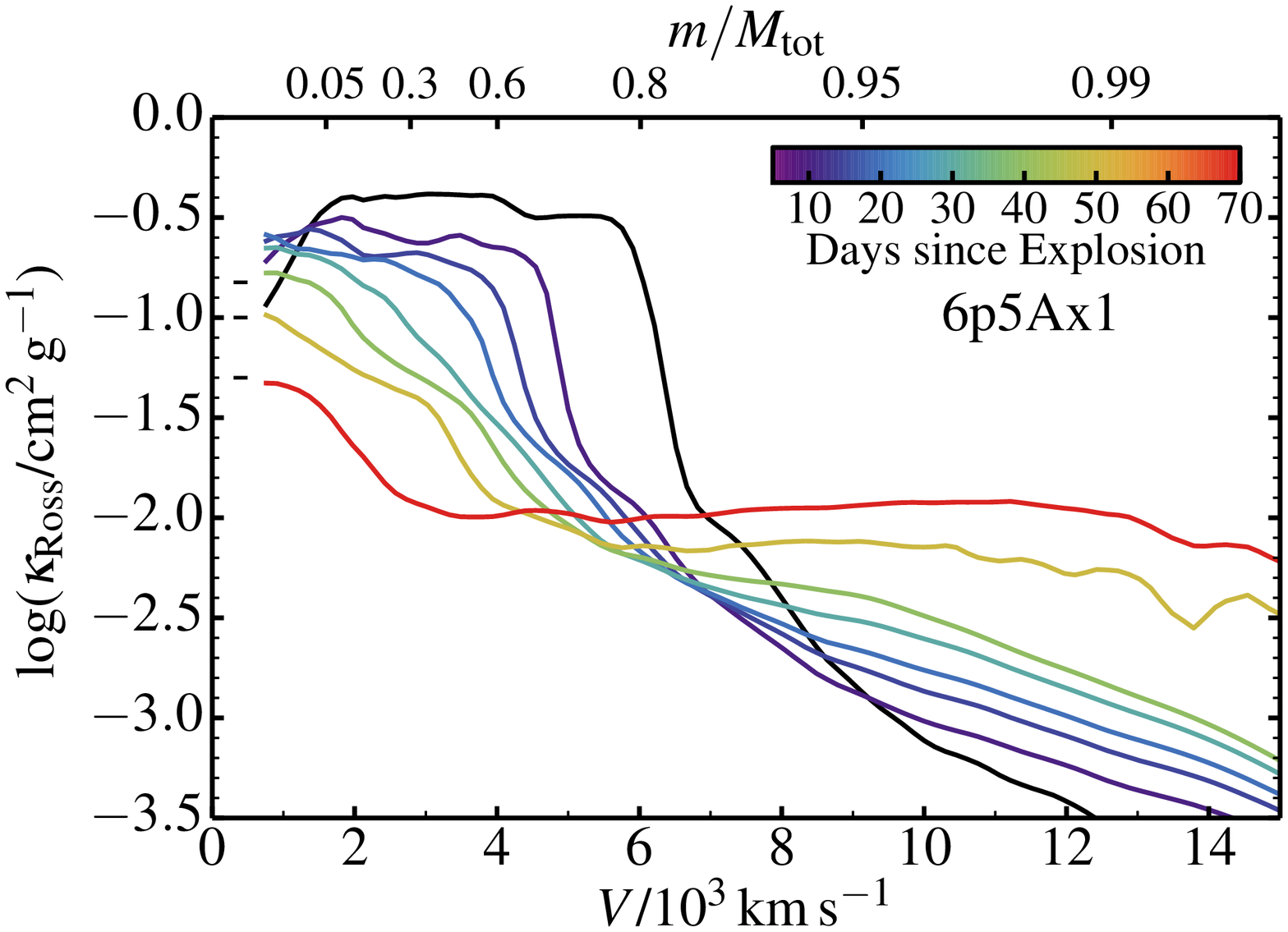,width=8.5cm}
\caption{Evolution of the Rosseland-mean opacity versus velocity and time
for model 3p65Ax1 and 6p5Ax1.
The times shown are 5, 10, 15, 20, 30, 40, 50, and 70\,d after explosion.
For reference, we add marks on the left to locate $\kappa_{\rm Ross}$ values of
0.05, 0.10, and 0.15\,cm$^2$\,g$^{-1}$ (to be compared to the values of
$\kappa_{\rm opt}$ used in the Arnett model).
\label{fig_kappa}
}
\end{figure}

The Arnett model makes a number of simplifications.
It assumes a progenitor star with a negligible radius. This is roughly equivalent to
dropping the term $E(t_0)$ in the Katz ratio.
In models 3p65Ax1 and 6p5Ax1, the total radiation energy stored within the ejecta
at 3\,d is 6--9$\times$\,10$^{47}$\,erg.
In practice, neglecting that term changes the ratio at the 1\% level,
so this simplification is adequate.
The Arnett model neglects ionization/excitation energy but this energy contribution is subdominant ---
it also affects the ratio at the 1\% level.
A feature of greater significance is the assumption of a fixed opacity.
Figure~\ref{fig_lc_valenti} shows how sensitive the Arnett model light curve is to different values
of this adopted opacity --- this an inherent limitation of the Arnett model since the value to
use is unknown. Figure~\ref{fig_kappa}
shows the complicated behaviour of the opacity in our SN Ibc simulations. In reality, the
opacity varies significantly both in space and time.
Assuming a uniform and fixed opacity tends to underestimate the opacity in the inner ejecta
and to overestimate it in the outer ejecta, which becomes essentially transparent very early on.
This may be the main reason for the offset between the Arnett model light curve and the
\cmfgen\ results.

\section{conclusions}
\label{sect_conc}

This paper is a follow-up to \citet{D15_SNIbc_I}, which focused on three models
of type IIb, Ib, and Ic. Here, we extend our analysis to a large grid of models for SNe IIb/Ib/Ic
that result from the terminal explosion of the mass donor in a close-binary system.
All our results are based on 1-D non-LTE time-dependent radiative-transfer simulations with \cmfgen.

With this grid, we cover ejecta masses in the range 1.7--5.2\,\msun, kinetic energies in the range
0.6--5.0$\times$10$^{51}$\,erg, and \iso{56}Ni masses in the range 0.05--0.30\,\msun.
Equipped with better physics (non-thermal processes, improved model atoms) compared
to \citet{dessart_11_wr}, this sample also covers a much bigger parameter space than the focused
study of \citet{d12_snibc} on chemical mixing.
The range of progenitor (mass, composition, radius) and explosion properties
(energy, \iso{56}Ni mass) is by definition limited but hopefully this range overlaps significantly with
the SNe IIb/Ib/Ic in Nature. Nonetheless, our approach allows a direct confrontation of observables
(e.g., bolometric luminosity) to initial progenitor/ejecta properties (ejecta mass, \iso{56}Ni mass).
Even if limited, this controlled experiment yields interesting insights into the properties of SN IIb/Ib/Ic
models, with, hopefully, relevance to SNe IIb/Ib/Ic in Nature.

A number of correlations emerge from this larger sample of models, which confirm some of
the results presented in \citet{D15_SNIbc_I}.
As expected, we find a strong correlation between the \iso{56}Ni mass and the maximum bolometric
luminosity and peak brightness. We provide fitted formulae over the \iso{56}Ni mass range 0.05--0.30\,\msun.
Our predictions at large \iso{56}Ni mass are influenced by the large ejecta masses of the corresponding models.
Ejecta with a larger kinetic energy mimic the behaviour of ejecta with a lower mass, confirming the
general degeneracy of light curve morphology for a given $E_{\rm kin}/M_{\rm e}$.
When considering the entire model set, however, we obtain a significant scatter in correlations
involving $E_{\rm kin}/M_{\rm e}$ (or $V_{\rm m}$).
Varying $E_{\rm kin}$ ($M_{\rm e}$) influences the rise time to maximum, the light curve width,
the early post-maximum decline, and the nebular decline rate.

Across a wide range of $E_{\rm kin}$, $M_{\rm e}$, and $M$(\iso{56}Ni), we find that the
rise time to maximum strongly correlates with the post-maximum decline. We provide a correlation
for the $R$ band. This property can help estimate the explosion time when early time observations are
lacking. It also alleviates the difficulty of securing such observations because SNe IIb/Ib/Ic are often very
faint one day after explosion and until \iso{56}Ni heating manifests itself.

A notorious problem for all type I SNe is the determination of the reddening. Indeed, unlike SNe II-Plateau,
continuum windows free of lines are lacking and the effects of line blanketing are much stronger at all times,
so a direct constraint on reddening is non trivial in type I SNe.
However, the properties (composition, ionisation, temperature) of the spectrum formation region
are quite similar in our models after maximum, which conspires to produce roughly the same
intrinsic colour. In our set, the distribution of the ($V-R$) colour at 10\,d after $R$-band maximum
has a mean of 0.33\,mag, with a standard deviation of 0.035\,mag.  Observationally, a comparable estimate
was made by \citet{drout_etal_11}.

Spectral line morphology evolves with time as the spectrum formation region recedes in mass space,
and therefore in velocity space. This is generally true for intermediate mass elements (C, O, Ca)
but there are notable exceptions. In SNe IIb, the Doppler velocity at maximum absorption in H$\alpha$
converges around maximum light to a value that corresponds to the velocity at the base of the H-rich
shell in the outer ejecta (the line absorption appears as a stationary dip blue-ward of the rest wavelength).
In our models, this velocity is always very large (in the range
8000--15000\,\kms) because at most 0.01\,\msun\ of H remains in the progenitor star at the time of explosion.

In most models we identify a reversal in the trajectory followed by the maximum absorption in the
He\one\,10830\,\AA\ line around the time of maximum brightness. This feature is unique to helium because
He\one\ is very sensitive to non-thermal excitation, which is intimately tied to the influence of $\gamma$-rays.
In the outer ejecta where the \iso{56}Ni mass fraction is small, this influence is weak early on,
but strengthens around maximum as the $\gamma$-ray mean free path becomes comparable to the SN radius.
The reversal affects the behaviour of the absorption seen around 10300\,\AA,
and it is due to He\one\ alone. Hence, this feature could serve as a test of the presence of helium in SNe Ic.
This reversal is, however, weak or absent in strongly mixed models.

Determining the mean expansion rate of SN ejecta is a critical step in constructing a suitable model of the event.
This is more meaningful than determining the photospheric velocity, which varies with time
and depends on the selected line. In \citet{D15_SNIbc_I}, we studied the evolution of the strongest
optical and near-IR lines and confronted the value of the Doppler velocity at maximum absorption with
the expansion rate $V_{\rm m}$ defined as $\sqrt{2E_{\rm kin}/M_{\rm e}}$.
The complicated processes that influence SN IIb/Ib/Ic spectra do not allow a very accurate
determination of  $V_{\rm m}$. Nonetheless, we find that at light curve maximum,
$V_{\rm m}$ is close to the measured value of the Doppler velocity at maximum absorption in He\one\,5875\,\AA\
in our SNe IIb/Ib models. For SNe Ic, one may use O\one\,7772\,\AA, although the correspondence deteriorates
with higher ejecta mass/energy.

Finally, we find that the Arnett model, originally designed for SNe Ia, is not very accurate when applied to SNe IIb/Ib/Ic.
There may be a combination of factors. We suspect the main shortcoming  is the assumption of a fixed opacity.
The Arnett model light curve depends strongly on that adopted opacity, which is somewhat arbitrarily supplied by the user.
There is in practice no good representative value, because the opacity varies with both ejecta location and age in SNe IIb/Ib/Ic.
Assuming a fixed average opacity systematically overestimates the true opacity of the outer regions
because these continuously recombine and turn transparent as they do so.
The large helium mass fraction contributes to the low opacity of the outer ejecta.
This transparent region grows in mass until the whole ejecta turns nebular.
In the inner ejecta, the opacity varies with time too, as the ionisation changes  and the sources of line blanketing evolve.
This complicated time and spatial dependences of the opacity are ignored in the Arnett model
and this simplification may compromise the rise time to maximum and the light curve width.

The next step with this work is to compare our models to observed SNe IIb, Ib, Ic.

\section*{acknowledgments}

LD acknowledges financial support from the European Community through an
International Re-integration Grant, under grant number PIRG04-GA-2008-239184,
and from ``Agence Nationale de la Recherche" grant ANR-2011-Blanc-SIMI-5-6-007-01.
DJH acknowledges support from STScI theory grant HST-AR-12640.01, and NASA theory grant NNX14AB41G.
SW was supported by NASA (NNX14AH34G) and the DOE High Energy Physics
Program (DE-SC-00010676).
SCY was supported by the Basic Science Research (2013R1A1A2061842) program through
the National Research Foundation of Korea (NRF).
This work was granted access to the HPC resources of CINES under the
allocations c2013046608 and c2014046608 made by GENCI (Grand Equipement National
de Calcul Intensif).
This work also utilised computing resources of the M\'esocentre SIGAMM,
hosted by the Observatoire de la C\^ote d'Azur, Nice, France.

\appendix

\section{Details on the radiative transfer calculations and the atomic data}
\label{sect_cmfgen}

The radiation code \cmfgen\ \citep{hm98}, is used to determine the atomic level populations.
In these calculations we assume
a 1-D homologous expansion, although we have routines available that relax this assumption.
We solve the transfer equation in the comoving-frame as discussed by \cite{HD12},
and we explicitly allow for the time dependence of the radiation field. To obtain the Eddington factors,
we perform a formal solution in which time dependence is neglected. As we explicitly treat the time dependence
with the moment equations, and since the Eddington factors are ratios of moments, this approximation
has only a minor influence on the calculations. The assumption of 1D is an approximation and there is
ample observational and theoretical evidence, especially for core collapse SNe, that multi-dimensional effects are important
\citep[see review by][]{2012PTEP.2012aA309J}. A related effect, that can influence the results, is mixing.
As the mixing is likely macroscopic, and not microscopic \citep{fryxell_mueller_arnett_91},
it cannot be easily treated in 1-D calculations. Given the huge computational costs associated with full
3-D non-LTE time-dependent calculations, and our still limited understanding of SN ejecta,
1-D calculations are an essential analysis tool.

We compute the observer's spectrum in two ways. First, we compute the spectrum with the \cmfgen\ calculation.
To do this we perform a Lorentz transformation of the outer boundary comoving-frame  intensities,
and then integrate them in the usual way to obtain the observed flux. In the second method we use a separate
code, \cmfflux\ \citep{BH05_2D}. In this code we first solve for the radiation field, similar to the \cmfgen\
calculation. However we include all bound-bound transitions in our model atoms, and typically use finer
frequency and spatial grids. The CMF calculation provides the mean intensity J which is used to compute
the electron scattering emissivity. We then perform a formal solution in the Observer's frame utilising
Lorentz transforms of the comoving emissivities and opacities. In general spectra computed by the two methods
are in excellent agreement.

An explicit assumption of our approach is that the light-travel time effects are small. We cannot accurately
model SN when rapid changes are occurring, such as at shock breakout. That is, we can not accurately
model the spectrum for events changing faster than a timescale of $R_{\rm phot}/c$.

An essential requirement of the calculations is accurate atomic data, and since we are performing
non-LTE calculations we require $gf$ values, energy levels, accurate wavelengths, collisional data,
photo-ionisation cross-sections, auto-ionising data,  and charge exchange cross-sections. The quality
of the available data varies considerably with the atomic species, and the complexity of the electronic
configuration. When available, we use accurate energy levels, as obtained from the NIST website
\citep{NIST_V5,NIST}  (sometimes indirectly from NIST as Robert Kurucz uses accurate energies
in his calculations when they are available). For Fe group elements not all levels are known (particularly
for highly excited levels, and for elements other than Fe), and in that case we use energy levels calculated by
Kurucz \citep{Kur09_ATD, Kur_web}. These energy levels lead to inaccurate wavelengths although in our
SNe calculations our results do not reveal a major problem when using inaccurate wavelengths. In the
\cmfflux\ calculations we have the ability to omit transitions with unknown wavelengths.

For H and He the atomic data is excellent. For CNO the data is also of high quality, and the data for
elements with atomic number up to $Z=$\,20 is also reasonable. In general reasonable atomic data
is available for Fe, but for the other Fe group elements much less data is available. Due to the complexity
of their electronic configuration, the data quality for Fe (and other Fe group elements when available)
is generally of lower quality than that of CNO.

Oscillator strengths are available from a wide a variety of sources  such as the Opacity Project \citep{Sea87_OP}
and the Fe project \citep{HBE93_IP}, and most are  theoretical. Because of their completeness, extensive
use is made of the calculations by  Kurucz \citep{Kur09_ATD, Kur_web}.
For some species oscillator strengths have been updated with values from NIST \citep{NIST_V5,NIST}.
When available, we use theoretical collisional data from the literature. For many species such data is
unavailable, and we use the approximate values obtained from the expression of van Regemorter
\citep{Reg62_col}. Photo-ionisation cross section of elements with $Z\le$\,26 (and $Z$ even) are
available through the Opacity Project \citep{Sea87_OP}, the Fe project \citep{HBE93_IP} and
through many other calculations.\footnote{For example, by the atomic data group at Ohio State University.
The data from S.\,N. Nahar is available online at http://www.astronomy.ohiostate.edu/$\sim$nahar/.}
A potential drawback
of these calculations is that the resonances are not at the correct wavelengths, and are treated in LS coupling.
This can affect spectral comparisons, and potentially could influence non-LTE calculations through
the incorrect treatment of overlapping spectral features. For many species, such as Co, accurate photo-ionisation
data is unavailable. In such cases we use approximations.

In addition to the primary sources listed above, the atomic data for the present calculations
were taken from
\citet{SL74},
\citet{Men83_col},
\citet{NS83_LTDR},
\citet{NS84_CNO_LTDR},
\citet{LP89_C_seq},
\citet{Topbase93},
\citet{NP93_Sil},
\citet{BKM93_Mg_seq},
\citet{ZP94_FeII_col},
\citet{NP94_FeII_phot},
\citet{MEL95_Al_seq},
\citet{1995A&A...293..953Z},
\citet{ZP95_FeIII_col},
\citet{Zha96_FeIII_col},
\citet{1999JPhB...32.5507F},
\citet{2002CoPhC.145..311S},
\citet{2002A&A...385..716T},
and \citet{2004A&A...420..763B}.

Due to the complexity of non-LTE calculations it is difficult to gauge the influence of poor atomic data
on the calculations. We do run sensitivity tests to the size of the atomic models, and in some cases to
the atomic data (for those cases where we have multiple data sets). For many SN models we get consistent
fits across many epochs, and since the sensitivities to the atomic data are changing with the SN epoch
(since the physical conditions are changing), this suggests that uncertainties in the atomic data are not
having a major influence (at least qualitatively). However we have had occurrences where the
neglect of certain atomic data has had a major influence on the calculations \citep{d14_tech}. 

\section{Model properties for the full sample}

In this section, we present a summary of the model properties.
Table~\ref{tab_ejecta_glob_appendix} gives some progenitor and ejecta masses
for the full grid of 27 models.
Table~\ref{tab_ejecta_mass_appendix} provides the ejecta yields for the dominant species.
Table~\ref{tab_ejecta_surf_appendix} gives the composition of some important species
in the outermost ejecta layers (i.e., at the progenitor surface).

\begin{table*}
\begin{center}
\caption{Progenitor and ejecta properties. The last three columns give the ejecta velocity that bounds 99\%
of the corresponding species total mass. The integration is done inwards in velocity space for H and He, and
outwards for \iso{56}Ni.
\label{tab_ejecta_glob_appendix}}
\begin{tabular}{
l@{\hspace{5mm}}c@{\hspace{5mm}}c@{\hspace{5mm}}c@{\hspace{5mm}}
c@{\hspace{5mm}}
c@{\hspace{5mm}}c@{\hspace{5mm}}c@{\hspace{5mm}}c@{\hspace{5mm}}
c@{\hspace{5mm}}c@{\hspace{5mm}}c@{\hspace{5mm}}c@{\hspace{5mm}}
}
\hline
Model    &    $M_{\rm i}$  &     $M_{\rm f}$ & $R_{\star}$  &$M_{\rm r}$ & $M_{\rm e}$ &        $Z$     &  $E_{\rm kin}$  &  $V_{\rm 99, H}$&  $V_{\rm 99, He}$&
$V_{\rm 99, Ni}$  \\
               &        [\msun]      &        [\msun]      & [cm] &        [\msun] &        [\msun] &     &   [B] &      [\kms]  &      [\kms]  &      [\kms]
\\
%         Minitial           Mfinal         Mremnant   Rstar       Mejecta                Z         Ekin [B]          V99 HYD           V99 HE         V99
%         NICK
\hline
3p0Cx1  &  18.0  &   3.00  & 1.10(12) &   1.29  &   1.71  &     0.02  &    0.62  &  1.19(4)  &   1.83(3)  &   6.07(3)  \\
3p0Cx2  &  18.0  &   3.00  & 1.10(12)  &   1.29  &   1.71  &     0.02  &    0.62  &  1.10(4)  &   1.64(3)  &   9.06(3)  \\
3p0Ax1  &  18.0  &   3.00  & 1.10(12)  &   1.27  &   1.73  &     0.02  &    1.25  &   1.68(4)  &   2.84(3)  &   6.91(3)  \\
3p0Ax2  &  18.0  &   3.00  & 1.10(12)  &   1.28  &   1.72  &     0.02  &    1.24  &   1.58(4)  &   2.47(3)  &   1.06(4)  \\
3p0Bx2  &  18.0  &   3.00  & 1.10(12)  &   1.27  &   1.73  &     0.02  &    2.50  &   2.30(4)  &   3.73(3)  &   1.21(4)  \\
3p65Cx1  & 16.0  &   3.65  & 1.17(12) &   1.46  &   2.19  &     0.02  &    0.61  &  9.11(3)  &   1.91(3)  &   6.01(3)  \\
3p65Cx2  & 16.0  &   3.65  & 1.17(12) &   1.47  &   2.18  &     0.02  &    0.61  &  8.43(3)  &   1.63(3)  &   8.61(3)  \\
3p65Ax1  & 16.0  &   3.65  & 1.17(12) &   1.42  &   2.23  &     0.02  &    1.24  &   1.32(4)  &   3.03(3)  &   6.85(3)  \\
3p65Ax2  & 16.0  &   3.65  & 1.17(12) &   1.43  &   2.22  &     0.02  &    1.22  &   1.24(4)  &   2.62(3)  &   1.01(4)  \\
3p65Bx2  & 16.0  &   3.65  & 1.17(12) &   1.42  &   2.23  &     0.02  &    2.47  &   1.83(4)  &   4.00(3)  &   1.19(4)  \\
3p65Dx2  & 16.0  &   3.65  & 1.17(12) &   1.42  &   2.23  &     0.02  &    5.08  &   2.68(4)  &   4.40(3)  &   1.38(4)  \\
4p64Cx1  & 18.0  &   4.64  & 1.10(12)  &   1.49  &   3.15  &    0.004  &    0.62  &  7.41(3)  &   2.03(3)  &   5.44(3)  \\
4p64Ax1  & 18.0  &   4.64  & 1.10(12)  &   1.53  &   3.11  &    0.004  &    1.22  &   1.03(4)  &   3.46(3)  &   6.66(3)  \\
4p64Ax2  & 18.0  &   4.64  & 1.10(12)  &   1.52  &   3.12  &    0.004  &    1.22  &   9.40(3)  &   2.79(3)  &   9.69(3)  \\
4p64Bx1  & 18.0  &   4.64  & 1.10(12)  &   1.46  &   3.18  &    0.004  &    2.45  &   1.45(4)  &   3.71(3)  &   7.78(3)  \\
4p64Bx2  & 18.0  &   4.64  & 1.10(12)  &   1.47  &   3.17  &    0.004  &    2.46  &   1.39(4)  &   4.48(3)  &   1.17(4)  \\
4p64Dx2  & 18.0  &   4.64  & 1.10(12)  &   1.43  &   3.21  &    0.004  &    5.13  &   2.08(4)  &   4.67(3)  &   1.38(4)  \\
5p11Ax1  & 60.0  &   5.11  & 5.19(10) &   1.57  &   3.54  &     0.02  &    1.25  &    \dots  &   1.69(3)  &   6.00(3)  \\
5p11Ax2  & 60.0  &   5.11  & 5.19(10) &   1.49  &   3.62  &     0.02  &    1.29  &    \dots  &   2.71(3)  &   9.18(3)  \\
5p11Bx1  & 60.0  &   5.11  & 5.19(10) &   1.49  &   3.62  &     0.02  &    2.49  &    \dots  &   1.59(3)  &   7.08(3)  \\
5p11Bx2  & 60.0  &   5.11  & 5.19(10) &   1.48  &   3.63  &     0.02  &    2.49  &    \dots  &   1.74(3)  &   1.10(4)  \\
6p5Ax1  &  25.0  &   6.50  & 2.01(11) &   1.53  &   4.97  &     0.02  &    1.26  &    \dots  &   3.26(3)  &   5.94(3)  \\
6p5Ax2  &  25.0  &   6.50  & 2.01(11) &   1.55  &   4.95  &     0.02  &    1.25  &    \dots  &   2.55(3)  &   8.60(3)  \\
6p5Bx1  &  25.0  &   6.50  & 2.01(11) &   1.55  &   4.95  &     0.02  &    2.42  &    \dots  &   1.96(3)  &   6.57(3)  \\
6p5Bx2  &  25.0  &   6.50  & 2.01(11) &   1.52  &   4.98  &     0.02  &    2.43  &    \dots  &   3.38(3)  &   1.02(4)  \\
6p5Gx1  &  25.0  &   6.50  & 2.01(11) &   1.36  &   5.14  &     0.02  &    5.28  &    \dots  &   2.89(3)  &   8.91(3)  \\
6p5Gx2  &  25.0  &   6.50  & 2.01(11) &   1.32  &   5.18  &     0.02  &    5.30  &    \dots  &   3.09(3)  &   1.22(4)  \\
\hline
\end{tabular}
\end{center}
\end{table*}
%
%3.024     1.30e38         19,600     1.10e12      1.67      1.48   1.32
%3.636     2.16e38         21,700     1.17e12      2.03      1.59   1.43
%4.628     3.77e38         25,600     1.10e12      2.74      1.67   1.50
%5.090     3.84e38         52,100     2.70e11      2.33      1.59   1.47

\begin{table*}
\begin{center}
    \caption{
Cumulative yields for our grid of models at $\sim$\,2\,d after explosion. For \iso{56}Ni, we give the original mass,
i.e., prior to decay.
\label{tab_ejecta_mass_appendix}}
\begin{tabular}{
l@{\hspace{5mm}}c@{\hspace{5mm}}c@{\hspace{5mm}}c@{\hspace{5mm}}
c@{\hspace{5mm}}c@{\hspace{5mm}}c@{\hspace{5mm}}c@{\hspace{5mm}}
c@{\hspace{5mm}}c@{\hspace{5mm}}c@{\hspace{5mm}}c@{\hspace{5mm}}
c@{\hspace{5mm}}c@{\hspace{5mm}}c@{\hspace{5mm}}
}
\hline
Model    &   H   &   He  &   C   &   N   &   O   &   Si  &    S  &   Ca  &   Fe  & \iso{56}Ni \\
         &        [\msun] &        [\msun] &        [\msun] &        [\msun] &        [\msun] &        [\msun] &        [\msun] &        [\msun] &
[\msun] &        [\msun]       \\
\hline
%              M H             M He              M C              M N              M O             M Si              M S             M Ca
%              M Fe         M  56 Ni
3p0Cx1  &   7.81(-4)  & 1.30(0)   & 5.48(-2)  & 1.15(-2)  & 1.18(-1)  & 6.21(-2)  & 2.48(-2)  & 2.36(-3)  & 2.05(-3)  & 3.87(-2) \\
3p0Cx2  &   7.81(-4)  & 1.30(0)   & 5.44(-2)  & 1.14(-2)  & 1.19(-1)  & 6.29(-2)  & 2.51(-2)  & 2.39(-3)  & 1.95(-3)  & 3.93(-2) \\
3p0Ax1  &   7.92(-4)  & 1.31(0)   & 5.48(-2)  & 1.15(-2)  & 1.24(-1)  & 5.63(-2)  & 2.12(-2)  & 2.60(-3)  & 2.54(-3)  & 5.78(-2) \\
3p0Ax2  &   7.73(-4)  & 1.31(0)   & 5.42(-2)  & 1.14(-2)  & 1.21(-1)  & 5.50(-2)  & 2.07(-2)  & 2.56(-3)  & 2.46(-3)  & 6.15(-2) \\
3p0Bx2  &   7.72(-4)  & 1.33(0)   & 5.39(-2)  & 1.15(-2)  & 1.19(-1)  & 4.93(-2)  & 2.00(-2)  & 3.32(-3)  & 2.78(-3)  & 7.02(-2) \\
3p65Cx1  &  4.74(-3)  & 1.47(0)   & 9.39(-2)  & 1.08(-2)  & 3.04(-1)  & 7.49(-2)  & 2.37(-2)  & 2.97(-3)  & 2.66(-3)  & 4.45(-2) \\
3p65Cx2  &  4.70(-3)  & 1.47(0)   & 9.31(-2)  & 1.08(-2)  & 3.04(-1)  & 7.57(-2)  & 2.39(-2)  & 3.00(-3)  & 2.64(-3)  & 4.50(-2) \\
3p65Ax1  &  4.99(-3)  & 1.49(0)   & 9.40(-2)  & 1.09(-2)  & 3.02(-1)  & 7.38(-2)  & 2.36(-2)  & 3.94(-3)  & 3.12(-3)  & 7.42(-2) \\
3p65Ax2  &  4.72(-3)  & 1.48(0)   & 9.37(-2)  & 1.08(-2)  & 3.01(-1)  & 7.30(-2)  & 2.33(-2)  & 3.91(-3)  & 3.13(-3)  & 7.66(-2) \\
3p65Bx2  &  4.95(-3)  & 1.51(0)   & 9.35(-2)  & 1.08(-2)  & 2.83(-1)  & 7.24(-2)  & 2.43(-2)  & 4.93(-3)  & 3.43(-3)  & 1.01(-1) \\
3p65Dx2  &  4.64(-3)  & 1.55(0)   & 9.33(-2)  & 1.07(-2)  & 2.51(-1)  & 7.64(-2)  & 2.86(-2)  & 5.48(-3)  & 3.60(-3)  & 1.05(-1) \\
4p64Cx1  &  1.66(-2)  & 1.72(0)   & 1.54(-1)  & 3.28(-3)  & 6.91(-1)  & 9.08(-2)  & 2.61(-2)  & 3.24(-3)  & 1.25(-3)  & 3.96(-2) \\
4p64Ax1  &  1.72(-2)  & 1.69(0)   & 1.48(-1)  & 3.23(-3)  & 6.85(-1)  & 9.35(-2)  & 2.65(-2)  & 3.58(-3)  & 1.42(-3)  & 6.20(-2) \\
4p64Ax2  &  1.70(-2)  & 1.69(0)   & 1.49(-1)  & 3.22(-3)  & 6.79(-1)  & 9.43(-2)  & 2.74(-2)  & 3.77(-3)  & 1.42(-3)  & 6.81(-2) \\
4p64Bx1  &  1.65(-2)  & 1.76(0)   & 1.52(-1)  & 3.26(-3)  & 6.75(-1)  & 1.01(-1)  & 2.66(-2)  & 4.90(-3)  & 2.44(-3)  & 8.76(-2) \\
4p64Bx2  &  1.67(-2)  & 1.75(0)   & 1.50(-1)  & 3.27(-3)  & 6.77(-1)  & 1.01(-1)  & 2.63(-2)  & 4.82(-3)  & 2.21(-3)  & 8.85(-2) \\
4p64Dx2  &  1.55(-2)  & 1.81(0)   & 1.50(-1)  & 3.28(-3)  & 6.44(-1)  & 1.18(-1)  & 3.42(-2)  & 6.15(-3)  & 2.29(-3)  & 1.09(-1) \\
5p11Ax1  &     0      & 3.15(-1)  & 8.92(-1)  &    0      & 1.42(0)   & 1.28(-1)  & 4.00(-2)  & 5.98(-3)  & 4.57(-3)  & 8.94(-2) \\
5p11Ax2  &     0      & 3.26(-1)  & 9.23(-1)  &    0      & 1.44(0)   & 1.28(-1)  & 4.04(-2)  & 6.13(-3)  & 4.69(-3)  & 9.46(-2) \\
5p11Bx1  &     0      & 3.37(-1)  & 8.86(-1)  &    0      & 1.40(0)   & 1.53(-1)  & 5.24(-2)  & 8.96(-3)  & 7.63(-3)  & 1.89(-1) \\
5p11Bx2  &     0      & 3.37(-1)  & 8.88(-1)  &    0      & 1.40(0)   & 1.53(-1)  & 5.21(-2)  & 8.88(-3)  & 7.83(-3)  & 1.88(-1) \\
%5p1Gx2  &   1.37(-6)  & 1.72(0)   & 4.66(-1)  & 9.10(-3)  & 1.10(0)   & 8.93(-2)  & 3.48(-2)  & 7.08(-3)  & 4.44(-3)  & 5.52(-2) \\
6p5Ax1  &      0      & 1.67(0)   & 4.13(-1)  & 7.59(-3)  & 1.57(0)   & 2.12(-1)  & 9.31(-2)  & 9.30(-3)  & 6.19(-3)  & 9.90(-2) \\
6p5Ax2  &      0      & 1.66(0)   & 4.11(-1)  & 7.52(-3)  & 1.57(0)   & 2.14(-1)  & 9.40(-2)  & 9.51(-3)  & 6.31(-3)  & 1.02(-1) \\
6p5Bx1  &      0      & 1.61(0)   & 3.94(-1)  & 7.21(-3)  & 1.53(0)   & 2.09(-1)  & 8.99(-2)  & 1.12(-2)  & 1.10(-2)  & 2.44(-1) \\
6p5Bx2  &      0      & 1.62(0)   & 3.99(-1)  & 7.21(-3)  & 1.54(0)   & 2.08(-1)  & 8.97(-2)  & 1.12(-2)  & 1.08(-2)  & 2.42(-1) \\
6p5Gx1  &      0      & 1.79(0)   & 3.93(-1)  & 7.56(-3)  & 1.53(0)   & 2.16(-1)  & 8.36(-2)  & 1.33(-2)  & 1.34(-2)  & 2.89(-1) \\
6p5Gx2  &      0      & 1.78(0)   & 4.00(-1)  & 7.53(-3)  & 1.55(0)   & 2.17(-1)  & 8.38(-2)  & 1.34(-2)  & 1.31(-2)  & 2.88(-1) \\
\hline
\end{tabular}
\end{center}
\end{table*}

\begin{table*}
\begin{center}
    \caption{
Mass fractions for some important species in the outermost mass shell for our grid of models.
This shell corresponds to the progenitor surface.
\label{tab_ejecta_surf_appendix}}
\begin{tabular}{
l@{\hspace{5mm}}c@{\hspace{5mm}}c@{\hspace{5mm}}c@{\hspace{5mm}}
c@{\hspace{5mm}}c@{\hspace{5mm}}c@{\hspace{5mm}}c@{\hspace{5mm}}
c@{\hspace{5mm}}c@{\hspace{5mm}}}
\hline
%             Xs H            Xs He             Xs C             Xs N             Xs O            Xs Si             Xs S            Xs Ca
%             Xs Fe
Model    &   $X_{\rm H,s}$ &  $X_{\rm He,s}$  &   $X_{\rm C,s}$ &   $X_{\rm N,s}$ &   $X_{\rm O,s}$ &   $X_{\rm Si,s}$ &   $X_{\rm S,s}$ &   $X_{\rm
Ca,s}$ &  $X_{\rm Fe,s}$  \\
\hline
3p0Cx1  &   7.268(-2)  & 9.088(-1)  & 2.009(-4)  & 1.330(-2)  & 4.019(-4)  & 7.348(-4)  & 2.969(-4)  & 4.579(-5)  & 9.086(-4)  \\
3p0Cx2  &   7.285(-2)  & 9.084(-1)  & 1.989(-4)  & 1.329(-2)  & 3.997(-4)  & 7.325(-4)  & 3.648(-4)  & 6.436(-5)  & 9.086(-4)  \\
3p0Ax1  &   7.167(-2)  & 9.096(-1)  & 2.009(-4)  & 1.329(-2)  & 4.008(-4)  & 7.346(-4)  & 3.648(-4)  & 6.437(-5)  & 9.086(-4)  \\
3p0Ax2  &   7.102(-2)  & 9.102(-1)  & 2.010(-4)  & 1.330(-2)  & 4.011(-4)  & 7.352(-4)  & 3.651(-4)  & 6.441(-5)  & 9.086(-4)  \\
3p0Bx2  &   6.987(-2)  & 9.116(-1)  & 2.009(-4)  & 1.329(-2)  & 3.998(-4)  & 7.347(-4)  & 2.969(-4)  & 4.578(-5)  & 9.086(-4)  \\
3p65Cx1  &  1.481(-1)  & 8.334(-1)  & 1.981(-4)  & 1.321(-2)  & 4.832(-4)  & 7.353(-4)  & 2.971(-4)  & 4.582(-5)  & 9.086(-4)  \\
3p65Cx2  &  1.479(-1)  & 8.334(-1)  & 1.988(-4)  & 1.319(-2)  & 4.846(-4)  & 7.344(-4)  & 3.647(-4)  & 6.435(-5)  & 9.086(-4)  \\
3p65Ax1  &  1.480(-1)  & 8.332(-1)  & 1.980(-4)  & 1.320(-2)  & 4.831(-4)  & 7.352(-4)  & 3.651(-4)  & 6.442(-5)  & 9.086(-4)  \\
3p65Ax2  &  1.479(-1)  & 8.334(-1)  & 1.978(-4)  & 1.319(-2)  & 4.826(-4)  & 7.344(-4)  & 3.647(-4)  & 6.435(-5)  & 9.086(-4)  \\
3p65Bx2  &  1.471(-1)  & 8.344(-1)  & 1.981(-4)  & 1.321(-2)  & 4.832(-4)  & 7.353(-4)  & 2.971(-4)  & 4.582(-5)  & 9.086(-4)  \\
3p65Dx2  &  1.449(-1)  & 8.366(-1)  & 1.979(-4)  & 1.319(-2)  & 4.817(-4)  & 7.346(-4)  & 2.968(-4)  & 4.578(-5)  & 9.086(-4)  \\
4p64Cx1  &  3.161(-1)  & 6.802(-1)  & 4.041(-5)  & 2.581(-3)  & 1.760(-4)  & 1.470(-4)  & 7.302(-5)  & 1.288(-5)  & 1.815(-4)  \\
4p64Ax1  &  3.161(-1)  & 6.802(-1)  & 4.051(-5)  & 2.581(-3)  & 1.760(-4)  & 1.470(-4)  & 7.302(-5)  & 1.288(-5)  & 1.815(-4)  \\
4p64Ax2  &  3.158(-1)  & 6.805(-1)  & 4.047(-5)  & 2.578(-3)  & 1.759(-4)  & 1.469(-4)  & 7.294(-5)  & 1.287(-5)  & 1.815(-4)  \\
4p64Bx1  &  3.151(-1)  & 6.812(-1)  & 4.051(-5)  & 2.581(-3)  & 1.750(-4)  & 1.470(-4)  & 7.302(-5)  & 1.288(-5)  & 1.815(-4)  \\
4p64Bx2  &  3.151(-1)  & 6.812(-1)  & 4.051(-5)  & 2.581(-3)  & 1.750(-4)  & 1.470(-4)  & 7.302(-5)  & 1.288(-5)  & 1.815(-4)  \\
4p64Dx2  &  3.091(-1)  & 6.872(-1)  & 4.081(-5)  & 2.581(-3)  & 1.680(-4)  & 1.470(-4)  & 7.302(-5)  & 1.288(-5)  & 1.815(-4)  \\
5p11Ax1  &     0     & 3.611(-1)  & 5.051(-1)  &    0     & 1.100(-1)  & 7.362(-4)  & 3.651(-4)  & 6.442(-5)  & 9.097(-4)  \\
5p11Ax2  &     0     & 3.611(-1)  & 5.051(-1)  &    0     & 1.100(-1)  & 7.362(-4)  & 3.651(-4)  & 6.442(-5)  & 9.097(-4)  \\
5p11Bx1  &     0     & 3.611(-1)  & 5.051(-1)  &    0     & 1.100(-1)  & 7.362(-4)  & 3.651(-4)  & 6.442(-5)  & 9.097(-4)  \\
5p11Bx2  &     0     & 3.611(-1)  & 5.051(-1)  &    0     & 1.100(-1)  & 7.362(-4)  & 3.651(-4)  & 6.442(-5)  & 9.097(-4)  \\
% 5p1Gx2  &   9.883(-5)  & 9.813(-1)  & 2.861(-4)  & 1.330(-2)  & 3.231(-4)  & 7.352(-4)  & 2.971(-4)  & 4.582(-5)  & 9.086(-4)  \\
6p5Ax1  &      0     & 9.813(-1)  & 4.147(-4)  & 1.309(-2)  & 3.098(-4)  & 7.345(-4)  & 3.647(-4)  & 6.435(-5)  & 9.086(-4)  \\
6p5Ax2  &      0     & 9.813(-1)  & 4.147(-4)  & 1.309(-2)  & 3.098(-4)  & 7.345(-4)  & 3.647(-4)  & 6.435(-5)  & 9.086(-4)  \\
6p5Bx1  &      0     & 9.813(-1)  & 4.147(-4)  & 1.309(-2)  & 3.098(-4)  & 7.345(-4)  & 3.647(-4)  & 6.435(-5)  & 9.086(-4)  \\
6p5Bx2  &      0     & 9.813(-1)  & 4.147(-4)  & 1.309(-2)  & 3.098(-4)  & 7.345(-4)  & 3.647(-4)  & 6.435(-5)  & 9.086(-4)  \\
6p5Gx1  &      0     & 9.813(-1)  & 4.147(-4)  & 1.309(-2)  & 3.098(-4)  & 7.345(-4)  & 3.647(-4)  & 6.435(-5)  & 9.086(-4)  \\
6p5Gx2  &      0     & 9.813(-1)  & 4.147(-4)  & 1.309(-2)  & 3.098(-4)  & 7.345(-4)  & 3.647(-4)  & 6.435(-5)  & 9.086(-4)  \\
\hline
\end{tabular}
\end{center}
\end{table*}

\section{Photometric characteristics}

Tables~\ref{tab_lc_mod1_appendix}--\ref{tab_lc_mod2_appendix}--\ref{tab_lc_mod3_appendix} give
a summary of the photometric properties of our grid of models, including the rise time,
value at maximum brightness, and post-maximum decline for the bolometric luminosity,
the luminosity falling between 1000\,\AA\ and 2.5\,$\mu$m ($L_{\rm UVOIR}$,
and the photometric bands $U$, $B$, $V$, $R$, $I$, $J$, $H$, and $K_{\rm s}$.

\begin{table*}
\begin{center}
    \caption{
    Some light curves properties of our models. For each entry, we give the rise to maximum, the value at peak,
and the magnitude change between peak and 15\,d later.
\label{tab_lc_mod1_appendix}}
\begin{tabular}{
l@{\hspace{3mm}}|
c@{\hspace{3mm}}c@{\hspace{3mm}}c@{\hspace{3mm}}|
c@{\hspace{3mm}}c@{\hspace{3mm}}c@{\hspace{3mm}}|
c@{\hspace{3mm}}c@{\hspace{3mm}}c@{\hspace{3mm}}|
c@{\hspace{3mm}}c@{\hspace{3mm}}c@{\hspace{3mm}}|
c@{\hspace{3mm}}c@{\hspace{3mm}}c@{\hspace{3mm}}
}
\hline
Model   & \multicolumn{3}{c}{$L_{\rm bol}$}   & \multicolumn{3}{c}{$L_{\rm UVOIR}$}   & \multicolumn{3}{c}{$U$}   & \multicolumn{3}{c}{$B$}   \\
               & $t_{\rm rise}$ & Max. & $\Delta M_{\rm 15}$ & $t_{\rm rise}$ & Max. & $\Delta M_{\rm 15}$ & $t_{\rm rise}$ & Max. & $\Delta M_{\rm 15}$ & $t_{\rm rise}$ & Max. & $\Delta M_{\rm 15}$ \\
               & [d]  & [erg\,s$^{-1}$] & [mag]  & [d]  & [erg\,s$^{-1}$] & [mag]    & [d]  &[mag] & [mag]  & [d]  & [mag] & [mag]  \\
\hline
% Model  & t_rise_lbol  &   & Max_lbol  &   DM15_lbol t_rise_luvoir  &  Max_luvoir  & DM15_luvoir  &   & t_rise_U  &   &   & Max_U  &   &   DM15_U  &   & t_rise_B  &   &   & Max_B  &   &   DM15_B
3p0Cx1	    &   2.530(1)   &   9.147(41)  &   6.837(-1)  &   2.499(1)   &   7.047(41)  &   7.784(-1)  &   2.366(1)   &  -1.494(1)   &   1.130(0)   &   2.378(1)   &  -1.587(1)   &   1.073(0)   \\
3p0Cx2	    &   2.375(1)   &   9.060(41)  &   7.124(-1)  &   2.359(1)   &   6.621(41)  &   7.309(-1)  &   2.221(1)   &  -1.449(1)   &   4.694(-1)  &   2.178(1)   &  -1.551(1)   &   6.891(-1)  \\
3p0Ax1	    &   2.296(1)   &   1.414(42)  &   7.221(-1)  &   2.263(1)   &   1.135(42)  &   8.486(-1)  &   2.092(1)   &  -1.577(1)   &   1.418(0)   &   2.141(1)   &  -1.655(1)   &   1.294(0)   \\
3p0Ax2	    &   2.228(1)   &   1.480(42)  &   7.401(-1)  &   2.214(1)   &   1.129(42)  &   7.936(-1)  &   2.178(1)   &  -1.535(1)   &   8.424(-1)  &   2.136(1)   &  -1.634(1)   &   1.009(0)   \\
3p0Bx2	    &   1.934(1)   &   1.818(42)  &   8.832(-1)  &   1.921(1)   &   1.417(42)  &   9.780(-1)  &   1.814(1)   &  -1.575(1)   &   8.646(-1)  &   1.829(1)   &  -1.674(1)   &   1.376(0)   \\
3p65Cx1	    &   2.826(1)   &   9.412(41)  &   5.636(-1)  &   2.770(1)   &   7.093(41)  &   6.352(-1)  &   2.437(1)   &  -1.484(1)   &   8.302(-1)  &   2.482(1)   &  -1.581(1)   &   8.027(-1)  \\
3p65Cx2	    &   2.665(1)   &   9.386(41)  &   5.389(-1)  &   2.630(1)   &   6.787(41)  &   5.361(-1)  &   2.559(1)   &  -1.449(1)   &   2.312(-1)  &   2.489(1)   &  -1.546(1)   &   4.728(-1)  \\
3p65Ax1	    &   2.631(1)   &   1.648(42)  &   6.157(-1)  &   2.579(1)   &   1.290(42)  &   6.956(-1)  &   2.387(1)   &  -1.576(1)   &   1.157(0)   &   2.410(1)   &  -1.662(1)   &   1.028(0)   \\
3p65Ax2	    &   2.487(1)   &   1.661(42)  &   5.455(-1)  &   2.455(1)   &   1.243(42)  &   5.896(-1)  &   2.189(1)   &  -1.539(1)   &   5.184(-1)  &   2.231(1)   &  -1.635(1)   &   6.811(-1)  \\
3p65Bx2	    &   2.209(1)   &   2.268(42)  &   6.983(-1)  &   2.180(1)   &   1.739(42)  &   7.631(-1)  &   2.022(1)   &  -1.590(1)   &   8.292(-1)  &   2.043(1)   &  -1.687(1)   &   1.022(0)   \\
3p65Dx2	    &   1.767(1)   &   2.770(42)  &   8.868(-1)  &   1.755(1)   &   2.210(42)  &   9.657(-1)  &   1.673(1)   &  -1.651(1)   &   1.250(0)   &   1.705(1)   &  -1.732(1)   &   1.504(0)   \\
4p64Cx1	    &   3.183(1)   &   7.623(41)  &   4.519(-1)  &   3.123(1)   &   5.572(41)  &   5.216(-1)  &   2.974(1)   &  -1.444(1)   &   6.655(-1)  &   2.931(1)   &  -1.542(1)   &   7.126(-1)  \\
4p64Ax1	    &   2.716(1)   &   1.318(42)  &   5.351(-1)  &   2.662(1)   &   9.871(41)  &   6.198(-1)  &   2.410(1)   &  -1.535(1)   &   1.051(0)   &   2.422(1)   &  -1.625(1)   &   1.014(0)   \\
4p64Ax2	    &   2.678(1)   &   1.427(42)  &   5.885(-1)  &   2.624(1)   &   1.034(42)  &   6.290(-1)  &   2.311(1)   &  -1.519(1)   &   4.769(-1)  &   2.363(1)   &  -1.599(1)   &   6.457(-1)  \\
4p64Bx1	    &   2.292(1)   &   2.008(42)  &   6.000(-1)  &   2.221(1)   &   1.586(42)  &   6.911(-1)  &   2.114(1)   &  -1.622(1)   &   1.487(0)   &   2.103(1)   &  -1.697(1)   &   1.296(0)   \\
4p64Bx2	    &   2.285(1)   &   2.098(42)  &   6.670(-1)  &   2.256(1)   &   1.552(42)  &   7.109(-1)  &   2.126(1)   &  -1.567(1)   &   7.612(-1)  &   2.115(1)   &  -1.661(1)   &   9.729(-1)  \\
4p64Dx2	    &   1.941(1)   &   2.707(42)  &   7.158(-1)  &   1.916(1)   &   2.050(42)  &   7.328(-1)  &   1.794(1)   &  -1.610(1)   &   1.008(0)   &   1.777(1)   &  -1.703(1)   &   1.070(0)   \\
5p11Ax1	    &   3.584(1)   &   1.590(42)  &   3.939(-1)  &   3.506(1)   &   1.182(42)  &   4.234(-1)  &   3.208(1)   &  -1.524(1)   &   6.838(-1)  &   3.256(1)   &  -1.633(1)   &   5.941(-1)  \\
5p11Ax2	    &   3.088(1)   &   1.703(42)  &   3.719(-1)  &   3.030(1)   &   1.253(42)  &   4.093(-1)  &   2.679(1)   &  -1.533(1)   &   5.063(-1)  &   2.747(1)   &  -1.631(1)   &   4.915(-1)  \\
5p11Bx1	    &   3.143(1)   &   3.570(42)  &   4.532(-1)  &   3.106(1)   &   2.898(42)  &   5.148(-1)  &   2.931(1)   &  -1.691(1)   &   1.054(0)   &   2.916(1)   &  -1.763(1)   &   8.763(-1)  \\
5p11Bx2	    &   2.919(1)   &   3.460(42)  &   4.310(-1)  &   2.860(1)   &   2.666(42)  &   4.844(-1)  &   2.474(1)   &  -1.651(1)   &   7.994(-1)  &   2.513(1)   &  -1.729(1)   &   7.407(-1)  \\
5p1Gx2	    &   2.020(1)   &   1.452(42)  &   7.732(-1)  &   1.994(1)   &   1.060(42)  &   7.960(-1)  &   1.866(1)   &  -1.490(1)   &   7.930(-1)  &   1.874(1)   &  -1.604(1)   &   1.001(0)   \\
6p5Ax1	    &   4.212(1)   &   1.495(42)  &   2.262(-1)  &   4.057(1)   &   1.079(42)  &   2.449(-1)  &   3.634(1)   &  -1.497(1)   &   2.554(-1)  &   3.668(1)   &  -1.606(1)   &   3.623(-1)  \\
6p5Ax2	    &   3.539(1)   &   1.623(42)  &   2.769(-1)  &   3.420(1)   &   1.169(42)  &   3.041(-1)  &   3.256(1)   &  -1.525(1)   &   4.026(-1)  &   3.129(1)   &  -1.609(1)   &   3.887(-1)  \\
6p5Bx1	    &   3.978(1)   &   3.934(42)  &   3.156(-1)  &   3.901(1)   &   3.074(42)  &   3.511(-1)  &   3.529(1)   &  -1.674(1)   &   6.020(-1)  &   3.581(1)   &  -1.753(1)   &   6.281(-1)  \\
6p5Bx2	    &   3.562(1)   &   3.856(42)  &   3.186(-1)  &   3.459(1)   &   2.916(42)  &   3.590(-1)  &   2.940(1)   &  -1.665(1)   &   5.378(-1)  &   2.980(1)   &  -1.732(1)   &   5.146(-1)  \\
6p5Gx1	    &   3.041(1)   &   5.261(42)  &   4.224(-1)  &   3.021(1)   &   4.232(42)  &   4.686(-1)  &   2.719(1)   &  -1.738(1)   &   8.600(-1)  &   2.745(1)   &  -1.801(1)   &   7.374(-1)  \\
6p5Gx2      &   2.905(1)   &   5.212(42)  &   4.214(-1)  &   2.848(1)   &   4.051(42)  &   4.731(-1)  &   2.608(1)   &  -1.711(1)   &   8.706(-1)  &   2.570(1)   &  -1.780(1)   &   8.509(-1)  \\
\hline
\end{tabular}
\end{center}
\end{table*}

\begin{table*}
\begin{center}
    \caption{Same as Table~\ref{tab_lc_mod1_appendix}, for now for the $V$, $R$, and $I$ bands.
\label{tab_lc_mod2_appendix}}
\begin{tabular}{
l@{\hspace{3mm}}|
c@{\hspace{3mm}}c@{\hspace{3mm}}c@{\hspace{3mm}}|
c@{\hspace{3mm}}c@{\hspace{3mm}}c@{\hspace{3mm}}|
c@{\hspace{3mm}}c@{\hspace{3mm}}c@{\hspace{3mm}}|
c@{\hspace{3mm}}c@{\hspace{3mm}}c@{\hspace{3mm}}|
c@{\hspace{3mm}}c@{\hspace{3mm}}c@{\hspace{3mm}}
}
\hline
Model    & \multicolumn{3}{c}{$V$}   & \multicolumn{3}{c}{$R$}   & \multicolumn{3}{c}{$I$}   \\
      & $t_{\rm rise}$ & Max. & $\Delta M_{\rm 15}$ & $t_{\rm rise}$ & Max. & $\Delta M_{\rm 15}$ & $t_{\rm rise}$ & Max. & $\Delta M_{\rm 15}$ \\
                & [d]  &[mag] & [mag]  & [d]  &[mag] & [mag]   & [d]  & [mag] & [mag]  \\
\hline
%   &   & t_rise_V  &   &   & Max_V  &   &   DM15_V  &   & t_rise_R  &   &   & Max_R  &   &   DM15_R  &   & t_rise_I  &   &   & Max_I  &   &   DM15_I
3p0Cx1	    &   2.472(1)   &  -1.654(1)   &   8.496(-1)  &   2.530(1)   &  -1.674(1)   &   7.056(-1)  &   2.736(1)   &  -1.680(1)   &   5.619(-1)  \\
3p0Cx2	    &   2.284(1)   &  -1.649(1)   &   8.160(-1)  &   2.350(1)   &  -1.678(1)   &   7.966(-1)  &   2.500(1)   &  -1.694(1)   &   6.794(-1)  \\
3p0Ax1	    &   2.237(1)   &  -1.709(1)   &   9.694(-1)  &   2.297(1)   &  -1.716(1)   &   6.774(-1)  &   2.643(1)   &  -1.710(1)   &   4.147(-1)  \\
3p0Ax2	    &   2.189(1)   &  -1.709(1)   &   9.159(-1)  &   2.210(1)   &  -1.727(1)   &   7.833(-1)  &   2.419(1)   &  -1.729(1)   &   5.728(-1)  \\
3p0Bx2	    &   1.894(1)   &  -1.732(1)   &   1.135(0)   &   1.903(1)   &  -1.748(1)   &   9.340(-1)  &   2.275(1)   &  -1.742(1)   &   7.092(-1)  \\
3p65Cx1	    &   2.731(1)   &  -1.655(1)   &   6.946(-1)  &   2.867(1)   &  -1.678(1)   &   6.146(-1)  &   3.064(1)   &  -1.694(1)   &   5.213(-1)  \\
3p65Cx2	    &   2.512(1)   &  -1.651(1)   &   5.876(-1)  &   2.610(1)   &  -1.680(1)   &   5.995(-1)  &   2.801(1)   &  -1.704(1)   &   5.346(-1)  \\
3p65Ax1	    &   2.551(1)   &  -1.723(1)   &   7.828(-1)  &   2.668(1)   &  -1.734(1)   &   6.099(-1)  &   2.972(1)   &  -1.740(1)   &   3.855(-1)  \\
3p65Ax2	    &   2.384(1)   &  -1.719(1)   &   6.861(-1)  &   2.483(1)   &  -1.739(1)   &   6.192(-1)  &   2.745(1)   &  -1.755(1)   &   4.828(-1)  \\
3p65Bx2	    &   2.180(1)   &  -1.755(1)   &   9.028(-1)  &   2.225(1)   &  -1.771(1)   &   7.666(-1)  &   2.484(1)   &  -1.775(1)   &   5.134(-1)  \\
3p65Dx2	    &   1.747(1)   &  -1.781(1)   &   1.101(0)   &   1.756(1)   &  -1.789(1)   &   9.144(-1)  &   2.448(1)   &  -1.777(1)   &   9.602(-1)  \\
4p64Cx1	    &   3.082(1)   &  -1.629(1)   &   5.895(-1)  &   3.208(1)   &  -1.653(1)   &   4.908(-1)  &   3.335(1)   &  -1.676(1)   &   3.852(-1)  \\
4p64Ax1	    &   2.625(1)   &  -1.692(1)   &   7.109(-1)  &   2.830(1)   &  -1.708(1)   &   5.696(-1)  &   3.032(1)   &  -1.728(1)   &   3.959(-1)  \\
4p64Ax2	    &   2.545(1)   &  -1.698(1)   &   7.175(-1)  &   2.675(1)   &  -1.723(1)   &   6.830(-1)  &   2.855(1)   &  -1.745(1)   &   5.184(-1)  \\
4p64Bx1	    &   2.245(1)   &  -1.739(1)   &   7.439(-1)  &   2.360(1)   &  -1.744(1)   &   5.079(-1)  &   2.621(1)   &  -1.759(1)   &   2.354(-1)  \\
4p64Bx2	    &   2.212(1)   &  -1.743(1)   &   8.557(-1)  &   2.288(1)   &  -1.761(1)   &   7.248(-1)  &   2.505(1)   &  -1.778(1)   &   4.733(-1)  \\
4p64Dx2	    &   1.889(1)   &  -1.774(1)   &   8.477(-1)  &   1.964(1)   &  -1.788(1)   &   7.378(-1)  &   2.192(1)   &  -1.803(1)   &   5.219(-1)  \\
5p11Ax1	    &   3.442(1)   &  -1.715(1)   &   4.745(-1)  &   3.616(1)   &  -1.734(1)   &   4.108(-1)  &   3.920(1)   &  -1.747(1)   &   2.772(-1)  \\
5p11Ax2	    &   2.947(1)   &  -1.722(1)   &   4.635(-1)  &   3.111(1)   &  -1.741(1)   &   4.293(-1)  &   3.390(1)   &  -1.759(1)   &   2.904(-1)  \\
5p11Bx1	    &   3.097(1)   &  -1.810(1)   &   5.142(-1)  &   3.287(1)   &  -1.811(1)   &   4.023(-1)  &   3.773(1)   &  -1.814(1)   &   2.034(-1)  \\
5p11Bx2	    &   2.793(1)   &  -1.808(1)   &   5.753(-1)  &   3.008(1)   &  -1.818(1)   &   4.540(-1)  &   3.518(1)   &  -1.830(1)   &   2.501(-1)  \\
5p1Gx2	    &   1.928(1)   &  -1.711(1)   &   1.008(0)   &   1.984(1)   &  -1.729(1)   &   8.876(-1)  &   2.247(1)   &  -1.739(1)   &   5.187(-1)  \\
6p5Ax1	    &   3.941(1)   &  -1.702(1)   &   2.804(-1)  &   4.274(1)   &  -1.731(1)   &   2.656(-1)  &   4.564(1)   &  -1.754(1)   &   2.064(-1)  \\
6p5Ax2	    &   3.351(1)   &  -1.707(1)   &   3.448(-1)  &   3.529(1)   &  -1.737(1)   &   2.999(-1)  &   3.691(1)   &  -1.764(1)   &   2.357(-1)  \\
6p5Bx1	    &   3.863(1)   &  -1.815(1)   &   3.708(-1)  &   4.228(1)   &  -1.830(1)   &   3.116(-1)  &   4.565(1)   &  -1.849(1)   &   2.034(-1)  \\
6p5Bx2	    &   3.426(1)   &  -1.808(1)   &   4.207(-1)  &   3.714(1)   &  -1.831(1)   &   3.539(-1)  &   4.026(1)   &  -1.853(1)   &   2.563(-1)  \\
6p5Gx1	    &   3.034(1)   &  -1.853(1)   &   5.216(-1)  &   3.259(1)   &  -1.856(1)   &   3.841(-1)  &   3.829(1)   &  -1.861(1)   &   2.065(-1)  \\
6p5Gx2      &   2.813(1)   &  -1.845(1)   &   5.403(-1)  &   3.034(1)   &  -1.861(1)   &   4.189(-1)  &   3.420(1)   &  -1.878(1)   &   2.454(-1)  \\
\hline
\end{tabular}
\end{center}
\end{table*}

\begin{table*}
\begin{center}
    \caption{
   Same as Table~\ref{tab_lc_mod1_appendix}, for now for the $J$, $H$, and $K$ bands.
\label{tab_lc_mod3_appendix}}
\begin{tabular}{
l@{\hspace{3mm}}|
c@{\hspace{3mm}}c@{\hspace{3mm}}c@{\hspace{3mm}}|
c@{\hspace{3mm}}c@{\hspace{3mm}}c@{\hspace{3mm}}|
c@{\hspace{3mm}}c@{\hspace{3mm}}c@{\hspace{3mm}}|
c@{\hspace{3mm}}c@{\hspace{3mm}}c@{\hspace{3mm}}|
c@{\hspace{3mm}}c@{\hspace{3mm}}c@{\hspace{3mm}}
}
\hline
\hline
Model    & \multicolumn{3}{c}{$J$}   & \multicolumn{3}{c}{$H$}   & \multicolumn{3}{c}{$K_{\rm s}$}   \\
      & $t_{\rm rise}$ & Max. & $\Delta M_{\rm 15}$ & $t_{\rm rise}$ & Max. & $\Delta M_{\rm 15}$ & $t_{\rm rise}$ & Max. & $\Delta M_{\rm 15}$ \\
                & [d]  &[mag] & [mag]  & [d]  &[mag] & [mag]   & [d]  & [mag] & [mag]  \\
\hline
%  &   & t_rise_J  &   &   & Max_J  &   &   DM15_J  &   & t_rise_H  &   &   & Max_H  &   &   DM15_H  &   & t_rise_K  &   &   & Max_K  &   &   DM15_K
3p0Cx1	      &   2.641(1)   &  -1.675(1)   &   4.906(-1)  &   2.806(1)   &  -1.693(1)   &   4.347(-1)  &   2.750(1)   &  -1.708(1)   &   4.098(-1)  \\
3p0Cx2	      &   2.284(1)   &  -1.692(1)   &   6.945(-1)  &   2.454(1)   &  -1.708(1)   &   6.869(-1)  &   2.454(1)   &  -1.729(1)   &   6.905(-1)  \\
3p0Ax1	      &   2.414(1)   &  -1.710(1)   &   4.692(-1)  &   2.669(1)   &  -1.722(1)   &   3.894(-1)  &   2.669(1)   &  -1.738(1)   &   4.142(-1)  \\
3p0Ax2	      &   2.199(1)   &  -1.730(1)   &   6.948(-1)  &   2.338(1)   &  -1.745(1)   &   6.082(-1)  &   2.315(1)   &  -1.772(1)   &   6.741(-1)  \\
3p0Bx2	      &   1.884(1)   &  -1.749(1)   &   7.121(-1)  &   2.398(1)   &  -1.755(1)   &   9.063(-1)  &   2.287(1)   &  -1.791(1)   &   9.108(-1)  \\
3p65Cx1	      &   3.020(1)   &  -1.685(1)   &   4.008(-1)  &   3.154(1)   &  -1.708(1)   &   3.562(-1)  &   3.050(1)   &  -1.717(1)   &   3.897(-1)  \\
3p65Cx2	      &   2.598(1)   &  -1.696(1)   &   5.156(-1)  &   2.731(1)   &  -1.719(1)   &   5.484(-1)  &   2.687(1)   &  -1.729(1)   &   5.315(-1)  \\
3p65Ax1	      &   2.751(1)   &  -1.731(1)   &   4.328(-1)  &   3.041(1)   &  -1.751(1)   &   2.992(-1)  &   2.862(1)   &  -1.764(1)   &   4.406(-1)  \\
3p65Ax2	      &   2.508(1)   &  -1.748(1)   &   4.589(-1)  &   2.661(1)   &  -1.766(1)   &   4.437(-1)  &   2.610(1)   &  -1.784(1)   &   5.417(-1)  \\
3p65Bx2	      &   2.156(1)   &  -1.778(1)   &   5.317(-1)  &   2.289(1)   &  -1.782(1)   &   4.064(-1)  &   2.337(1)   &  -1.812(1)   &   5.785(-1)  \\
3p65Dx2	      &   1.680(1)   &  -1.789(1)   &   6.537(-1)  &   2.584(1)   &  -1.800(1)   &   1.342(0)   &   2.460(1)   &  -1.818(1)   &   1.322(0)   \\
4p64Cx1	      &   3.424(1)   &  -1.673(1)   &   3.000(-1)  &   3.459(1)   &  -1.699(1)   &   2.334(-1)  &   3.351(1)   &  -1.701(1)   &   2.975(-1)  \\
4p64Ax1	      &   2.945(1)   &  -1.723(1)   &   4.045(-1)  &   3.075(1)   &  -1.749(1)   &   2.848(-1)  &   2.902(1)   &  -1.755(1)   &   4.206(-1)  \\
4p64Ax2	      &   2.730(1)   &  -1.744(1)   &   5.186(-1)  &   2.855(1)   &  -1.766(1)   &   4.412(-1)  &   2.744(1)   &  -1.776(1)   &   5.659(-1)  \\
4p64Bx1	      &   2.394(1)   &  -1.756(1)   &   4.317(-1)  &   2.758(1)   &  -1.779(1)   &   3.898(-1)  &   2.439(1)   &  -1.787(1)   &   4.550(-1)  \\
4p64Bx2	      &   2.310(1)   &  -1.779(1)   &   6.013(-1)  &   2.534(1)   &  -1.798(1)   &   5.722(-1)  &   2.398(1)   &  -1.814(1)   &   6.702(-1)  \\
4p64Dx2	      &   1.828(1)   &  -1.801(1)   &   6.921(-1)  &   2.262(1)   &  -1.812(1)   &   7.200(-1)  &   2.011(1)   &  -1.835(1)   &   7.650(-1)  \\
5p11Ax1	      &   3.786(1)   &  -1.746(1)   &   3.279(-1)  &   3.977(1)   &  -1.772(1)   &   2.557(-1)  &   3.668(1)   &  -1.768(1)   &   3.500(-1)  \\
5p11Ax2	      &   3.324(1)   &  -1.759(1)   &   3.149(-1)  &   3.373(1)   &  -1.788(1)   &   2.741(-1)  &   3.241(1)   &  -1.781(1)   &   3.361(-1)  \\
5p11Bx1	      &   3.046(1)   &  -1.803(1)   &   3.511(-1)  &   3.766(1)   &  -1.824(1)   &   2.917(-1)  &   3.351(1)   &  -1.823(1)   &   3.778(-1)  \\
5p11Bx2	      &   3.080(1)   &  -1.815(1)   &   3.231(-1)  &   3.388(1)   &  -1.845(1)   &   3.113(-1)  &   3.167(1)   &  -1.839(1)   &   3.723(-1)  \\
5p1Gx2	      &   2.023(1)   &  -1.750(1)   &   6.498(-1)  &   2.161(1)   &  -1.761(1)   &   7.059(-1)  &   2.023(1)   &  -1.764(1)   &   6.957(-1)  \\
6p5Ax1	      &   4.655(1)   &  -1.752(1)   &   2.208(-1)  &   4.769(1)   &  -1.777(1)   &   1.514(-1)  &   4.610(1)   &  -1.776(1)   &   2.127(-1)  \\
6p5Ax2	      &   3.882(1)   &  -1.761(1)   &   2.103(-1)  &   3.844(1)   &  -1.785(1)   &   1.797(-1)  &   3.729(1)   &  -1.789(1)   &   2.093(-1)  \\
6p5Bx1	      &   4.112(1)   &  -1.824(1)   &   2.451(-1)  &   4.581(1)   &  -1.850(1)   &   1.937(-1)  &   4.249(1)   &  -1.850(1)   &   2.555(-1)  \\
6p5Bx2	      &   3.804(1)   &  -1.836(1)   &   2.390(-1)  &   4.205(1)   &  -1.860(1)   &   2.350(-1)  &   3.840(1)   &  -1.867(1)   &   2.675(-1)  \\
6p5Gx1	      &   2.877(1)   &  -1.839(1)   &   3.291(-1)  &   3.567(1)   &  -1.866(1)   &   3.206(-1)  &   3.290(1)   &  -1.865(1)   &   3.210(-1)  \\
6p5Gx2        &   2.868(1)   &  -1.858(1)   &   2.786(-1)  &   3.420(1)   &  -1.881(1)   &   3.518(-1)  &   3.125(1)   &  -1.891(1)   &   3.571(-1)  \\
\hline
\end{tabular}
\end{center}
\end{table*}

\label{lastpage}

\section{Expression of the bolometric luminosity}
\label{appendix_valenti}

In Section~\ref{sect_arnett}, we plot the bolometric luminosity
from the Arnett model and compare it to the \cmfgen\ results.
\citet{arnett_82} provides an expression for the bolometric luminosity
that includes the contribution from \iso{56}Ni alone.
\citet{valenti_08_03jd} extend this expression to also treat \iso{56}Co decay,
but we find two errors in their expressions. Below, we provide the various terms
entering the expression that we use for the bolometric luminosity in the Arnett model.

For the mean lifetimes  and decay energies of \iso{56}Ni and \iso{56}Co, we use \citep{nadyozhin_94,valenti_08_03jd}:
\begin{equation}
    \tau_{\rm Ni} = 6.0749 / \log{2}  \,\,\, {\rm d} \, , \\
\end{equation}
\begin{equation}
    \tau_{\rm Co} = 77.233 / \log{2}  \,\,\, {\rm d} \, , \\
\end{equation}
\begin{equation}
    \epsilon_{\rm Ni} = 3.9 \times\ 10^{10}\,{\rm erg}\,{\rm g}^{-1}\,{\rm s}^{-1} \, ,  \\
\end{equation}
\begin{equation}
    \epsilon_{\rm Co} = 6.78 \times 10^9 \,{\rm erg}\,{\rm g}^{-1}\,{\rm s}^{-1} \, ,  \\
\end{equation}

\citet{arnett_82} defines the time scale $\tau_{\rm m}$ as
\begin{equation}
    \tau_{\rm m} = \left( {\frac{\kappa_{\rm opt}}{\beta c}} \sqrt{\frac{10 M_{\rm e}^3}{3E_{\rm kin}}} \right)^{1/2}   \\
\end{equation}
and from \citet{valenti_08_03jd} we use:
\begin{equation}
    \beta =13.8 . \\
\end{equation}

Using Equation~31 of \citet{arnett_82}:
\begin{equation}
   \Lambda(x,y) = \exp(-x^2) \int_0^x   2 z \exp(-2zy + z^2) dz  \,\, ,\\
\end{equation}
and the following definitions for $x$, $y$, $s$, and $w$,
\begin{equation}
    x = t / \tau_{\rm m}  \,\, , \\
\end{equation}
\begin{equation}
    y = 0.5 \tau_{\rm m} / \tau_{\rm Ni} \,\, ,\\
\end{equation}
\begin{equation}
    s = 0.5 \tau_{\rm m} / \tau_{\rm Co}\,\, ,  \\
\end{equation}
\begin{equation}
    w = \tau_{\rm Co} / (\tau_{\rm Co} - \tau_{\rm Ni}) \,\, ,\\
\end{equation}
we find that the bolometric luminosity in the Arnett model is:
\begin{equation}
     L_{\rm bol}(t) = M_{\rm Ni} \left( \left[\epsilon_{\rm Ni} - \omega \epsilon_{\rm Co}\right] \Lambda(x,y)  + \omega  \epsilon_{\rm Co} \Lambda(x,s) \right)  \,\, ,
\end{equation}
where $M_{\rm Ni}$ is the initial \iso{56}Ni mass. This expression assumes full $\gamma$-ray trapping.


\begin{thebibliography}{76}
\expandafter\ifx\csname natexlab\endcsname\relax\def\natexlab#1{#1}\fi

\bibitem[{{Anderson} {et~al.}(2010){Anderson}, {Covarrubias}, {James}, {Hamuy},
  \& {Habergham}}]{anderson_etal_10}
{Anderson}, J.~P., {Covarrubias}, R.~A., {James}, P.~A., {Hamuy}, M., \&
  {Habergham}, S.~M. 2010, \mnras, 407, 2660

\bibitem[{{Anderson} {et~al.}(2012){Anderson}, {Habergham}, {James}, \&
  {Hamuy}}]{anderson_etal_12}
{Anderson}, J.~P., {Habergham}, S.~M., {James}, P.~A., \& {Hamuy}, M. 2012,
  \mnras, 424, 1372

\bibitem[{{Anderson} \& {James}(2008)}]{anderson_james_08}
{Anderson}, J.~P. \& {James}, P.~A. 2008, \mnras, 390, 1527

\bibitem[{{Anderson} \& {James}(2009)}]{anderson_james_09}
---. 2009, \mnras, 399, 559

\bibitem[{{Arcavi} {et~al.}(2010){Arcavi}, {Gal-Yam}, {Kasliwal}, {Quimby},
  {Ofek}, {Kulkarni}, {Nugent}, {Cenko}, {Bloom}, {Sullivan}, {Howell},
  {Poznanski}, {Filippenko}, {Law}, {Hook}, {J{\"o}nsson}, {Blake}, {Cooke},
  {Dekany}, {Rahmer}, {Hale}, {Smith}, {Zolkower}, {Velur}, {Walters},
  {Henning}, {Bui}, {McKenna}, \& {Jacobsen}}]{arcavi+10}
{Arcavi}, I., {Gal-Yam}, A., {Kasliwal}, M.~M., {Quimby}, R.~M., {Ofek}, E.~O.,
  {Kulkarni}, S.~R., {Nugent}, P.~E., {Cenko}, S.~B., {Bloom}, J.~S.,
  {Sullivan}, M., {Howell}, D.~A., {Poznanski}, D., {Filippenko}, A.~V., {Law},
  N., {Hook}, I., {J{\"o}nsson}, J., {Blake}, S., {Cooke}, J., {Dekany}, R.,
  {Rahmer}, G., {Hale}, D., {Smith}, R., {Zolkower}, J., {Velur}, V.,
  {Walters}, R., {Henning}, J., {Bui}, K., {McKenna}, D., \& {Jacobsen}, J.
  2010, \apj, 721, 777

\bibitem[{{Arnett}(1982)}]{arnett_82}
{Arnett}, W.~D. 1982, \apj, 253, 785

\bibitem[{{Bautista}(2004)}]{2004A&A...420..763B}
{Bautista}, M.~A. 2004, \aap, 420, 763

\bibitem[{{Bersten} {et~al.}(2012){Bersten}, {Benvenuto}, {Nomoto}, {Ergon},
  {Folatelli}, {Sollerman}, {Benetti}, {Botticella}, {Fraser}, {Kotak},
  {Maeda}, {Ochner}, \& {Tomasella}}]{bersten_etal_12_11dh}
{Bersten}, M.~C., {Benvenuto}, O.~G., {Nomoto}, K., {Ergon}, M., {Folatelli},
  G., {Sollerman}, J., {Benetti}, S., {Botticella}, M.~T., {Fraser}, M.,
  {Kotak}, R., {Maeda}, K., {Ochner}, P., \& {Tomasella}, L. 2012, \apj, 757,
  31

\bibitem[{{Bianco} {et~al.}(2014){Bianco}, {Modjaz}, {Hicken}, {Friedman},
  {Kirshner}, {Bloom}, {Challis}, {Marion}, {Wood-Vasey}, \&
  {Rest}}]{bianco_etal_14}
{Bianco}, F.~B., {Modjaz}, M., {Hicken}, M., {Friedman}, A., {Kirshner}, R.~P.,
  {Bloom}, J.~S., {Challis}, P., {Marion}, G.~H., {Wood-Vasey}, W.~M., \&
  {Rest}, A. 2014, \apjs, 213, 19

\bibitem[{{Blinnikov} {et~al.}(1998){Blinnikov}, {Eastman}, {Bartunov},
  {Popolitov}, \& {Woosley}}]{blinnikov_94_93j}
{Blinnikov}, S.~I., {Eastman}, R., {Bartunov}, O.~S., {Popolitov}, V.~A., \&
  {Woosley}, S.~E. 1998, \apj, 496, 454

\bibitem[{{Blondin} {et~al.}(2013){Blondin}, {Dessart}, {Hillier}, \&
  {Khokhlov}}]{blondin_etal_13}
{Blondin}, S., {Dessart}, L., {Hillier}, D.~J., \& {Khokhlov}, A.~M. 2013,
  \mnras, 429, 2127

\bibitem[{{Busche} \& {Hillier}(2005)}]{BH05_2D}
{Busche}, J.~R. \& {Hillier}, D.~J. 2005, \aj, 129, 454

\bibitem[{{Butler} {et~al.}(1993){Butler}, {Mendoza}, \&
  {Zeippen}}]{BKM93_Mg_seq}
{Butler}, K., {Mendoza}, C., \& {Zeippen}, C.~J. 1993, Journal of Physics B
  Atomic Molecular Physics, 26, 4409

\bibitem[{{Claeys} {et~al.}(2011){Claeys}, {de Mink}, {Pols}, {Eldridge}, \&
  {Baes}}]{claeys+13}
{Claeys}, J.~S.~W., {de Mink}, S.~E., {Pols}, O.~R., {Eldridge}, J.~J., \&
  {Baes}, M. 2011, \aap, 528, A131

\bibitem[{{Crowther}(2013)}]{crowther_13}
{Crowther}, P.~A. 2013, \mnras, 428, 1927

\bibitem[{{Cunto} {et~al.}(1993){Cunto}, {Mendoza}, {Ochsenbein}, \&
  {Zeippen}}]{Topbase93}
{Cunto}, W., {Mendoza}, C., {Ochsenbein}, F., \& {Zeippen}, C.~J. 1993, \aap,
  275, L5

\bibitem[{{Dessart} \& {Hillier}(2010)}]{dh10_87a}
{Dessart}, L. \& {Hillier}, D.~J. 2010, \mnras, 405, 2141

\bibitem[{{Dessart} {et~al.}(2014){Dessart}, {Hillier}, {Blondin}, \&
  {Khokhlov}}]{d14_tech}
{Dessart}, L., {Hillier}, D.~J., {Blondin}, S., \& {Khokhlov}, A. 2014, \mnras,
  441, 3249

\bibitem[{{Dessart} {et~al.}(2012){Dessart}, {Hillier}, {Li}, \&
  {Woosley}}]{d12_snibc}
{Dessart}, L., {Hillier}, D.~J., {Li}, C., \& {Woosley}, S. 2012, \mnras, 424,
  2139

\bibitem[{{Dessart} {et~al.}(2011){Dessart}, {Hillier}, {Livne}, {Yoon},
  {Woosley}, {Waldman}, \& {Langer}}]{dessart_11_wr}
{Dessart}, L., {Hillier}, D.~J., {Livne}, E., {Yoon}, S.-C., {Woosley}, S.,
  {Waldman}, R., \& {Langer}, N. 2011, \mnras, 414, 2985

\bibitem[{{Dessart} {et~al.}(2015){Dessart}, {Hillier}, {Woosley}, {Livne},
  {Waldman}, {Yoon}, \& {Langer}}]{D15_SNIbc_I}
{Dessart}, L., {Hillier}, D.~J., {Woosley}, S., {Livne}, E., {Waldman}, R.,
  {Yoon}, S.-C., \& {Langer}, N. 2015, \mnras, 453, 2189

\bibitem[{{Drout} {et~al.}(2011){Drout}, {Soderberg}, {Gal-Yam}, {Cenko},
  {Fox}, {Leonard}, {Sand}, {Moon}, {Arcavi}, \& {Green}}]{drout_etal_11}
{Drout}, M.~R., {Soderberg}, A.~M., {Gal-Yam}, A., {Cenko}, S.~B., {Fox},
  D.~B., {Leonard}, D.~C., {Sand}, D.~J., {Moon}, D.-S., {Arcavi}, I., \&
  {Green}, Y. 2011, \apj, 741, 97

\bibitem[{{Eldridge} {et~al.}(2013){Eldridge}, {Fraser}, {Smartt}, {Maund}, \&
  {Crockett}}]{eldridge_presn_13}
{Eldridge}, J.~J., {Fraser}, M., {Smartt}, S.~J., {Maund}, J.~R., \&
  {Crockett}, R.~M. 2013, \mnras, 436, 774

\bibitem[{{Eldridge} {et~al.}(2008){Eldridge}, {Izzard}, \&
  {Tout}}]{eldridge_etal_08}
{Eldridge}, J.~J., {Izzard}, R.~G., \& {Tout}, C.~A. 2008, \mnras, 384, 1109

\bibitem[{{Ensman} \& {Woosley}(1988)}]{ensman_woosley_88}
{Ensman}, L.~M. \& {Woosley}, S.~E. 1988, \apj, 333, 754

\bibitem[{{Ergon} {et~al.}(2014){Ergon}, {Sollerman}, {Fraser}, {Pastorello},
  {Taubenberger}, {Elias-Rosa}, {Bersten}, {Jerkstrand}, {Benetti},
  {Botticella}, {Fransson}, {Harutyunyan}, {Kotak}, {Smartt}, {Valenti},
  {Bufano}, {Cappellaro}, {Fiaschi}, {Howell}, {Kankare}, {Magill}, {Mattila},
  {Maund}, {Naves}, {Ochner}, {Ruiz}, {Smith}, {Tomasella}, \&
  {Turatto}}]{ergon_14_11dh}
{Ergon}, M., {Sollerman}, J., {Fraser}, M., {Pastorello}, A., {Taubenberger},
  S., {Elias-Rosa}, N., {Bersten}, M., {Jerkstrand}, A., {Benetti}, S.,
  {Botticella}, M.~T., {Fransson}, C., {Harutyunyan}, A., {Kotak}, R.,
  {Smartt}, S., {Valenti}, S., {Bufano}, F., {Cappellaro}, E., {Fiaschi}, M.,
  {Howell}, A., {Kankare}, E., {Magill}, L., {Mattila}, S., {Maund}, J.,
  {Naves}, R., {Ochner}, P., {Ruiz}, J., {Smith}, K., {Tomasella}, L., \&
  {Turatto}, M. 2014, \aap, 562, A17

\bibitem[{{Fernley} {et~al.}(1999){Fernley}, {Hibbert}, {Kingston}, \&
  {Seaton}}]{1999JPhB...32.5507F}
{Fernley}, J.~A., {Hibbert}, A., {Kingston}, A.~E., \& {Seaton}, M.~J. 1999,
  Journal of Physics B Atomic Molecular Physics, 32, 5507

\bibitem[{{Filippenko} {et~al.}(1990){Filippenko}, {Porter}, \&
  {Sargent}}]{filippenko_87M_90}
{Filippenko}, A.~V., {Porter}, A.~C., \& {Sargent}, W.~L.~W. 1990, \aj, 100,
  1575

\bibitem[{{Fryxell} {et~al.}(1991{\natexlab{a}}){Fryxell}, {Arnett}, \&
  {Mueller}}]{fryxell_mueller_arnett_91}
{Fryxell}, B., {Arnett}, D., \& {Mueller}, E. 1991{\natexlab{a}}, \apj, 367,
  619

\bibitem[{{Fryxell} {et~al.}(1991{\natexlab{b}}){Fryxell}, {Arnett}, \&
  {M\"{u}ller}}]{fryxell_etal_91}
{Fryxell}, B., {Arnett}, D., \& {M\"{u}ller}, E. 1991{\natexlab{b}}, \apj, 367,
  619

\bibitem[{{Groh} {et~al.}(2013){Groh}, {Meynet}, {Georgy}, \&
  {Ekstr{\"o}m}}]{groh_13_presn}
{Groh}, J.~H., {Meynet}, G., {Georgy}, C., \& {Ekstr{\"o}m}, S. 2013, \aap,
  558, A131

\bibitem[{{Harkness} {et~al.}(1987){Harkness}, {Wheeler}, {Margon}, {Downes},
  {Kirshner}, {Uomoto}, {Barker}, {Cochran}, {Dinerstein}, {Garnett}, \&
  {Levreault}}]{harkness_ib_87}
{Harkness}, R.~P., {Wheeler}, J.~C., {Margon}, B., {Downes}, R.~A., {Kirshner},
  R.~P., {Uomoto}, A., {Barker}, E.~S., {Cochran}, A.~L., {Dinerstein}, H.~L.,
  {Garnett}, D.~R., \& {Levreault}, R.~M. 1987, \apj, 317, 355

\bibitem[{{Hillier} \& {Dessart}(2012)}]{HD12}
{Hillier}, D.~J. \& {Dessart}, L. 2012, \mnras, 424, 252

\bibitem[{{Hillier} \& {Miller}(1998)}]{hm98}
{Hillier}, D.~J. \& {Miller}, D.~L. 1998, \apj, 496, 407

\bibitem[{{Hummer} {et~al.}(1993){Hummer}, {Berrington}, {Eissner}, {Pradhan},
  {Saraph}, \& {Tully}}]{HBE93_IP}
{Hummer}, D.~G., {Berrington}, K.~A., {Eissner}, W., {Pradhan}, A.~K.,
  {Saraph}, H.~E., \& {Tully}, J.~A. 1993, \aap, 279, 298

\bibitem[{{Janka} {et~al.}(2012){Janka}, {Hanke}, {H{\"u}depohl}, {Marek},
  {M{\"u}ller}, \& {Obergaulinger}}]{2012PTEP.2012aA309J}
{Janka}, H.-T., {Hanke}, F., {H{\"u}depohl}, L., {Marek}, A., {M{\"u}ller}, B.,
  \& {Obergaulinger}, M. 2012, Progress of Theoretical and Experimental
  Physics, 2012, 010000

\bibitem[{{Katz} {et~al.}(2013){Katz}, {Kushnir}, \& {Dong}}]{katz_13_56ni}
{Katz}, B., {Kushnir}, D., \& {Dong}, S. 2013, ArXiv:1301.6766

\bibitem[{{Kelly} \& {Kirshner}(2012)}]{kelly_kirshner_12}
{Kelly}, P.~L. \& {Kirshner}, R.~P. 2012, \apj, 759, 107

\bibitem[{{Kim} {et~al.}(2015){Kim}, {Yoon}, \& {Koo}}]{kim_snibc_prog}
{Kim}, H.-J., {Yoon}, S.-C., \& {Koo}, B.-C. 2015, \apj, 809, 131

\bibitem[{{Kramida} {et~al.}(2012){Kramida}, {Ralchenko}, {Reader}, \& {NIST
  ASD Team}}]{NIST_V5}
{Kramida}, A.~E., {Ralchenko}, Y., {Reader}, J., \& {NIST ASD Team}. 2012,
  {NIST Atomic Spectra Database (version 5.0)}

\bibitem[{{Kurucz}(2009)}]{Kur09_ATD}
{Kurucz}, R.~L. 2009, in American Institute of Physics Conference Series, Vol.
  1171, American Institute of Physics Conference Series, ed. {I.~Hubeny,
  J.~M.~Stone, K.~MacGregor, \& K.~Werner}, 43--51

\bibitem[{{Kurucz}(2010)}]{Kur_web}
{Kurucz}, R.~L. 2010

\bibitem[{{Liu} {et~al.}(2015){Liu}, {Modjaz}, {Bianco}, \&
  {Graur}}]{liu_snibc_15}
{Liu}, Y.-Q., {Modjaz}, M., {Bianco}, F.~B., \& {Graur}, O. 2015,
  ArXiv:1510.08049

\bibitem[{{Lucy}(1991)}]{lucy_91}
{Lucy}, L.~B. 1991, \apj, 383, 308

\bibitem[{{Luo} \& {Pradhan}(1989)}]{LP89_C_seq}
{Luo}, D. \& {Pradhan}, A.~K. 1989, Journal of Physics B Atomic Molecular
  Physics, 22, 3377

\bibitem[{{Mendoza}(1983)}]{Men83_col}
{Mendoza}, C. 1983, in IAU Symposium, Vol. 103, Planetary Nebulae, ed.
  {D.~R.~Flower}, 143--172

\bibitem[{{Mendoza} {et~al.}(1995){Mendoza}, {Eissner}, {LeDourneuf}, \&
  {Zeippen}}]{MEL95_Al_seq}
{Mendoza}, C., {Eissner}, W., {LeDourneuf}, M., \& {Zeippen}, C.~J. 1995,
  Journal of Physics B Atomic Molecular Physics, 28, 3485

\bibitem[{{Modjaz} {et~al.}(2011){Modjaz}, {Kewley}, {Bloom}, {Filippenko},
  {Perley}, \& {Silverman}}]{modjaz_etal_11}
{Modjaz}, M., {Kewley}, L., {Bloom}, J.~S., {Filippenko}, A.~V., {Perley}, D.,
  \& {Silverman}, J.~M. 2011, \apjl, 731, L4

\bibitem[{{Nadyozhin}(1994)}]{nadyozhin_94}
{Nadyozhin}, D.~K. 1994, \apjs, 92, 527

\bibitem[{{Nahar} \& {Pradhan}(1993)}]{NP93_Sil}
{Nahar}, S.~N. \& {Pradhan}, A.~K. 1993, Journal of Physics B Atomic Molecular
  Physics, 26, 1109

\bibitem[{{Nahar} \& {Pradhan}(1994)}]{NP94_FeII_phot}
---. 1994, Journal of Physics B Atomic Molecular Physics, 27, 429

\bibitem[{{Nussbaumer} \& {Storey}(1983)}]{NS83_LTDR}
{Nussbaumer}, H. \& {Storey}, P.~J. 1983, \aap, 126, 75

\bibitem[{{Nussbaumer} \& {Storey}(1984)}]{NS84_CNO_LTDR}
---. 1984, \aaps, 56, 293

\bibitem[{{Podsiadlowski} {et~al.}(1992){Podsiadlowski}, {Joss}, \&
  {Hsu}}]{podsiadlowski_92}
{Podsiadlowski}, P., {Joss}, P.~C., \& {Hsu}, J.~J.~L. 1992, \apj, 391, 246

\bibitem[{{Ralchenko} {et~al.}(2010){Ralchenko}, {Kramida}, {Reader}, \& {NIST
  ASD Team}}]{NIST}
{Ralchenko}, Y., {Kramida}, A.~E., {Reader}, I., \& {NIST ASD Team}. 2010, NIST
  Atomic Spectra Database (version 3.1.5)

\bibitem[{{Sanders} {et~al.}(2012){Sanders}, {Soderberg}, {Levesque}, {Foley},
  {Chornock}, {Milisavljevic}, {Margutti}, {Berger}, {Drout}, {Czekala}, \&
  {Dittmann}}]{sanders_etal_12}
{Sanders}, N.~E., {Soderberg}, A.~M., {Levesque}, E.~M., {Foley}, R.~J.,
  {Chornock}, R., {Milisavljevic}, D., {Margutti}, R., {Berger}, E., {Drout},
  M.~R., {Czekala}, I., \& {Dittmann}, J.~A. 2012, \apj, 758, 132

\bibitem[{{Sauer} {et~al.}(2006){Sauer}, {Mazzali}, {Deng}, {Valenti},
  {Nomoto}, \& {Filippenko}}]{sauer_etal_06}
{Sauer}, D.~N., {Mazzali}, P.~A., {Deng}, J., {Valenti}, S., {Nomoto}, K., \&
  {Filippenko}, A.~V. 2006, \mnras, 369, 1939

\bibitem[{{Seaton}(1987)}]{Sea87_OP}
{Seaton}, M.~J. 1987, Journal of Physics B Atomic Molecular Physics, 20, 6363

\bibitem[{{Shine} \& {Linsky}(1974)}]{SL74}
{Shine}, R.~A. \& {Linsky}, J.~L. 1974, \solphys, 39, 49

\bibitem[{{Smith} {et~al.}(2011){Smith}, {Li}, {Filippenko}, \&
  {Chornock}}]{smith_etal_11}
{Smith}, N., {Li}, W., {Filippenko}, A.~V., \& {Chornock}, R. 2011, \mnras,
  412, 1522

\bibitem[{{Sukhbold} {et~al.}(2015){Sukhbold}, {Ertl}, {Woosley}, {Brown}, \&
  {Janka}}]{tuguldur_ccsn_15}
{Sukhbold}, T., {Ertl}, T., {Woosley}, S.~E., {Brown}, J.~M., \& {Janka}, H.-T.
  2015, ArXiv:1510.04643

\bibitem[{{Sunderland} {et~al.}(2002){Sunderland}, {Noble}, {Burke}, \&
  {Burke}}]{2002CoPhC.145..311S}
{Sunderland}, A.~G., {Noble}, C.~J., {Burke}, V.~M., \& {Burke}, P.~G. 2002,
  Computer Physics Communications, 145, 311

\bibitem[{{Tachiev} \& {Froese Fischer}(2002)}]{2002A&A...385..716T}
{Tachiev}, G.~I. \& {Froese Fischer}, C. 2002, \aap, 385, 716

\bibitem[{{Ugliano} {et~al.}(2012){Ugliano}, {Janka}, {Marek}, \&
  {Arcones}}]{ugliano_ccsn_12}
{Ugliano}, M., {Janka}, H.-T., {Marek}, A., \& {Arcones}, A. 2012, \apj, 757,
  69

\bibitem[{{Valenti} {et~al.}(2008){Valenti}, {Benetti}, {Cappellaro}, {Patat},
  {Mazzali}, {Turatto}, {Hurley}, {Maeda}, {Gal-Yam}, {Foley}, {Filippenko},
  {Pastorello}, {Challis}, {Frontera}, {Harutyunyan}, {Iye}, {Kawabata},
  {Kirshner}, {Li}, {Lipkin}, {Matheson}, {Nomoto}, {Ofek}, {Ohyama}, {Pian},
  {Poznanski}, {Salvo}, {Sauer}, {Schmidt}, {Soderberg}, \&
  {Zampieri}}]{valenti_08_03jd}
{Valenti}, S., {Benetti}, S., {Cappellaro}, E., {Patat}, F., {Mazzali}, P.,
  {Turatto}, M., {Hurley}, K., {Maeda}, K., {Gal-Yam}, A., {Foley}, R.~J.,
  {Filippenko}, A.~V., {Pastorello}, A., {Challis}, P., {Frontera}, F.,
  {Harutyunyan}, A., {Iye}, M., {Kawabata}, K., {Kirshner}, R.~P., {Li}, W.,
  {Lipkin}, Y.~M., {Matheson}, T., {Nomoto}, K., {Ofek}, E.~O., {Ohyama}, Y.,
  {Pian}, E., {Poznanski}, D., {Salvo}, M., {Sauer}, D.~N., {Schmidt}, B.~P.,
  {Soderberg}, A., \& {Zampieri}, L. 2008, \mnras, 383, 1485

\bibitem[{{van Regemorter}(1962)}]{Reg62_col}
{van Regemorter}, H. 1962, \apj, 136, 906

\bibitem[{{Wheeler} {et~al.}(1987){Wheeler}, {Harkness}, {Barker}, {Cochran},
  \& {Wills}}]{wheeler_ibc_87}
{Wheeler}, J.~C., {Harkness}, R.~P., {Barker}, E.~S., {Cochran}, A.~L., \&
  {Wills}, D. 1987, \apjl, 313, L69

\bibitem[{{Wheeler} \& {Levreault}(1985)}]{wheeler_levreault_85}
{Wheeler}, J.~C. \& {Levreault}, R. 1985, \apjl, 294, L17

\bibitem[{{Wongwathanarat} {et~al.}(2015){Wongwathanarat}, {Mueller}, \&
  {Janka}}]{wongwathanarat_15_3d}
{Wongwathanarat}, A., {Mueller}, E., \& {Janka}, H.-T. 2015, \aap, 577, A48

\bibitem[{{Woosley} {et~al.}(1995){Woosley}, {Langer}, \&
  {Weaver}}]{woosley_etal_95}
{Woosley}, S.~E., {Langer}, N., \& {Weaver}, T.~A. 1995, \apj, 448, 315

\bibitem[{{Yoon} {et~al.}(2010){Yoon}, {Woosley}, \& {Langer}}]{yoon_etal_10}
{Yoon}, S., {Woosley}, S.~E., \& {Langer}, N. 2010, \apj, 725, 940

\bibitem[{{Yoon} {et~al.}(2012){Yoon}, {Gr{\"a}fener}, {Vink}, {Kozyreva}, \&
  {Izzard}}]{yoon_presn_12}
{Yoon}, S.-C., {Gr{\"a}fener}, G., {Vink}, J.~S., {Kozyreva}, A., \& {Izzard},
  R.~G. 2012, \aap, 544, L11

\bibitem[{{Zhang}(1996)}]{Zha96_FeIII_col}
{Zhang}, H. 1996, \aaps, 119, 523

\bibitem[{{Zhang} \& {Pradhan}(1994)}]{ZP94_FeII_col}
{Zhang}, H.~L. \& {Pradhan}, A.~K. 1994, VizieR Online Data Catalog, 329, 30953

\bibitem[{{Zhang} \& {Pradhan}(1995{\natexlab{a}})}]{1995A&A...293..953Z}
---. 1995{\natexlab{a}}, \aap, 293

\bibitem[{{Zhang} \& {Pradhan}(1995{\natexlab{b}})}]{ZP95_FeIII_col}
---. 1995{\natexlab{b}}, Journal of Physics B Atomic Molecular Physics, 28,
  3403

\end{thebibliography}
\end{document}